\documentclass[lettersize,9pt]{article}
\usepackage[margin=1in]{geometry}
\usepackage{subfigure}
\usepackage{amsfonts}
\usepackage{graphicx}
\usepackage{epstopdf}
\usepackage{color}
\usepackage{algorithmic}
\usepackage[title]{appendix}

\usepackage{amsmath}
\usepackage{amssymb}
\graphicspath{{figures/}}

\renewcommand\theequation{\thesection.\arabic{equation}}

\ifpdf
  \DeclareGraphicsExtensions{.eps,.pdf,.png,.jpg}
\else
  \DeclareGraphicsExtensions{.eps}
\fi

\newtheorem{remark}{Remark}[section]

\renewcommand{\v}[1]{\mathbf{#1}}

\newcommand{\eps}{\varepsilon}
\newcommand{\mc}{\mathcal}
\newcommand{\mrm}{\mathrm}
\newcommand{\R}{\mathbb{R}}

\newcommand{\ubar}{\bar{u}}
\newcommand{\qdisk}{\Omega_+}

\newcommand{\wtwoh}{\mathbf{W}_{2H}}

\title{Weakly Nonlinear Analysis of Peanut-Shaped Deformations for
  Localized Spots of Singularly Perturbed Reaction-Diffusion Systems}
\author{ Tony Wong\thanks{Dept. of Mathematics, Univ. of British
    Columbia, Vancouver, B.C., Canada.} \and Michael
  J. Ward\footnotemark[1]\,\, \thanks{corresponding author,
    \texttt{ward@math.ubc.ca}}}

\begin{document}

\maketitle

\begin{abstract}
  Spatially localized 2-D spot patterns occur for a wide variety of
  two component reaction-diffusion systems in the singular limit of a
  large diffusivity ratio. Such localized, far-from-equilibrium,
  patterns are known to exhibit a wide range of different
  instabilities such as breathing oscillations, spot annihilation, and
  spot self-replication behavior. Prior numerical simulations of the
  Schnakenberg and Brusselator systems have suggested that a localized
  peanut-shaped linear instability of a localized spot is the
  mechanism initiating a fully nonlinear spot self-replication event.
  From a development and implementation of a weakly nonlinear theory
  for shape deformations of a localized spot, it is shown through a
  normal form amplitude equation that a peanut-shaped linear
  instability of a steady-state spot solution is always subcritical
  for both the Schnakenberg and Brusselator reaction-diffusion
  systems. The weakly nonlinear theory is validated by using the
  global bifurcation software {\em pde2path} [H.~Uecker et al.,
  Numerical Mathematics: Theory, Methods and Applications, {\bf 7}(1),
  (2014)] to numerically compute an unstable, non-radially symmetric,
  steady-state spot solution branch that originates from a
  symmetry-breaking bifurcation point.
\end{abstract}



\section{Introduction}\label{sec:intro}

Spatially localized patterns arise in a diverse range of applications
including, the ferrocyanide-iodate-sulphite (FIS) reaction
(cf.~\cite{LMPS}, \cite{ls}), the chloride-dioxide-malonic acid
reaction (cf.~\cite{blobs}), certain electronic gas discharge systems
\cite{as1}, fluid-convection phenomena \cite{K}, and the emergence of
plant root hair cells mediated by the plant hormone auxin
(cf.~\cite{brena}), among others. One qualitatively novel feature in
many of these settings is the observation that spatially localized
spot-type patterns can undergo a seemingly spontaneous
self-replication process.  

Many of these observed localized patterns, most notably those in
chemical physics and biology, are modeled by nonlinear
reaction-diffusion (RD) systems. In \cite{pearson}, where the
two-component Gray-Scott RD model was used to qualitatively model the
FIS reaction, full PDE simulations revealed a wide variety of highly
complex spatio-temporal localized patterns including, self-replicating
spot patterns, stripe patterns, and labyrinthian space-filling curves
(see also \cite{rpd}, \cite{MO1} and \cite{MO2}). This numerical study
showed convincingly that in the fully nonlinear regime a two-component
RD system with seemingly very simple reaction kinetics can admit
highly intricate solution behavior, which cannot be described by a
conventional Turing stability analysis (cf.~\cite{turing}) of some
spatially uniform base state. For certain three-component RD
  systems in the limit of small diffusivity, Nishiura
  et.~al. (cf.~\cite{NTU}, \cite{TSN}) showed from PDE simulations and
  a weakly nonlinear bifurcation analysis that a subcritical
  peanut-shaped instability of a localized radially symmetric spot
  plays a key role in understanding the dynamics of traveling spot
  solutions.  These previous studies, partially motivated by the
  pioneering numerical study of \cite{pearson}, have provided the
  impetus for developing new theoretical approaches to analyze some of
  the novel dynamical behaviors and instabilities of localized
  patterns in RD systems in the ``far-from-equilibrium'' regime
  \cite{nishiura}. A survey of some novel phenomena and theoretical
approaches associated with localized pattern formation problems are
given in \cite{vanag}, \cite{nishiura} and \cite{K}. The main goal of
this paper is to use a weakly nonlinear analysis to study the onset of
spot self-replication for certain two-component RD systems in the
so-called ``semi-strong'' regime, characterized by a large diffusivity
ratio between the solution components.

The derivation of amplitude, or normal form, equations using a
multi-scale perturbation analysis is a standard approach for
characterizing the weakly nonlinear development of small amplitude
patterns near bifurcation points. It has been used with considerable
success in physical applications, such as in hydrodynamic stability
theory and materials science (cf.~\cite{ch}, \cite{w}) and in
biological and chemical modeling through RD
systems defined in planar spatial domains and on the sphere
(cf.~\cite{w}, \cite{M1}, \cite{M2}, \cite{callahan}). However, in
certain applications, the effectiveness of normal form theory is
limited owing to the existence of subcritical bifurcations
(cf.~\cite{callahan}) or the need for an extreme fine-tuning of the
model parameters in order to be within the range of validity of the
theory (cf.~\cite{wh}). In contrast to the relative ease in undertaking a
weakly nonlinear theory for an RD system near a Turing bifurcation
point of the linearization around a spatially uniform or patternless
state, it is considerably more challenging to implement such a theory
for spatially localized steady-state patterns. This is owing to the
fact that the linearization of the RD system around a spatially
localized spot solution leads to a singularly perturbed eigenvalue
problem in which the underlying linearized operator is spatially
heterogeneous.  In addition, various terms in the multi-scale
expansion that are needed to derive the amplitude equation involve
solving rather complicated spatially inhomogeneous boundary value
problems. In this direction, a weakly nonlinear analysis of temporal
amplitude oscillations (breathing instabilities) of 1-D spike patterns
was developed for a class of generalized Gierer-Meinhardt (GM) models in
\cite{veer} and for the Gray-Scott and Schnakenberg models in
\cite{gomez}. A criterion for whether these oscillations, emerging
from a Hopf bifurcation point of the linearization, are subcritical or
supercritical was derived. A related weakly nonlinear analysis for
competition instabilities of 1-D steady-state spike patterns for the
GM and Schnakenberg models, resulting from a zero-eigenvalue
crossing of the linearization, was developed in \cite{comp}. Finally,
for a class of coupled bulk-surface RD systems, a weakly nonlinear
analysis for Turing, Hopf, and codimension-two Turing-Hopf
bifurcations of a patterned base-state was derived in
\cite{bulk_memb}.

The focus of this paper is to develop and implement a weakly nonlinear
theory to analyze branching behavior associated with peanut-shaped
deformations of a locally radially symmetric steady-state spot
solution for certain singularly perturbed RD systems.  Previous
numerical simulations of the Schnakenberg and Brusselator RD systems
in \cite{KWW}, \cite{RRW} and \cite{TW2016} (see also \cite{TW}) have
indicated that a non-radially symmetric peanut-shape deformation of
the spot profile can, in certain cases, trigger a fully nonlinear spot
self-replication event. The parameter threshold for the onset of this
shape deformation linear instability has been calculated in \cite{KWW}
and \cite{RRW} for the Schnakenberg and Brusselator models,
respectively. We will extend this linear theory by using a multi-scale
perturbation approach to derive a normal form amplitude equation
characterizing the local branching behavior associated with
peanut-shaped instabilities of the spot profile. From a numerical
evaluation of the coefficients in this amplitude equation we will show
that a peanut-shaped instability of the spot profile is always
subcritical for both the Schnakenberg and Brusselator models. This
theoretical result supports the numerical findings in \cite{KWW},
\cite{RRW} and \cite{TW2016} that a peanut-shaped instability of a
localized spot is the trigger for a fully nonlinear spot-splitting
event, and it solves an open problem discussed in the survey article
\cite{ward}.

\begin{figure}[htbp]
\centering
 \includegraphics[width=0.24\textwidth]{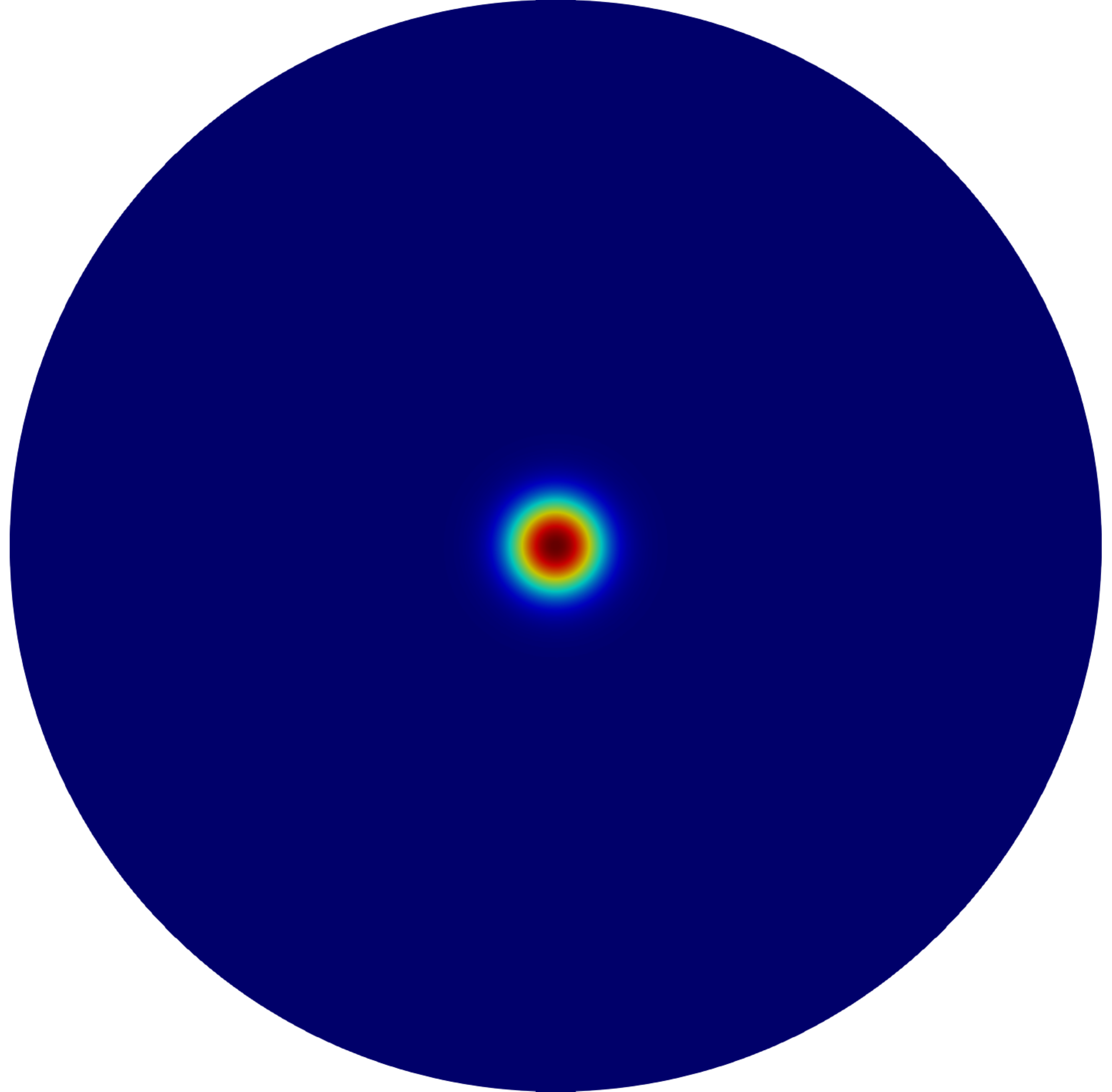} 
 \includegraphics[width=0.24\textwidth]{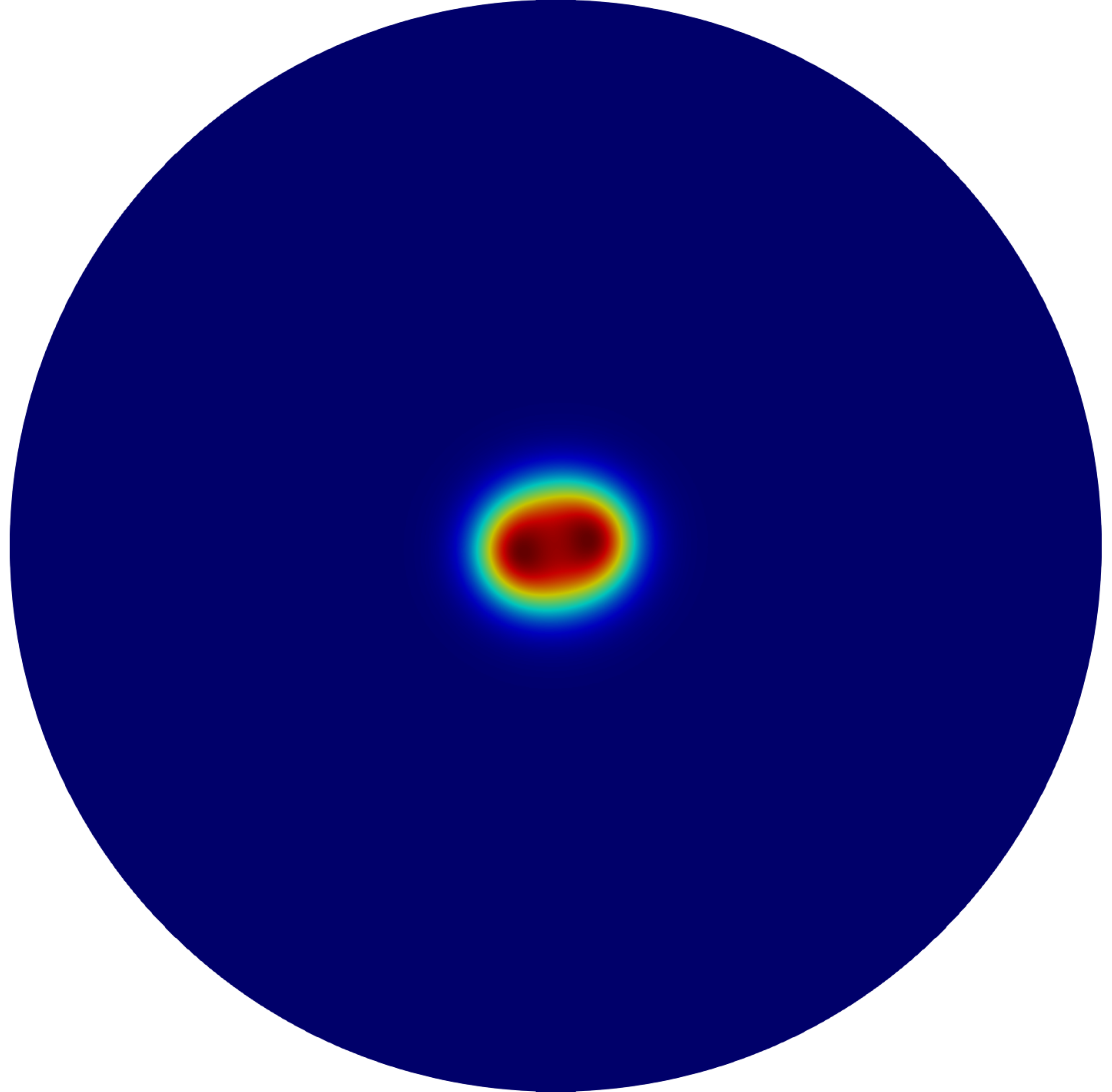}
 \includegraphics[width=0.24\textwidth]{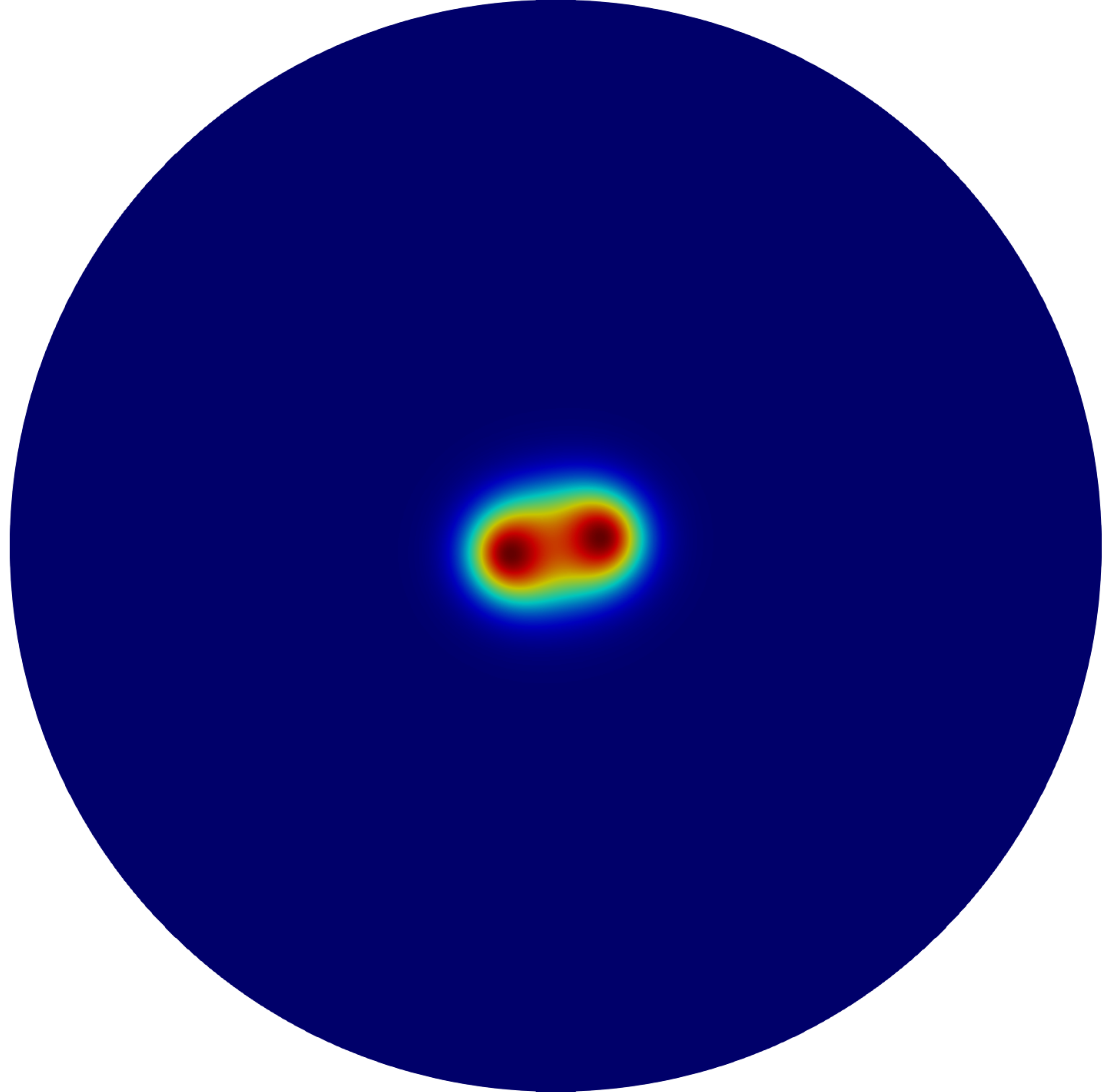}
 \includegraphics[width=0.24\textwidth]{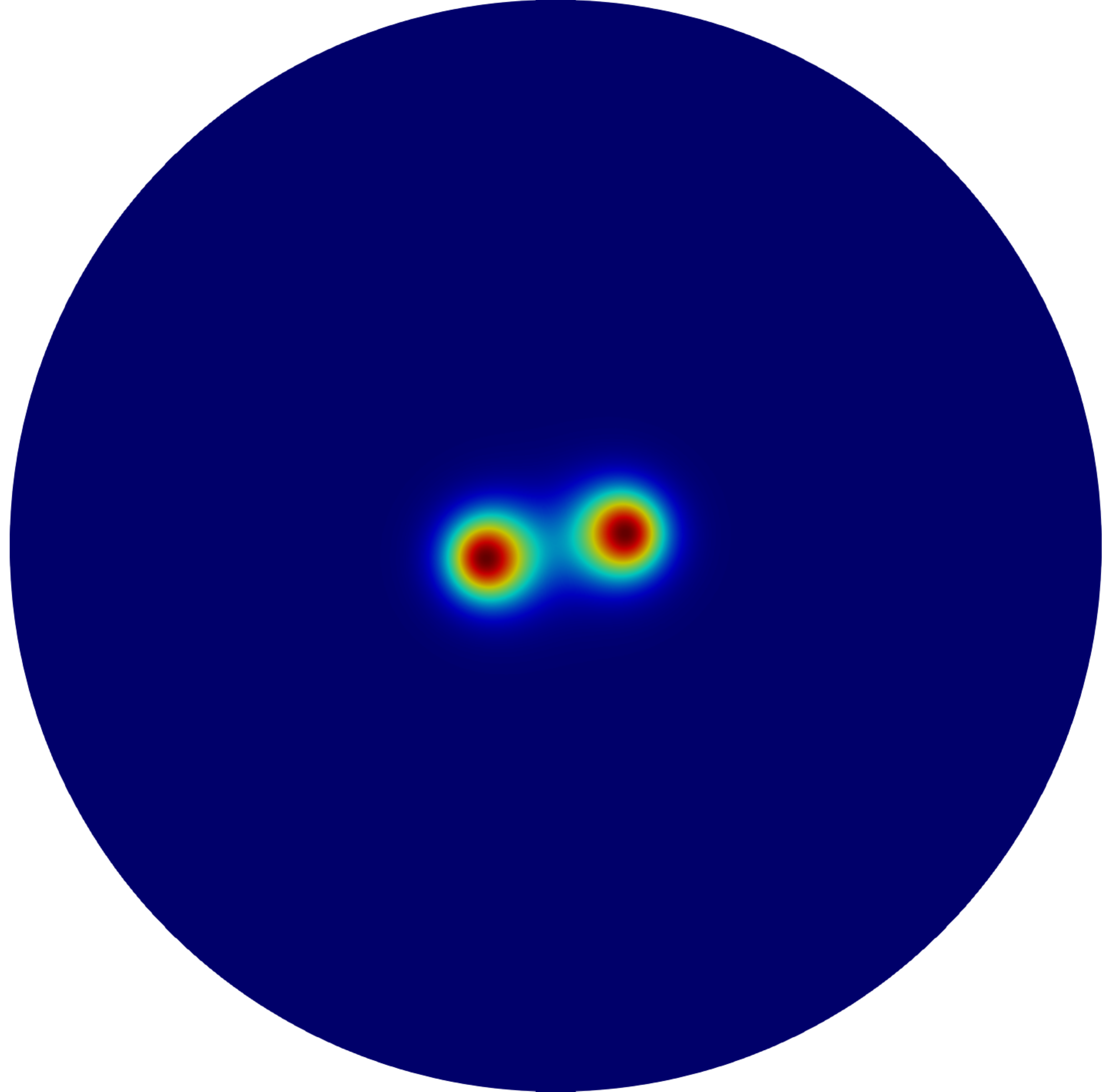}
 \caption{Contour plot of $v$ from a numerical solution of the
   Schnakenberg RD system \eqref{intro:pde} in the unit disk at four
   different times showing a spot self-replication event as the
   feed-rate $a$ is slowly increased past the peanut-shape instability
   threshold $a_c\approx 8.6$ of a localized spot. Parameters are
   $D = 1$, $\tau = 1$, $\eps = 0.03$ and
   $a = \mbox{min}(8.6 + 0.06 \,t, 10)$. Left: $t=2$.  Left-Middle:
   $t=68$. Right-middle: $t=74$. Right:
   $t=82$.}\label{fig:spot_replicate}
\end{figure}

The dimensionless Schnakenberg model in the two-dimensional unit disk
$\Omega = \{ \v{x} : |\v{x}| \leq 1 \}$ is formulated as
\begin{equation}\label{intro:pde}
v_t = \eps^2 \Delta v - v + u v^2, \quad \quad
\tau u_t = D \Delta u + a - \eps^{-2} uv^2, \quad \v{x} \in \Omega\,,
\end{equation}
with $\partial_n v = \partial_n u = 0$ on $\partial\Omega$. Here
$\eps\ll 1$, $D=\mc{O}(1)$, $\tau=\mc{O}(1)$, and the constant $a>0$
is called the feed-rate. For a spot centered at the origin of the
disk, the contour plot in Fig.~\ref{fig:spot_replicate} of $v$ at
different times, as computed numerically from \eqref{intro:pde}, shows
a spot self-replication event as the feed-rate $a$ is slowly ramped
above the threshold value $a_{c}\approx 8.6$. At this threshold value
of $a$ the spot profile becomes unstable to a peanut-shaped
deformation (see \S \ref{sec:linstab} for the linear stability
analysis).

Rigorous analytical results for the existence and linear stability of
localized spot patterns for the Schnakenberg model \eqref{intro:pde}
in the large $D$ regime $D=\mc{O}(\nu^{-1})$, where
$\nu={-1/\log\eps}$, are given in \cite{ww_schnak} and for the related
Gray-Scott model in \cite{wgs3} (see \cite{wei-book} for a survey of
such rigorous results). For the regime $D=\mc{O}(1)$, a hybrid
analytical-numerical approach, which has the effect of summing all
logarithmic terms in powers of $\nu$, was developed in \cite{KWW} to
construct quasi-equilibrium patterns, to analyze their linear stability
properties, and to characterize slow spot dynamics. An extension of this
hybrid methodology applied to other RD systems was given in \cite{CW2011},
\cite{RRW}, \cite{TW}, \cite{TW2016} and \cite{brena}, and is surveyed
in \cite{ward}.

We remark that the mechanism underlying the self-replication of 1-D
localized patterns is rather different than the more conventional
symmetry-breaking mechanism that occurs in 2-D. In a one-dimensional
domain, the self-replication behavior of spike patterns has been
interpreted in terms of a nearly-coinciding hierarchical saddle-node
global bifurcation structure of branches of multi-spike equilibria,
together with the existence of a dimple-shaped eigenfunction of the
linearization near the saddle-node point (see \cite{nu1}, \cite{enu},
\cite{dgk1}, \cite{kww_gm}, \cite{u}, \cite{kww_gs}, \cite{MO1},
\cite{rpd} and the references therein).

The outline of this paper is as follows.  For the Schnakenberg RD
model \eqref{intro:pde}, in \S \ref{sec:equil} we use the method of
matched asymptotic expansions to construct a steady-state, locally
radially symmetric, spot solution centered at the origin of the unit
disk. In \S \ref{sec:linstab} we perform a linear stability analysis
for non-radially symmetric perturbations of this localized
steady-state, and we numerically compute the threshold conditions for
the onset of a peanut-shaped instability of a localized spot. Although
much of this steady-state and linear stability theory has been
described previously in \cite{KWW}, it provides the required
background context for describing the new weakly nonlinear theory in
\S \ref{sec:amp}. More specifically, in \S \ref{sec:amp} we develop
and implement a weakly nonlinear analysis to characterize the
branching behavior associated with peanut-shaped instabilities of a
localized spot. From a numerical evaluation of the coefficients in the
resulting normal form amplitude equation we show that a peanut-shaped
deformation of a localized spot is subcritical. By using the
bifurcation software {\em pde2path} \cite{pde2path}, the weakly
nonlinear theory is validated in \S \ref{sec:validate} by numerically
computing an unstable non-radially symmetric steady-state spot
solution branch that emerges from the peanut-shaped linear stability
threshold of a locally radially symmetric spot solution. In \S
\ref{sec:bruss} we perform a similar multi-scale asymptotic reduction
to derive an amplitude equation characterizing the weakly nonlinear
development of peanut-shaped deformations of a localized spot for the
Brusselator RD model, originally introduced in \cite{PL}. From a
numerical evaluation of the coefficients in this amplitude equation,
which depend on a parameter in the Brusselator reaction-kinetics, it
is shown that peanut-shaped linear instabilities are always
subcritical. This theoretical result predicting subcriticality is
again validated using {\em pde2path} \cite{pde2path}. In \S
\ref{sec:disc} we summarize a few key qualitative features of our
hybrid analytical-numerical approach to derive the amplitude equation,
and we discuss a few possible extensions of this work.

\section{Asymptotic construction of steady state solution}\label{sec:equil}

\begin{figure}[htbp]
\begin{center}
\includegraphics[height=4.2cm,width=0.5\textwidth]{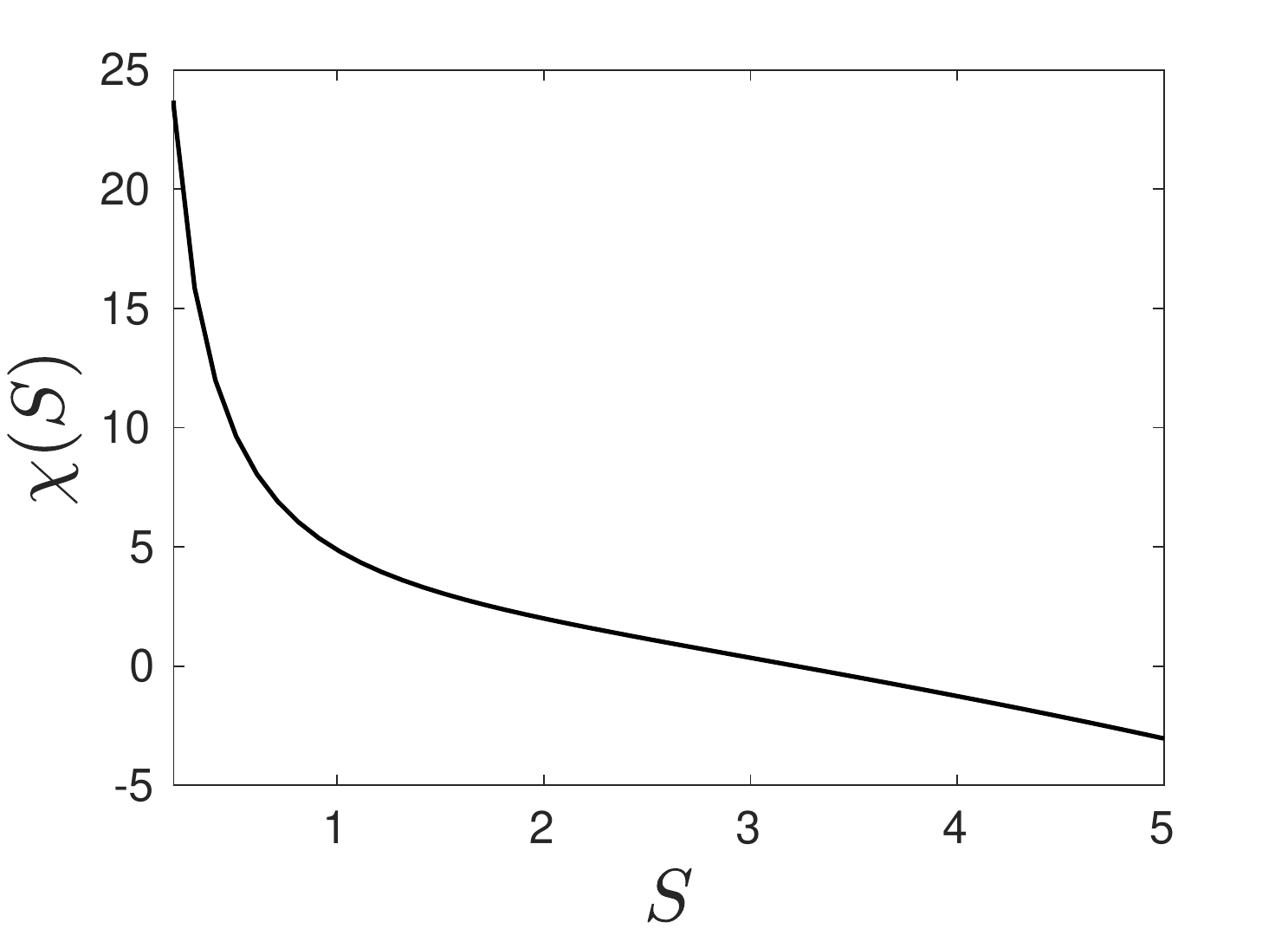}
\caption{Numerical result for $\chi$ versus the source strength
  parameter $S$, as computed numerically from the BVP
  \eqref{steady:core_problem}.}\label{fig:chi}
\end{center}
\end{figure}

We use the method of matched asymptotic expansions to construct a
steady-state single spot solution centered at $\v{x}_0 = \v{0}$ in the
unit disk. In the inner region near $\v{x}=0$, we set
\begin{equation}
  v = \sqrt{D} \, V(\v{y})\,, \quad u = {U(\v{y})/\sqrt{D}} \,,
  \quad \mbox{where} \quad \v{y} = \eps^{-1} \v{x}\,.
\end{equation}
In the inner region, for $\v{y}\in \R^2$, the steady-state problem is
\begin{equation}\label{steady:UV}
\Delta_\v{y} V - V + U V^2 = 0\,, \qquad
\Delta_\v{y} U - U V^2 + \frac{a \eps^2}{\sqrt{D}} = 0\,.
\end{equation}
We seek a radially symmetric solution in the form
$V = V_0(\rho) + o(1)$ and $U = U_0(\rho) + o(1)$, where
$\rho = |\v{y}|$. Upon neglecting the $\mc{O}(\eps^2)$ terms, we obtain
the {\em core problem}
\begin{equation}\label{steady:core_problem}
\begin{split}
  &\Delta_\rho V_0 - V_0 + U_0 V_0^2 = 0\,, \quad
  \Delta_\rho U_0 - U_0 V_0^2 = 0\,,  \quad \mbox{where } \Delta_{\rho}\equiv \partial_{\rho\rho}+\rho^{-1}\partial_{\rho} \,, \\
  &U_0^{\prime}(0) = V_0^{\prime}(0) = 0\,; \quad V_0 \to 0\,, \quad U_0 \sim
  S \log\rho + \chi(S) + o(1)\,, \quad \mbox{as} \,\,\, \rho \to \infty\,,
\end{split}
\end{equation}
In particular, we must allow $U_0$ to have far-field
  logarithmic growth whose strength is characterized by the parameter
  $S>0$, which will be determined below (see \eqref{steady:S_2}) in
  terms of the feed rate parameter $a$. The $\mc{O}(1)$ term in the
  far-field behavior depends on $S$, and is denoted by $\chi(S)$. It
  must be computed numerically from the BVP
  \eqref{steady:core_problem}. A plot of the numerically-computed
  $\chi$ versus $S$ is shown in Fig.~\ref{fig:chi}.  By integrating
the $U_0$ equation in \eqref{steady:core_problem}, we obtain the
identity that
\begin{eqnarray}\label{steady:S}
S = \int_0^\infty U_0 V_0^2 \, \rho \, \mrm{d} \rho\,.
\end{eqnarray}

In the limit $\eps \to 0$, the term $\eps^{-2} uv^2$ in the outer region
can be represented, in the sense of distributions, as a Dirac source
term using the correspondence rule
\begin{equation}
  \eps^{-2} u v^2 \to 2 \pi \sqrt{D} \left(\int_0^{\infty} U_0 V_0^2 \rho \, d
    \rho \right) \, \delta(\v{x}) = 2 \pi S \sqrt{D} \, \delta(\v{x}) \,,
\end{equation}
where \eqref{steady:S} was used. As a result, the outer problem for
$u$ in \eqref{intro:pde} is
\begin{equation}\label{steady:u}
  \Delta u = -\frac{a}{D} + \frac{2\pi S }{\sqrt{D}} \delta(\v{x})\,,
  \quad \v{x} \in \Omega\,; \qquad \partial_n u = 0\,, \quad
  \v{x} \in \partial\Omega \,.
\end{equation}
We integrate \eqref{steady:u} over the disk and use the
Divergence theorem and $|\Omega|=\pi$, to get
\begin{equation} \label{steady:S_2}
S = \frac{a |\Omega|}{2 \pi \sqrt{D}} = \frac{a}{2\sqrt{D}} \,.
\end{equation}

To represent the solution to \eqref{steady:u} we introduce the Neumann
Green's function $G(\v{x};\v{x}_0)$ for the unit disk, which is defined
uniquely by
\begin{equation}
\begin{split}
  &\Delta G = \frac{1}{|\Omega|} - \delta(\v{x} - \v{x}_0)\,, \quad
  \v{x} \in \Omega\,; \qquad \partial_n G = 0\,, \quad
  \v{x} \in \partial\Omega; \\ &\int_{\Omega} G \, d\v{x} = 0\,,
  \quad G \sim -\frac{1}{2\pi} \log|\v{x} - \v{x}_0| + R_0 + o(1)\,,
  \quad \mbox{as} \quad \v{x} \to \v{x}_0\,,
\end{split}
\end{equation}
where $R_0$ is the regular part of the Green's function. The solution
to \eqref{steady:u} is
\begin{equation}\label{steady:usol}
u = -\frac{2\pi S}{\sqrt{D}} G(\v{x};\v{0}) + \ubar \,,
\end{equation}
where $\ubar$ is a constant to be determined below by asymptotic
matching the inner and outer solutions. The Neumann Green's function
with singularity at the origin is
\begin{equation}\label{steady:green_origin}
  G(\v{x};\v{0}) = -\frac{1}{2\pi} \log|\v{x}| + \frac{|\v{x}|^2}{4\pi} -
  \frac{3}{8\pi}\,.
\end{equation}
Therefore, by using \eqref{steady:green_origin} in \eqref{steady:usol},
the outer solution $u$ satisfies
\begin{equation}\label{steady:matching1}
  u = \frac{S}{\sqrt{D}} \log |\v{x}| + \frac{3S}{4\sqrt{D}} + \ubar +
  \mc{O}(|\v{x}|^2)\,, \quad \mbox{as} \quad \v{x}\to \v{0}\,.
\end{equation}
By using the far-field behavior of the inner solution $U$ in
\eqref{steady:core_problem}, we obtain for $\rho\gg 1$ that
\begin{equation}\label{steady:matching2}
u = \frac{U}{\sqrt{D}} 
\sim \frac{1}{\sqrt{D}} \left[ S \log |\v{x}| + \frac{S}{\nu} + \chi(S) \right]\,,
\quad \mbox{where} \quad \nu \equiv -\frac{1}{\log\eps}\,.
\end{equation}
From an asymptotic matching of \eqref{steady:matching1} and
\eqref{steady:matching2}, we identity $\ubar$ as
\begin{equation}\label{steady:ubar}
  \ubar = \frac{1}{\sqrt{D}} \left( \chi(S) + \frac{S}{\nu} -
    \frac{3S}{4}\right)\,.
\end{equation}
Upon substituting \eqref{steady:ubar} and \eqref{steady:green_origin} into
\eqref{steady:usol} we conclude that the outer solution is 
\begin{equation}\label{steady:u_sol}
  u = \frac{1}{\sqrt{D}} \left(S \log|\v{x}| - \frac{S|\v{x}|^2}{2} +
    \chi(S) + \frac{S}{\nu}\right)\,, \qquad \mbox{where} \quad
     S = \frac{a}{2\sqrt{D}} \,.
\end{equation}

\begin{remark}\label{sch:equil_eps2}
  Our asymptotic approximation of matching the core solution
    to the outer solution effectively sums all the logarithmic term in
    the expansion in powers of $\nu$. (see \cite{KWW} and the references
    therein). Since the spot is centered at the origin of the unit
  disk, there is no ${\mathcal O}(\eps)$ term in the local behavior
  near $\v{x}=0$ of the outer solution. More specifically, setting
  $\v{x}=\eps y$, the outer solution \eqref{steady:u_sol} yields
\begin{equation}\label{steady:u_sol_exp}
u \sim \frac{1}{\sqrt{D}} \left(S \log|\v{y}| + \chi(S) - \frac{S\eps^2 |\v{y}|^2}
    {2} \right)\,,
\end{equation}
as we approach the inner region, which yields an unmatched
${\mathcal O}(\eps^2)$ term. Together with \eqref{steady:UV}, this
implies that the steady-state inner solution has the asymptotics
$V\sim V_0 + {\mathcal O}(\eps^2)$ and
$U\sim U_0 + {\mathcal O}(\eps^2)$. This estimate is needed below in
our weakly nonlinear analysis. In contrast, when a spot is not
centered at its steady-state location, the correction to $V_0$ and
$U_0$ in the inner expansion is ${\mathcal O}(\eps)$ and is determined by
the gradient of the regular part of the Green's function.
\end{remark}

\section{Linear stability analysis}\label{sec:linstab}

In this section, we perform a linear stability analysis of the
steady-state one-spot solution in the unit disk. For convenience, we
will represent a column vector by the notation
$(u_1, u_2) \mbox{ or } \begin{pmatrix} u_1 \\ u_2 \end{pmatrix}.$ For
a steady-state spot centered at the origin, we will formulate the
linearized stability problem in the quarter disk, defined by
$\qdisk = \{\v{x} = (x,y)\,: \,\, |\v{x}| < 1, \, x \geq 0, \, y \geq
0\}$.

Let $v_e, \, u_e$ be the steady-state spot solution centered at the
origin. We introduce the perturbation
\begin{equation}
v = v_e + e^{\lambda t} \phi\,, \quad u = u_e + e^{\lambda t} \eta\,,
\end{equation}
into \eqref{intro:pde} and linearize. This leads to the singularly
perturbed eigenvalue problem
\begin{equation}
  \eps^2 \Delta \phi - \phi + 2 u_e v_e \phi + v_e^2 \eta = \lambda \phi\,,
  \qquad D \Delta \eta - \eps^{-2} (2 u_e v_e \phi + v_e^2 \eta) =
  \tau \lambda \eta\,,
\end{equation}
with $\partial_n \phi = \partial_n \eta = 0$ on $\partial\Omega$. 

In the inner region near $\v{x}=\v{0}$ we introduce
\begin{equation}
\begin{pmatrix} \phi \\ \eta\end{pmatrix} = \mrm{Re}(e^{im\theta}) \begin{pmatrix}
  \Phi(\rho) \\ N(\rho)/D \end{pmatrix}\,, \quad \mbox{where} \quad
\rho = |\v{y}| = \eps|\v{x}|\,, \quad \theta=\mbox{arg}(\v{y})\,,
\end{equation}
with $m=2,3,\ldots$.  With $v_e \sim \sqrt{D}V_0$ and
$u_e \sim {U_0/\sqrt{D}}$, we neglect the $\mc{O}(\eps^2)$ terms to
obtain the eigenvalue problem 
\begin{equation}\label{linstab:eig}
  \mc{L}_m \begin{pmatrix} \Phi \\ N \end{pmatrix} + \begin{pmatrix}
    -1 + 2U_0 V_0 & V_0^2 \\ -2 U_0 V_0 & -V_0^2 \end{pmatrix}
  \begin{pmatrix} \Phi \\ N \end{pmatrix} = \lambda
  \begin{pmatrix} 1 & 0 \\ 0 & 0 \end{pmatrix}
  \begin{pmatrix} \Phi \\ N\end{pmatrix}\,,
\end{equation}
where the operator $\mc{L}_m$ is defined by
$ \mc{L}_m \v{\Phi} = \partial_{\rho\rho} \v{\Phi} + \rho^{-1}
\partial_{\rho} \v{\Phi} - m^2 \rho^{-2} \v{\Phi}$.  We seek
eigenfunctions of \eqref{bruss:loc_eig} with $\Phi\to 0$ and $N\to 0$
as $\rho\to\infty$. An unstable eigenvalue of this spectral problem
satisfying $\mbox{Re}(\lambda)>0$ corresponds to a non-radially
symmetric spot-deformation instability.

\begin{figure}[htbp]
	\centering
\includegraphics[height=4.2cm,width=0.45\textwidth]{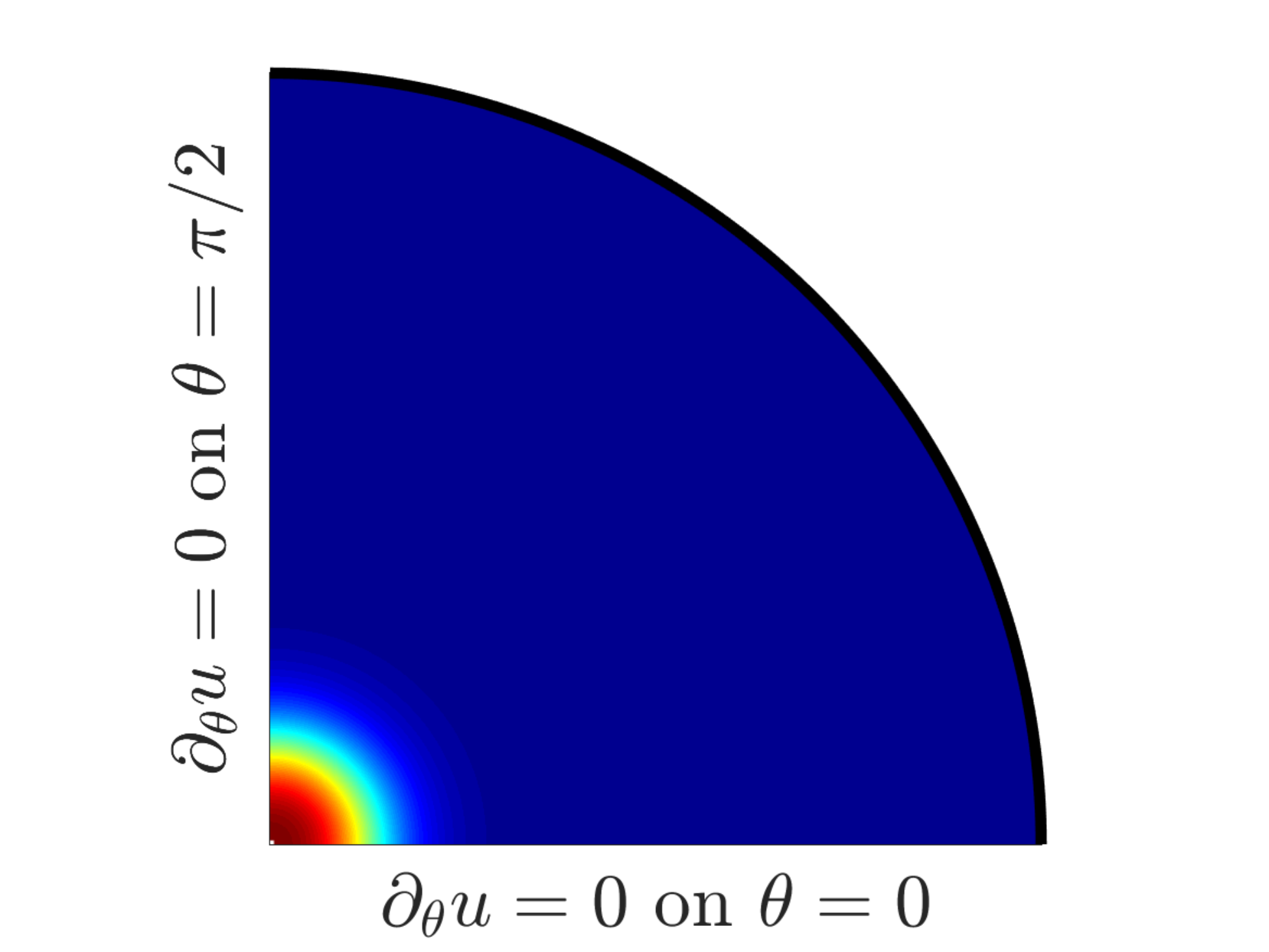}
\includegraphics[height=4.4cm,width=0.48\textwidth]{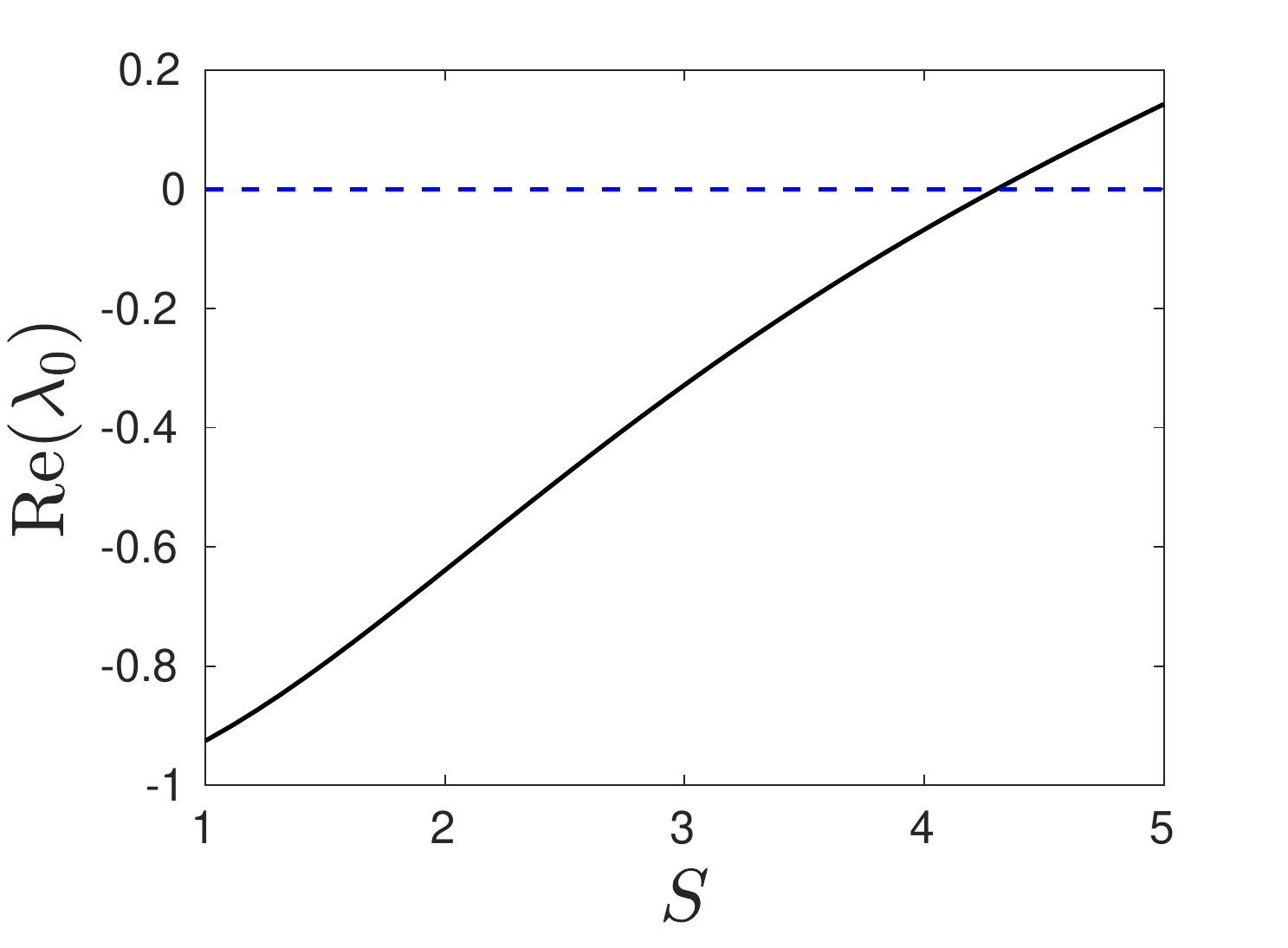}
\caption{Left panel: Plot of the quarter-disk geometry for the
  linearized stability problem with a steady-state spot centered at
  the origin when $S=S_c$. Right panel: Plot of the numerically
  computed real part of the eigenvalue $\lambda_0$ with the largest
  real part to \eqref{linstab:eig} for angular mode $m=2$. We compute
  $\mrm{Re}(\lambda_0) = 0$ (dotted line) when
  $S = S_c \approx 4.3022$ (see also \cite{KWW}).}\label{fig:eig_sc}
\end{figure}

For each angular mode $m=2,3,\ldots$, the eigenvalue $\lambda_0$ of
\eqref{linstab:eig} with the largest real part is a function of the
source strength $S$. To determine $\lambda_0$ we discretize
\eqref{linstab:eig} as done in \cite{KWW} to obtain a
finite-dimensional generalized eigenvalue problem. We calculate
$\lambda_0$ numerically from this discretized problem, with the
results shown in the right panel of Fig.~\ref{fig:eig_sc}. In the
left panel of Fig.~\ref{fig:eig_sc} we show the quarter-disk geometry.

\begin{figure}[htbp]
	\includegraphics[width=0.48\textwidth]{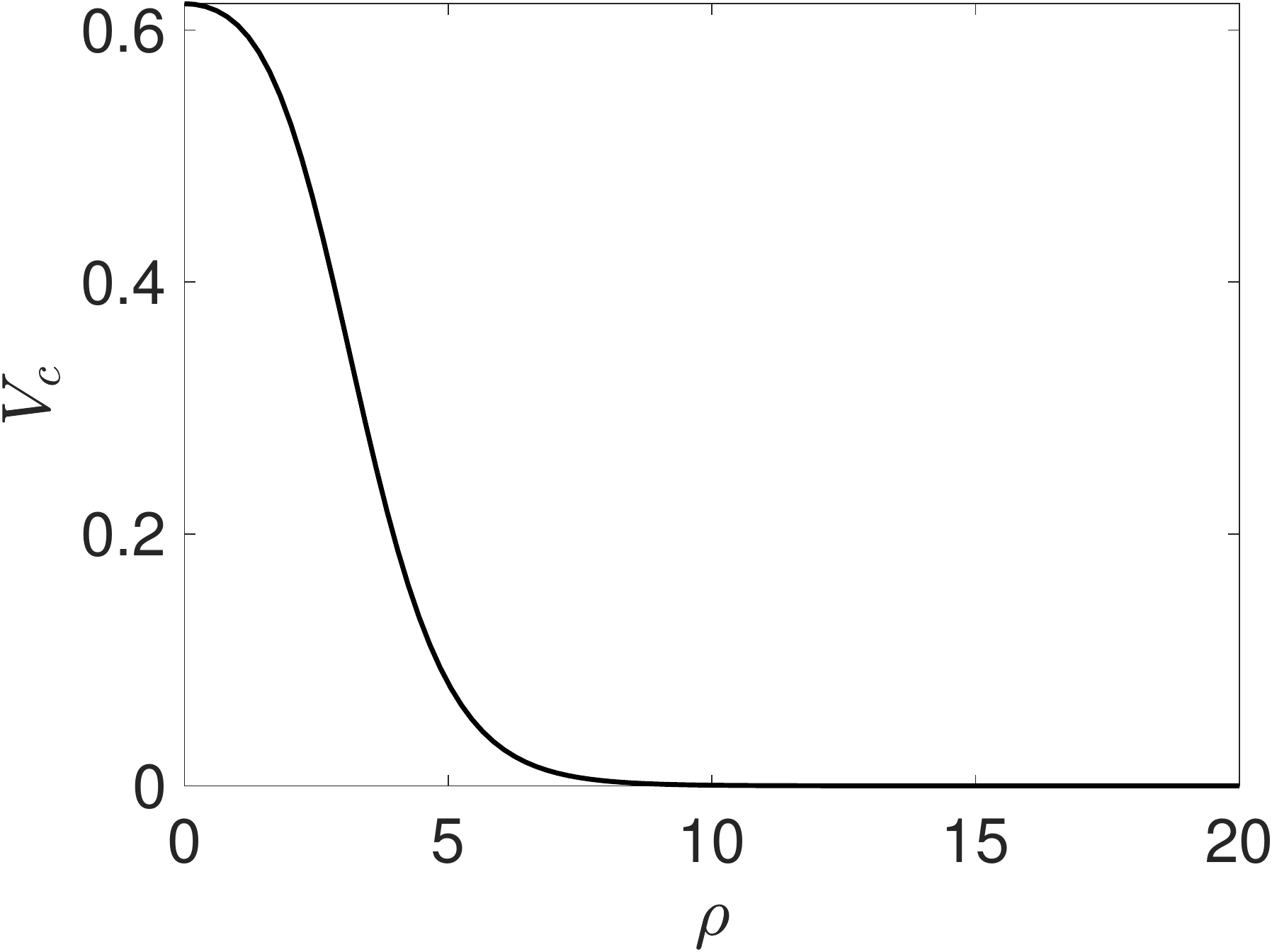} \hfill
	\includegraphics[width=0.48\textwidth]{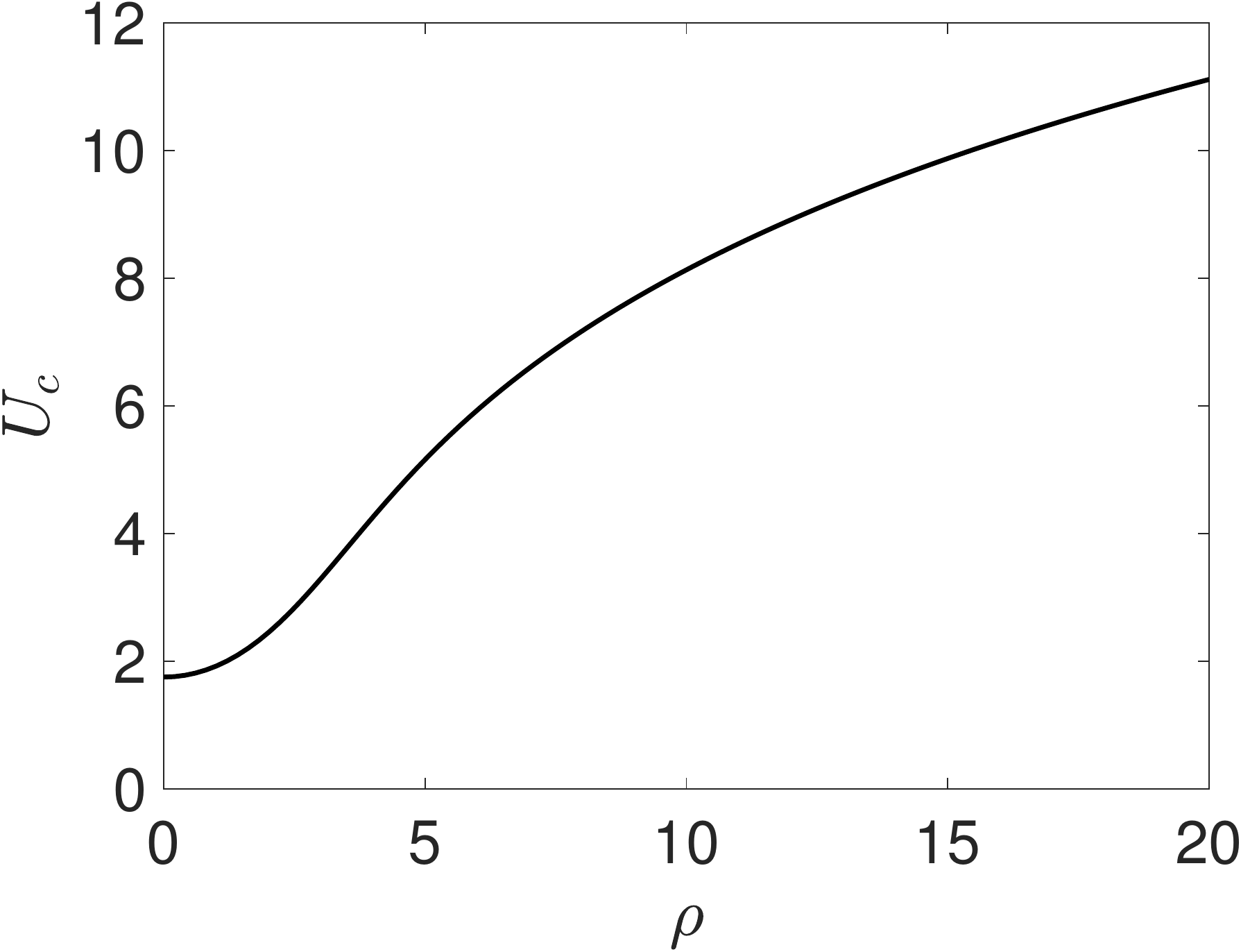}
	\caption{Numerical solution to \eqref{steady:core_problem}
          at the peanut-splitting threshold $S=S_c\approx 4.3022$.
          Left panel: $V_c=V_0(\rho ;S_c)$. Right panel: $U_c=U_0(\rho;S_c)$.}
        \label{fig:core}
\end{figure}

For the angular mode $m=2$, we find that $\mrm{Re}(\lambda_0) = 0$
when $S = S_c \approx 4.3022$, which agrees with the result first
obtained in \cite{KWW}. At this critical value of $S$, we define
\begin{equation}\label{linstab:Vc_Uc_Mc}
  V_c(\rho) \equiv V_0(\rho\,;S_c), \quad U_c(\rho) \equiv U_0(\rho\,; S_c)\,,
  \quad  M_c \equiv
  \begin{pmatrix} -1 + 2U_c V_c & V_c^2 \\ -2 U_c V_c & -V_c^2 \end{pmatrix}\,,
\end{equation}
so that there exists a non-trivial solution, labeled by
$\v{\Phi}_c \equiv (\Phi_c, N_c)$, to
\begin{equation}\label{linstab:eigc}
\mc{L}_2 \v{\Phi}_c + M_c \v{\Phi}_c = \v{0}\,.
\end{equation}
For $m=2$, we have that $\Phi_c \to 0$ exponentially as
$\rho\to \infty$ and $N_c = \mc{O}(\rho^{-2})$ as $\rho \to
\infty$. As such, we impose $\partial_\rho N_c \sim -2{N_c/\rho}$ for
$\rho\gg 1$. Since \eqref{linstab:eigc} is a linear homogeneous
system, the solution is unique up to a multiplicative constant. We
normalize the solution to \eqref{linstab:eigc} using the condition
\begin{equation}\label{linstab:normalization}
\int_0^{\infty} \Phi_c^2 \, \rho \, d\rho = 1\,.
\end{equation}
A plot of the numerically computed inner solution $V_c$ and $U_c$ is
shown in Fig.~\ref{fig:core}.

Next, for $S = S_c$, it follows that there exists a nontrivial solution
$ \v{\Phi_c}^* = (\Phi_c^*, N_c^*)$ to the adjoint problem
\begin{equation}\label{linstab:adjsol}
  \mc{L}_2 \v{\Phi_c^*} + M_c^T \v{\Phi_c^*} = \v{0}\,, \quad
  \Phi_c^* \to 0\,, \quad \partial_{\rho}N_c^* \sim -\frac{2N_c^{*}}{\rho}
  \quad \mbox{as}\quad \rho \to \infty\,,
\end{equation}
for which we impose the convenient normalization condition
$\int_0^\infty (\Phi_c^*)^2 \rho \, d\rho = 1$.

\begin{figure}[htbp]
	\includegraphics[width=0.48\textwidth]{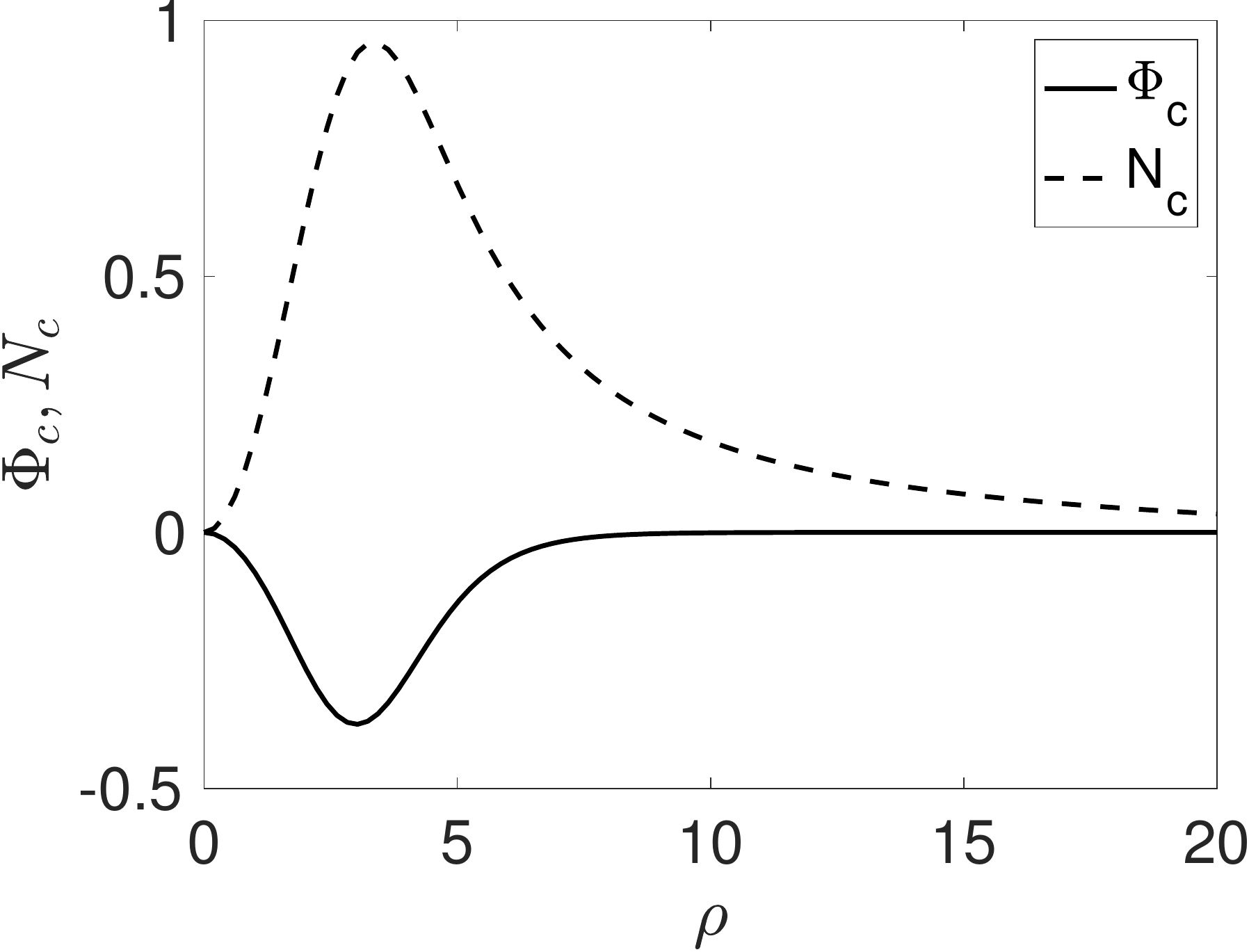} \hfill
	\includegraphics[width=0.48\textwidth]{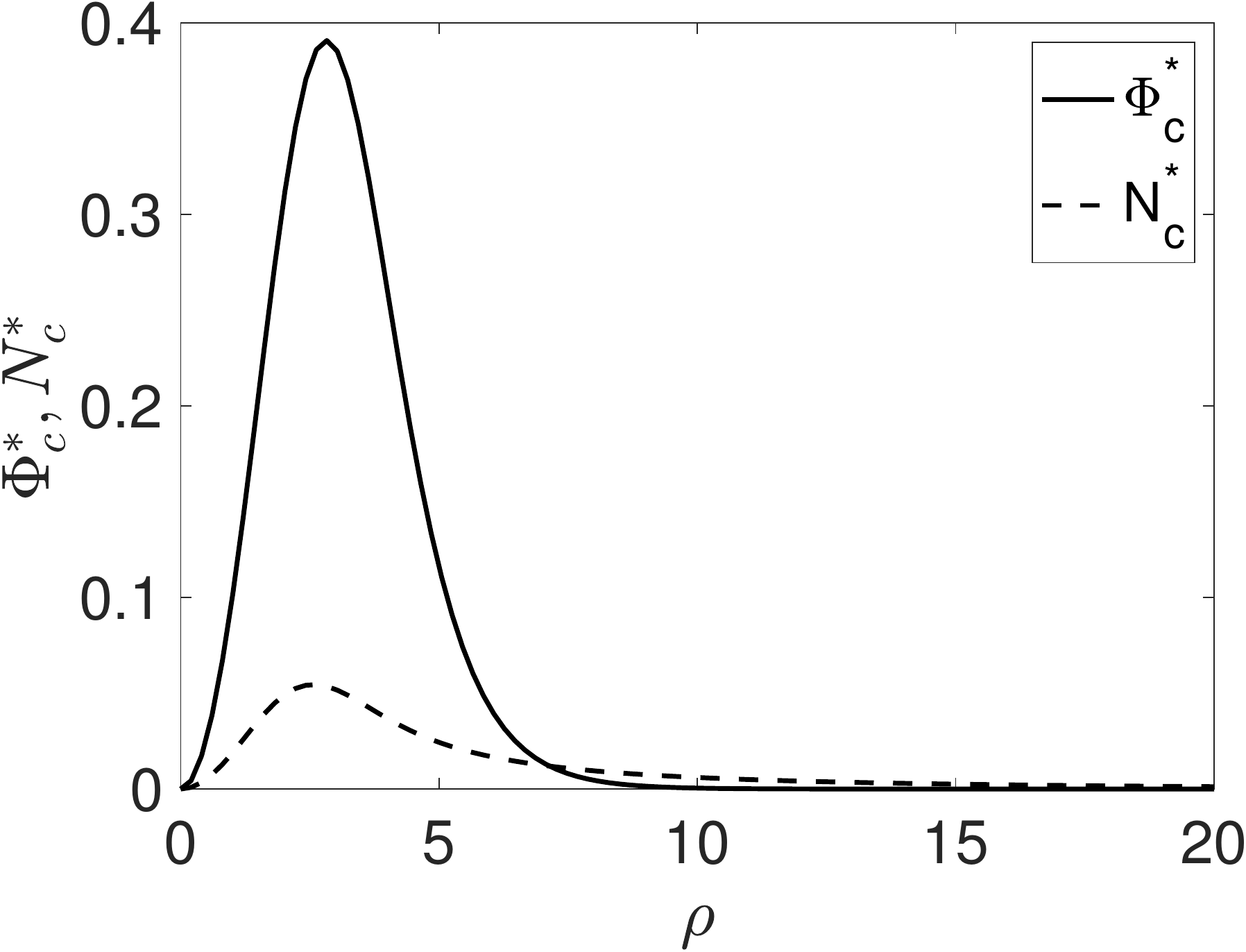}
	\caption{The numerically computed null vector and the adjoint
          satisfying \eqref{linstab:eigc} and \eqref{linstab:adjsol},
          respectively. Left panel: $\Phi_c$ and $N_c$ versus
          $\rho$. Right panel: $\Phi_c^*$ and $N_c^*$ versus
          $\rho$.}\label{fig:null}
\end{figure}

In Fig.~\ref{fig:null} we plot the numerically computed nullvector
$\Phi_c$ and $N_c$, satisfying \eqref{linstab:eigc}, as well as the
adjoint $\Phi_c^*$ and $N_c^*$, satisfying \eqref{linstab:adjsol}.

\subsection{Eigenvalue of splitting perturbation theory}

In this subsection we calculate the change in the eigenvalue
associated with the mode $m=2$ shape deformation when $S$ is slightly
above $S_c$. This calculation is needed to clearly identify the linear
term in the amplitude equation for peanut-splitting instabilities, as
derived below in \S \ref{sec:amp} using a weakly nonlinear analysis.

We denote $V_0(\rho\,;S)$ and $U_0(\rho\,;S)$ as the solution to the core
problem \eqref{steady:core_problem}. The linearized eigenproblem
associated with the angular mode $m=2$ is given by
\begin{equation}\label{linstab:perbeig}
  \mc{L}_2 \v{\Phi} + M \v{\Phi} = \lambda B \v{\Phi}\,, \quad \mbox{where}
\quad  M = \begin{pmatrix} -1 + 2U_0 V_0 & V_0^2 \\ -2 U_0 V_0 & -V_0^2
  \end{pmatrix}\,, \quad B = \begin{pmatrix} 1 & 0 \\ 0 & 0 \end{pmatrix}
  \,.
\end{equation}
When $S = S_c\,$, we have
$V_c = V_0(\rho \,; S_c), \, U_c = U_0(\rho \, ;S_c)$ and $M = M_c\,$,
for which $\lambda = 0$ is an eigenvalue in
\eqref{linstab:perbeig}. We now calculate the change in the eigenvalue
$\lambda$ when
\begin{equation}
S = S_c + \sigma^2\,, \qquad \mbox{where} \quad \sigma \ll 1\,.
\end{equation}

For convenience, we introduce the short hand notation 
\begin{equation*}
  \partial_S V_c = \partial_S V_0 \mid_{S=S_c}\,, \qquad
  \partial_S U_c = \partial_S U_0 \mid_{S=S_c}\,.
\end{equation*}
We first expand the core solution for $\sigma \ll 1$ as
\begin{equation}\label{linstab:pert1}
  V_0 = V_c + \sigma^2 \partial_S V_c + \ldots\,, \quad U_0 = U_c +
  \sigma^2 \partial_S U_c + \ldots \,,
\end{equation}
so that the perturbation to the matrix $M$ is
\begin{equation}\label{linstab:pert2}
  M = M_c + \sigma^2 M_1 + \ldots\,, \quad \mbox{with} \quad M_1 =
  \begin{pmatrix} 2\partial_S(U_c V_c) & \partial_S(V_c^2) \\
    -2\partial_S(U_c V_c) & -\partial_S(V_c^2) \end{pmatrix} \,,
\end{equation}
where we write $\partial_S(V_c U_c)=\partial_S(V_0 U_0)\vert_{S=S_c}$ and
$\partial_S(V_c^2)=\partial_S(V_0^2)\vert_{S=S_c}$.

Next, we expand the eigenpair for $\sigma \ll 1$ as
\begin{equation}\label{linstab:pert3}
  \lambda = \sigma^2 \lambda_1 + \ldots\,, \quad
  \begin{pmatrix} \Phi \\ N \end{pmatrix} =
  \begin{pmatrix} \Phi_c \\ N_c \end{pmatrix} + \sigma^2 \begin{pmatrix}
    \Phi_1 \\ N_1 \end{pmatrix} + \ldots \,.
\end{equation}
We substitute \eqref{linstab:pert1}, \eqref{linstab:pert2} and
\eqref{linstab:pert3} into \eqref{linstab:perbeig}. The $\mc{O}(1)$
terms yield \eqref{linstab:eigc}, while from the $\mc{O}(\sigma)$
terms we obtain that $\v{\Phi}_1 = (\Phi_1, N_1)$ satisfies
\begin{equation}\label{linstab:phi1}
\mc{L}_2 \v{\Phi}_1 + M_c \v{\Phi}_1 = - (\lambda_1 B + M_1) \v{\Phi}_c \,.
\end{equation}
Upon taking the inner product between \eqref{linstab:phi1} and the
adjoint solution defined in \eqref{linstab:adjsol}, we have
\begin{equation}
  \int_0^\infty \v{\Phi}_c^* \cdot \left( \mc{L}_2 \v{\Phi}_1 + M_c \v{\Phi}_1
  \right) \rho \, d\rho = \int_0^\infty \v{\Phi}_c^* \cdot
  \left[\partial_\rho(\rho \, \partial_\rho \v{\Phi}_1) - \frac{4}{\rho} \,
    \v{\Phi}_1 + \rho M_c \v{\Phi}_1 \right]  d\rho \,,
\end{equation}
where we have used
$\rho \mc{L}_2 \v{\Phi}_1 = \rho \left[\rho^{-1}(\rho \,
  \partial_\rho\v{\Phi})_{1\rho} - \rho^{-2}\v{\Phi}_1\right] =
\partial_\rho(\rho\,\partial_\rho \v{\Phi}_1) -
4\rho^{-1}\v{\Phi}_1\,.$ By using integration-by-parts twice, the
identity $\lim\limits_{\rho \to 0} \rho \, \v{\Phi}_1
\left(\partial_\rho \v{\Phi}_{1} \right) = 0$, and decay at
infinity, we obtain
\begin{equation*}\label{linstab:temp}
\begin{split}
  \int_0^\infty \v{\Phi}_c^* \cdot (\mc{L}_2 \v{\Phi}_1 + M_c \v{\Phi}_1 )
  \rho \, d\rho &= \int_0^\infty \v{\Phi}_1 \cdot (\mc{L}_2 \v{\Phi}_c^*) \rho
  \, d\rho + \int_0^\infty \v{\Phi}_c^* \cdot (M_c \v{\Phi}_1) \rho \, d\rho \\
    &= \int_0^\infty\left[-\v{\Phi}_1 \cdot (M_c^T \v{\Phi}_c^*) +
    \v{\Phi}_c^* \cdot (M_c \v{\Phi}_1) \right] \rho \, d\rho = 0\,.
\end{split}
\end{equation*}

Together with \eqref{linstab:phi1}, we have derived the solvability
condition
\begin{equation}\label{linstab:solvability}
  \int_0^\infty \v{\Phi}_c^* \cdot (\mc{L}_2 \v{\Phi}_1 + M_c \v{\Phi}_1) \rho
  \, d\rho = \int_0^\infty \v{\Phi}_c^*  \cdot
  \left[ (\lambda_1 B - M_1) \v{\Phi}_c \right] \rho \, d\rho = 0\,.
\end{equation}
By solving for $\lambda$, and then rearranging the resulting expression,
we obtain that
\begin{equation}\label{linstab:lambda1}
  \lambda_1 = \frac{\int_0^\infty \left[ 2 \Phi_c \partial_S (U_c V_c)
      + N_c \partial_S (V_c)^2 
      \right](\Phi_c^* - N_c^*) \rho \, d\rho}
  {\int_0^\infty \Phi_c^* \Phi_c \, \rho \, d\rho} \,.
\end{equation}
From a numerical quadrature of the integrals in
\eqref{linstab:lambda1}, which involves the numerical solution to
\eqref{linstab:Vc_Uc_Mc}, \eqref{linstab:eigc} and
\eqref{linstab:adjsol}, we calculate that $\lambda_1 \approx
0.2174$. Therefore, when $S=S_c+\sigma^2$ for $\sigma\ll 1$ we
conclude that $\lambda\sim 0.2174 \sigma^2$.

\begin{remark}
  As shown in \cite{KWW} for the Schnakenburg model, as $a$ is
  increased the first non-radially symmetric mode to go unstable is
  the $m=2$ peanut-splitting mode, which occurs when
  $S=\Sigma_2\approx 4.3022$. Higher modes first go unstable at larger
  values of $S$, denoted by $\Sigma_m$. From Table 1 of \cite{KWW},
  these critical values of $S$ are $\Sigma_3\approx 5.439$,
  $\Sigma_4\approx 6.143$, $\Sigma_5\approx 6.403$ and
  $\Sigma_6\approx 6.517$. Since our weakly nonlinear analysis will
  focus only on a neighbourhood of $\Sigma_2$, the higher modes
  $m\geq 3$ are all linearly stable in this neighbourhood.
\end{remark}

\section{Amplitude equation for the Schnakenberg model}\label{sec:amp}

In this section we derive the amplitude equation associated with the
peanut-splitting linear stability threshold for the Schnakenberg
model. This amplitude equation will show that this spot
shape-deformation instability is subcritical.

To do so, we first introduce a small perturbation around the linear
stability threshold $S_c$ given by $S = S_c + \kappa \sigma^2$, where
$\kappa = \pm 1$. In this way, the obtain the Taylor expansion
$\chi(S) = \chi(S_c) + \kappa \chi^{\prime}(S_c) \sigma^2 +
\mc{O}(\sigma^4)$.  Then, we introduce a slow time scale
$T = \sigma^2 t$. As such, the inner problem in terms of
$V = {v/\sqrt{D}}$ and $U = \sqrt{D} u$ for $\v{y}\in \R^2$ is
\begin{subequations}\label{amp:sc_all}
  \begin{equation}\label{amp:pde_inner}
  \sigma^2 V_T = \Delta_\v{y} V - V + UV^2\,, \qquad
  \frac{\sigma^2 \eps^2 \tau }{D} U_T = \Delta_\v{y} U - UV^2 +
  \frac{a\eps^2}{\sqrt{D}}\,,
\end{equation}
for which we impose $V\to 0$ exponentially as $\rho\to\infty$, while
\begin{equation}\label{amp:pde_far_field}
  U \sim \left(S_c + \kappa \sigma^2\right) \log \rho +
  \chi(S_c) + \sigma^2 \left[\kappa \chi^{\prime}(S_c) + \mc{O}(1)\right] +
  \ldots \,, \quad \mbox{as} \quad \rho = |\v{y}| \to \infty\,.
\end{equation}
\end{subequations}

In \eqref{amp:sc_all}, we expand $V = V(\rho, \phi, T)$ and
$U = U(\rho, \phi, T)$ as
\begin{equation}\label{amp:expansion}
  V = V_0 + \sigma V_1 + \sigma^2 V_2 + \sigma^3 V_3 + \ldots\,, \quad
  U = U_0 + \sigma U_1 + \sigma^2 U_2 + \sigma^3 U_3 + \ldots\,,
\end{equation}
where $V_0, \, U_0$ is the radially symmetry core solution, satisfying
\eqref{steady:core_problem}. Furthermore, we assume that
\begin{equation}\label{amp:sig_eps}
\sigma^3 \gg \mc{O}(\eps^2)\,,
\end{equation}
so that the $\mc{O}(\eps^2)$ terms in \eqref{amp:pde_inner} are
asymptotically smaller than terms of order $\mc{O}(\sigma^k)$ for
$k \leq 3$.

\begin{remark}
  The error in our asymptotic construction is $\mc{O}(\eps^2)$ for a
  spot that is centered at its equilibrium location (see Remark
  \ref{sch:equil_eps2}). We need the scaling assumption
  \eqref{amp:sig_eps} to ensure that the higher order in $\eps$
  approximation of the steady-state is smaller than the approximation
  error involved in deriving the amplitude equation. For a spot
  pattern in a quasi-equilibrium state, the error in the construction
  of the steady-state is $\mc{O}(\eps)$, which renders our analysis
  invalid for quasi-equilibrium patterns. We refer to the discussion
  section \S \ref{sec:disc} where this issue is elaborated further.
\end{remark}

We then substitute \eqref{amp:expansion} into \eqref{amp:sc_all} and
collect powers of $\sigma$. From the $\mc{O}(1)$ terms, we obtain that
$V_0$ and $U_0$ satisfy
\begin{subequations}\label{amp:V0_U0}
\begin{align}
  \Delta_\rho &V_0 - V_0 + U_0 V_0^2 = 0\,, \quad
                \Delta_\rho U_0 - U_0 V_0^2 = 0\,, \\
              &V_0 \to 0, \quad U_0 \sim S_c \log \rho + \mc{O}(1)\,,
          \quad \mbox{as} \quad \rho \to \infty\,. \label{amp:U0_far_field}
\end{align}
\end{subequations}
From the far-field condition \eqref{amp:U0_far_field}, we can identify
that $V_0$ and $U_0$ are the core solution with $S=S_c$. In other
words, we have
\begin{equation}\label{amp:matching0}
V_0 = V_c(\rho)\,, \quad U_0 = U_c(\rho)\,.
\end{equation}
From collecting $\mc{O}(\sigma)$ terms, and setting $V_0=V_c$ and
$U_0=U_c$, we find that $\v{V}_1 = (V_1, U_1)$
satisfies
\begin{equation}\label{amp:V1}
  \Delta_\v{y} \v{V}_1 + M_c \, \v{V}_1 = \v{0}\,, \quad \mbox{where} \quad
  M_c = \begin{pmatrix} -1 + 2 U_c V_c & V_c^2 \\ -2 U_c V_c & -V_c^2
  \end{pmatrix}\,.
\end{equation}
We conclude that $\v{V}_1$ is related to the eigenfunction solution to
\eqref{linstab:eigc}. We introduce the amplitude function $A = A(T)$, while
writing $\v{V}_1$ as
\begin{equation}\label{amp:matching1}
\v{V}_1 = A \cos(2\phi) \begin{pmatrix} \Phi_c \\ N_c \end{pmatrix}\,,
\end{equation}
where $\Phi_c$ and $N_c$ satisfy \eqref{linstab:eigc} with normalization
\eqref{linstab:normalization}.

\begin{remark}\label{sch:quart}
  In our linear stability analysis in the quarter-disk it
  is only the angular factor $\cos(2\phi)$ in \eqref{amp:matching1},
  as opposed to the alternative choice of $\sin(2\phi)$, that
  satisfies the no-flux conditions for $V$ and $U$ at
  $\phi=0,{\pi/2}$. In this way, our domain restriction to the
  quarter-disk ensures a one-dimensional null-space for
  \eqref{amp:V1}.
\end{remark}

By collecting $\mc{O}(\sigma^2)$ terms we readily obtain that
$\v{V}_2 = (V_2, U_2)$ on $\v{y}\in \R^2$ satisfies 
\begin{subequations}
\begin{equation}\label{amp:V2}
\Delta_\v{y} \v{V}_2 + M_c \v{V}_2 = F_2 \, \v{q} \,,
\end{equation}
where we have defined $F_2$ and $\v{q}$ by
\begin{equation}
  F_2 \equiv  2 V_c V_1 U_1 + U_c V_1^2\,, \qquad \v{q} \equiv
  \begin{pmatrix} -1 \\ 1 \end{pmatrix}\,.
\end{equation}
\end{subequations}
By using \eqref{amp:matching1} for $V_1$ and $U_1$, together with the
identity $2 \cos^2 \phi = 1 + \cos (2\phi)$, we can write $F_2$ as
\begin{equation}\label{amp:F20}
  F_2 = A^2 F_{20} + A^2 F_{20} \cos (4\phi)\,, \qquad
  F_{20} = \frac{1}{2}\left(U_c \Phi_c^2 + 2 V_c \Phi_c N_c\right) \,.
\end{equation}
This suggests a decomposition of the solution to \eqref{amp:V2} in the form
\begin{equation}\label{amp:V2_decomposition}
\v{V}_2 = \v{V}_{20}(\rho) + A^2 \, \v{V}_{24}(\rho) \cos (4\phi)\,,
\end{equation}
where the problems for $\v{V}_{20}$ and $\v{V}_{24}$ are formulated below.

Firstly, we define $\v{V}_{24} = (V_{24}, U_{24})$ to be the radially
symmetric solution to
\begin{subequations}\label{amp:V24_all}
\begin{equation}\label{amp:V24}
\mc{L}_4 \v{V}_{24} + M_{c} \v{V}_{24} = F_{20} \, \v{q} \,,
\end{equation}
where
$\mc{L}_m \v{V}_{24} = \partial_{\rho\rho} \v{V}_{24} + \rho^{-1}
\partial_{\rho} \v{V}_{24} - m^2 \rho^{-2} \v{V}_{24}$, for which we
can impose that
\begin{equation}
  V_{24} \to 0\,, \quad U_{24} = \mc{O}(\rho^{-4})
  \,\quad \longrightarrow \quad \partial_\rho \,
  U_{24} \sim -\frac{4}{\rho} \, U_{24} \,, \quad \mbox{as}\quad
  \rho \to \infty\,.
\end{equation}
\end{subequations}
Next, we define $\v{V}_{20} = (V_{20}, U_{20})$ to be the solution to
\begin{subequations}\label{amp:V20_all}
\begin{equation}\label{amp:V20}
\Delta_\rho \v{V}_{20} + M_c \v{V}_{20} = A^2 F_{20} \, \v{q} \,.
\end{equation}
We can impose $V_{20} \to 0$ exponentially as $\rho \to \infty$. As
indicated in \eqref{amp:pde_far_field}, we have
\begin{equation}
  U_2 \sim \kappa \log \rho + \mc{O}(1)\,, \quad \mbox{as} \quad
  \rho \to \infty\,. 
\end{equation}
\end{subequations}
Since $U_{24} =\mc{O}(\rho^{-4})\ll 1$ as $\rho\to\infty$, we must
have $U_{20} \sim \kappa \log \rho + \mc{O}(1)$.

Next, we decompose $\v{V}_{20}$ by first observing that
$\wtwoh \equiv (\partial_S V_c, \partial_S U_c)$ is a radial solution
to the homogeneous problem
\begin{equation}\label{amp:wtwoh}
  \Delta_\rho \wtwoh + M_c \wtwoh = \v{0}\,, \quad
  \wtwoh \sim \left(0, \log \rho + \chi^{\prime}(S_c)\right)\,, \quad
  \mbox{as} \quad \rho \to \infty\,.
\end{equation}
This suggests that it is convenient to introduce the following
decomposition to isolate the two sources of inhomogeneity in
\eqref{amp:V20_all}:
\begin{equation}\label{amp:V20_decomposition}
\v{V}_{20} = \kappa \wtwoh + A^2 \hat{\v{V}}_{20}\,,
\end{equation}
where $\hat{\v{V}}_{20} = (\hat{V}_{20}, \hat{U}_{20})$ is taken to be
the radial solution to
\begin{equation}\label{amp:Vhat20}
  \Delta_\rho \hat{\v{V}}_{20} + M_c \hat{\v{V}}_{20} = F_{20} \,
  \v{q}\,, \quad \hat{V}_{20} \to 0\,, \quad
  \partial_\rho \hat{U}_{20} \to 0\,, \quad \mbox{as} \quad \rho \to \infty \,.
\end{equation}
In Appendix \ref{sec:far_field} we discuss in detail the derivation of
the far-field condition for $\hat{U}_{20}$ imposed in
\eqref{amp:Vhat20}.  Moreover, since
$\hat{U}_{20}\to U_{20\infty}\neq 0$ as $\rho\to \infty$, at the end
of Appendix \ref{sec:far_field} we show how this fact can be accounted
for in a simple modification of the outer solution given in
\eqref{steady:u_sol}.

\begin{figure}[htbp]
	\includegraphics[width=0.48\textwidth]{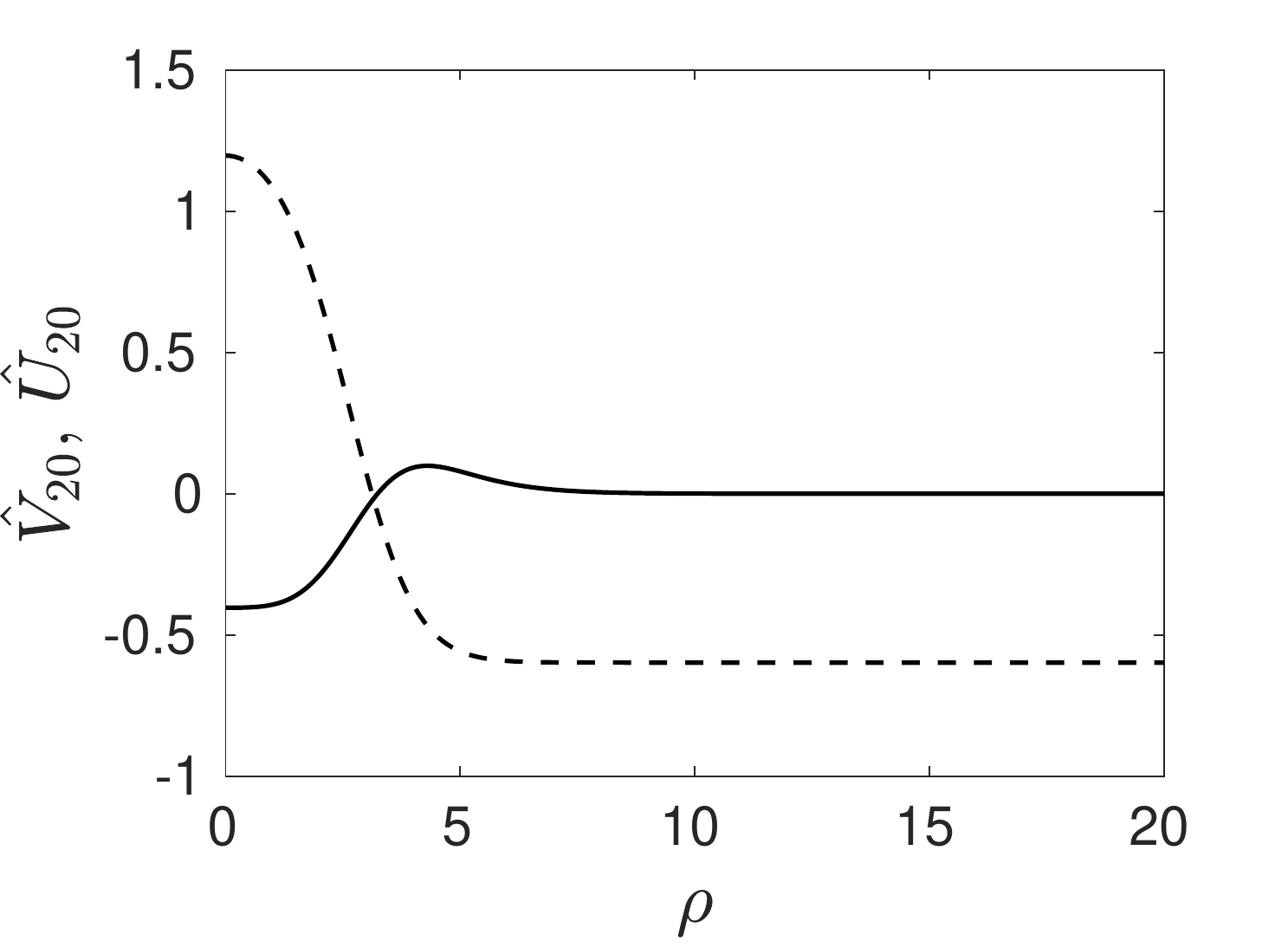} \hfill
	\includegraphics[width=0.48\textwidth]{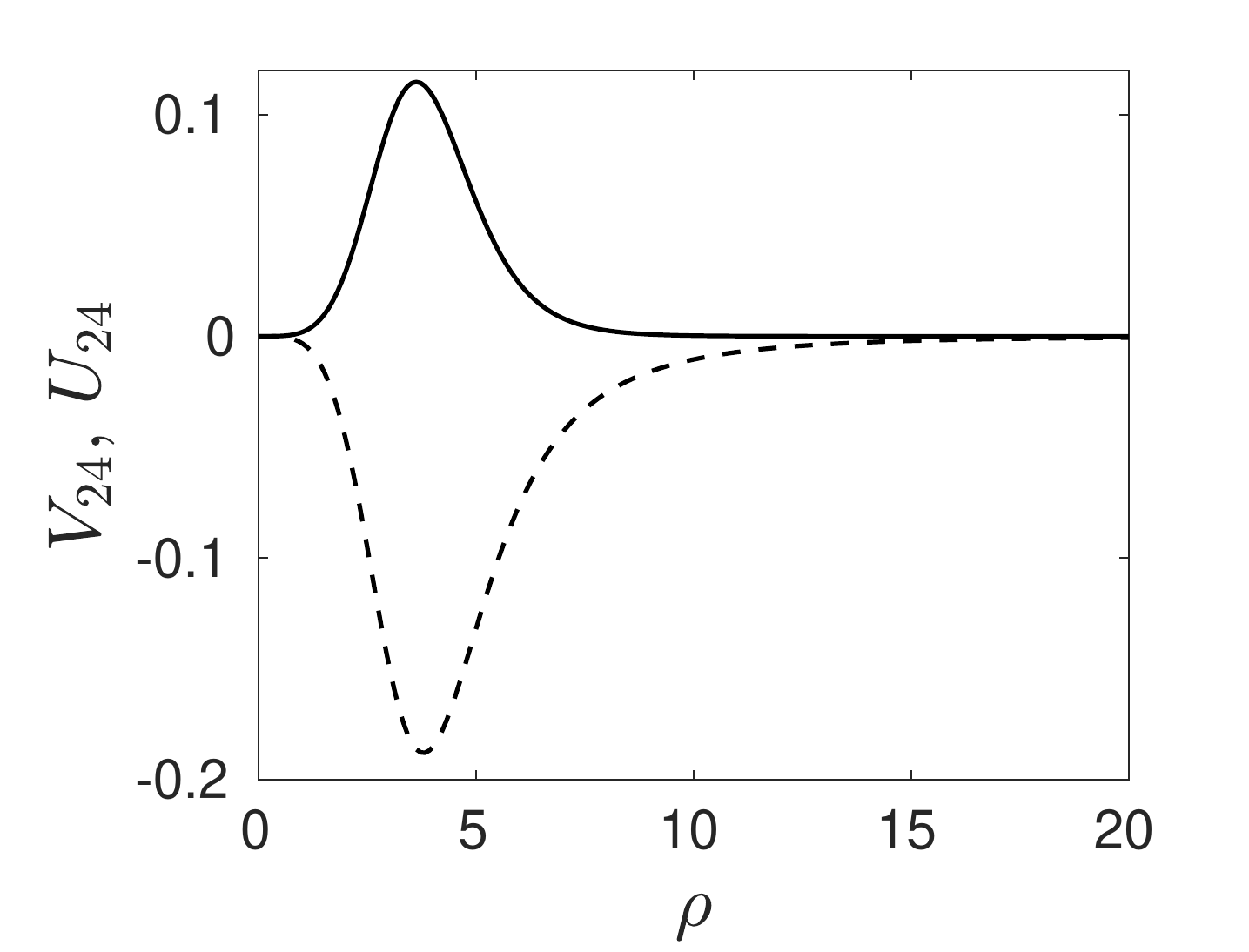}
	\caption{Left panel: Plot of the numerical solution for
          $\hat{V}_{20}$ (solid line) and $\hat{U}_{20}$ (dashed
          line).  Right panel: Plot of the numerical solution for
          $V_{24}$ (solid line) and $U_{24}$ (dashed line).}
        \label{fig:order_two}
\end{figure}

In view of \eqref{amp:V20_decomposition} and \eqref{amp:V24_all}, the
solution to \eqref{amp:V2}, as written in \eqref{amp:V2_decomposition},
is 
\begin{equation}\label{amp:V2fullsol}
  \v{V}_2 = \kappa \wtwoh + A^2
  \left[ \hat{\v{V}}_{20} + \v{V}_{24} \cos(4\phi) \right]\,.
\end{equation}
In the left and right panels of Fig.~\ref{fig:order_two} we plot the
numerically computed solution to \eqref{amp:Vhat20} and
\eqref{amp:V24_all}, respectively.

The solvability condition, which yields the amplitude equation for $A$,
arises from the $\mc{O}(\sigma^3)$ problem. At this order, we find that
$\v{V}_3 = (V_3, U_3)$ satisfies
\begin{subequations}\label{amp:V3_all}
\begin{equation}\label{amp:V3}
\Delta_\v{y} \v{V}_3 + M_c \v{V}_3 = F_3 \, \v{q} + \partial_T V_1 \, \v{e}_1\,,
\end{equation}
where we have defined $F_3$ and $\v{e}_1$ by
\begin{equation}\label{amp:F3_e1}
  F_3 \equiv 2 V_c V_1 U_2 + U_1 V_1^2 + 2 V_c U_1 V_2 + 2 U_c V_1 V_2\,, \qquad
  \v{e}_1 \equiv \begin{pmatrix}
    1 \\ 0 \end{pmatrix}\,.
\end{equation}
\end{subequations}
Upon substituting \eqref{amp:matching1} and \eqref{amp:V2fullsol} into
$F_3$, we can write $F_3$ in \eqref{amp:F3_e1} in terms of a truncated
Fourier cosine expansion as
\begin{subequations}\label{amp:F3_all}
\begin{equation}\label{amp:F3}
F_3 = (\kappa g_1 A + g_2 A^3) \cos(2\phi) + g_3 A^3 \cos(6\phi)\,,
\end{equation}
where $g_1, \, g_2$ and $g_3$ are defined by
\begin{align}
g_1 &=  2\Phi_c \partial_S(V_c U_c) + N_c \partial_S(V_c^2)\,, \label{amp:g1}\\
  g_2 &= 2 V_c \Phi_c \hat{U}_{20} + V_c \Phi_c U_{24} + \frac{3}{4}
        \Phi_c^2 N_c + (V_c N_c + U_c \Phi_c) (2 \hat{V}_{20} + V_{24})\,,
        \label{amp:g2} \\
  g_3 &= \frac{1}{4} N_c \Phi_c^2 + V_c \Phi_c U_{24} +
        (V_c N_c + U_c \Phi_c) V_{24}\,. \label{amp:g3}
\end{align}
\end{subequations}

In this way, the solution $\v{V}_3 = (V_3,U_3)$ to \eqref{amp:V3} satisfies
\begin{equation}\label{amp:V3_nodal}
  \Delta \v{V}_3 + M_c \v{V}_3 = (\kappa g_1 A + g_2 A^3) \cos(2\phi) \,
  \v{q} + g_3 A^3 \cos(6\phi) \, \v{q} + A^{\prime} \Phi_c \cos(2\phi) \,
  \v{e}_1\,,
\end{equation}
where $A^{\prime}\equiv {dA/dT}$.  The right-hand side of this expression
suggests that we decompose $\v{V}_3$ as
\begin{subequations}
\begin{equation}\label{amp:V3_decomposition}
\v{V}_3 = \v{W}_2(\rho) \cos(2\phi) + \v{W}_6(\rho) \cos(6\phi)\,,
\end{equation}
so that from \eqref{amp:V3_nodal} we obtain that $\v{W}_2$ and
$\v{W}_6$ are radial solutions to
\begin{align}
  \mc{L}_2 \v{W}_2 + M_c \v{W}_2 &= (\kappa g_1 A + g_2 A^3) \, \v{q} +
                  A^{\prime}\Phi_c \, \v{e}_1\,, \label{amp:w2}\\
\mc{L}_6 \v{W}_6 + M_c \v{W}_6 &= g_3 A^3 \v{q}\,. \label{amp:w6}
\end{align}
\end{subequations}

We now impose a solvability condition for the solution to
\eqref{amp:w2}. Recall from \eqref{linstab:adjsol} that there is a
non-trivial solution $\v{\Phi}_c^* = (\Phi_c^*, N_c^*)$ to
$\mc{L}_2 \v{\Phi}_c^* + M_c^T \v{\Phi}_c^* = \v{0}$.

As in the derivation of the eigenvalue expansion in
\eqref{linstab:solvability}, we have
\begin{equation}
  \int_0^{\infty} \v{\Phi}_c^* \cdot (\mc{L}_2 \v{W}_2 + M_c \v{W}_2) \,
  \rho \, d \rho = 0\,.
\end{equation}
This yields that 
\begin{equation}
  \int_0^\infty \left(\v{\Phi}_c^* \cdot \v{q} \right) (\kappa g_1 A + g_2 A^3)
  \, \rho \, \mathrm{d} \rho = -A^{\prime} \int_0^\infty \v{\Phi}_c^* \cdot
  (\Phi_c \, \v{e}_1) \, \rho \, d \rho \,,
\end{equation}

so that upon using $\v{e}_1=(1,0)$ and $\v{q}=(-1,1)$, we solve for  $A^{\prime}$ to obtain
\begin{equation}
  -A^{\prime} \int_0^\infty \Phi_c \Phi_c^* \, \rho \, \mrm{d} \rho =
  \int_0^\infty (\kappa g_1 A + g_2 A^3) (N_c^* - \Phi_c^*) \, \rho
  \, \mrm{d} \rho \,.
\end{equation}

By rearranging this expression we conclude that
\begin{equation}
  \frac{dA}{dT} = \left[\frac{\kappa \int_0^\infty g_1 (\Phi_c^* - N_c^*) \,
      \rho \, \mrm{d}\rho}{\int_0^\infty \Phi_c \Phi_c^*
      \, \rho \, \mrm{d}\rho} \right] A + \left[\frac{\int_0^\infty g_2
      (\Phi_c^* - N_c^*) \, \rho \, \mrm{d} \rho}{\int_0^\infty \Phi_c \Phi_c^*
      \, \rho \mrm{d}\rho} \right] A^3 \,.
\end{equation}

In summary, the normal form of the amplitude equation is given by
\begin{subequations}
\begin{equation}\label{amp:normalform0}
  \frac{dA}{dT} = \kappa c_1 A + c_3 A^3\,, \qquad \mbox{with}
  \quad T = \sigma^2 t \,,
\end{equation}
where $c_1$ and $c_3$ are given by
\begin{equation}\label{amp:c1_c3}
  c_1 = \frac{\int_0^\infty g_1 (\Phi_c^* - N_c^*) \, \rho \,
    \mrm{d}\rho}{\int_0^\infty \Phi_c \Phi_c^* \, \rho \, \mrm{d}\rho}\,, \qquad
  c_3 = \frac{\int_0^\infty g_2 (\Phi_c^* - N_c^*) \, \rho \,\mrm{d} \rho}
  {\int_0^\infty \Phi_c \Phi_c^* \, \rho \mrm{d}\rho}\,,
\end{equation}
\end{subequations}
and $g_1$ and $g_2$ are given in \eqref{amp:g1} and \eqref{amp:g2},
respectively.
By comparing our expression for $c_1$ in \eqref{amp:c1_c3} with
\eqref{linstab:lambda1} we conclude that
$c_1=\lambda_1\approx 0.2174$, where $\lambda_1$ is the eigenvalue for
the mode $m=2$ instability, as derived in \eqref{linstab:lambda1} when
$S=S_c+\sigma^2$ with $\sigma\ll 1$. Moreover, from a numerical
quadrature we calculate that $c_3 \approx 0.1224$.

Multiplying both sides of \eqref{amp:normalform0} by $\sigma$ and using
the time scale transformation $\frac{d}{dT} = \sigma^{-2}
\frac{d}{dt}$, the amplitude equation \eqref{amp:normalform0} in terms
of $\tilde{A} \equiv \sigma A$ is
\begin{equation}\label{amp:normalform}
\frac{d\tilde{A}}{dt} = \kappa \sigma^2 c_1 \tilde{A} + c_3 \tilde{A}^3\,.
\end{equation}
Since $c_1, c_3$ are numerically found to be positive, the non-zero
steady small amplitude $\tilde{A}_0$ in \eqref{amp:normalform} exists
only when $\kappa = -1$. In this case, we have
\begin{equation}\label{amp:small_amplitude}
  \tilde{A}_0 = \sqrt{\frac{c_1(S_c - S)}{c_3}} \,, \quad \mbox{for} \quad
  S < S_c \,.
\end{equation}

\begin{remark}\label{sch:consistent}
  By our assumption $\sigma^3 \gg \mc{O}(\eps^2)$, we conclude that our weakly
  nonlinear analysis is valid only when
  $S_c - S = \sigma^2 \gg \mc{O}(\eps^{4/3})$.
\end{remark}

\subsection{Numerical validation of the amplitude equation}\label{sec:validate}

In this subsection we numerically verify the asymptotic approximation
of the steady-state in \eqref{amp:small_amplitude} as obtained from our
amplitude equation. Our approach is to compute the norm difference
between the radially symmetric spot solution and its associated
bifurcating solution branch originating from the zero eigenvalue
crossing of the peanut-shape instability. To do so, we revisit the
expansion scheme \eqref{amp:expansion} with $V_0 = V_c$ and
$\sigma V_1 = \sigma A \cos(2\phi) \Phi_c = \tilde{A}
\cos(2\phi) \Phi_c$ for $S = S_c + \kappa \sigma^2$ with
$\sigma \ll 1$. This yields the steady-state prediction
\begin{equation}\label{amp:expansion2}
  V(\v{y}\,;S) = V_c(\rho) + \tilde{A} \, \Phi_c(\rho)
  \cos(2\phi) + \mc{O}(\sigma^2)\,,
\end{equation}
with $|\v{y}|=\rho$. We also expand the radially symmetric one-spot
inner solution for $S=S_c+\kappa \sigma^2$ as 
\begin{equation}\label{amp:expansion3}
  V_0(\rho\,;S) = V_0(\rho\,;S_c) + \kappa \sigma^2
  \left[\partial_S V_0(\rho\,;S) \right]\vert_{S=S_c} + \ldots =
  V_c(\rho) + \mc{O}(\sigma^2)\,.
\end{equation}

Let $r = |\v{x}| = \eps \,\rho$. We define the $L_2$-function norm in
the quarter disk by
\begin{equation*}
  ||v|| = \left[\int_0^{\pi/2} \int_{0}^{1} v(r,\phi)^2 r \, dr \, d\phi
  \right]^{1/2}= \eps\left[\int_0^{\pi/2} \int_0^{1/\eps} v(\rho,\phi)^2 \rho \,
    d\rho \, d\phi\right]^{1/2} \,.
\end{equation*}
Let $v(r,\phi\,; S) = V(\v{y}\,;S)$ and
$v_0(r,\phi) = V_0(\rho\,; S)$. From \eqref{amp:expansion2} and
\eqref{amp:expansion3}, we have
\begin{equation}\label{sc:dnorm_all}
\begin{split}
  ||v-v_0||^2 &= \eps^2 \int_0^{\pi/2} \int_0^{1/\eps} \left[ \tilde{A} \,
    \Phi_c(\rho) \cos(2\phi) \right]^2 \, \rho \, d\rho \, d\phi +
  \mc{O}(\eps^2\sigma^3)\,, \\
  &= \eps^2 \tilde{A}^2 \int_0^{\pi/2} \cos^2(2\phi) d\phi \left(
  \int_0^{1/\eps} \Phi_c^2(\rho)\rho \, d\rho \right) + \mc{O}(\eps^2\sigma^3)\,.
\end{split}
\end{equation}
Then, by using the normalization condition
\eqref{linstab:normalization}, together with the steady-state
amplitude in \eqref{amp:small_amplitude}, our theoretical prediction
from the weakly nonlinear analysis for the non-radially symmetric
solution branch is that for $S_c-S=\sigma^2 \gg \mc{O}(\eps^{4/3})$,
we have
\begin{equation}\label{amp:vnormdiff}
  ||v - v_0|| \sim \frac{\eps}{2} \sqrt{\frac{\pi c_1 (S_c-S)}{c_3}}\,,
  \quad \mbox{as} \quad \sigma \to 0^+\,, \,\, \eps \to 0^+ \,,
\end{equation}
where $c_1\approx 0.2174$ and $c_3 \approx 0.1224$.   

\begin{figure}[htbp]
\centering
\includegraphics[height=4.3cm,width=0.45\textwidth]{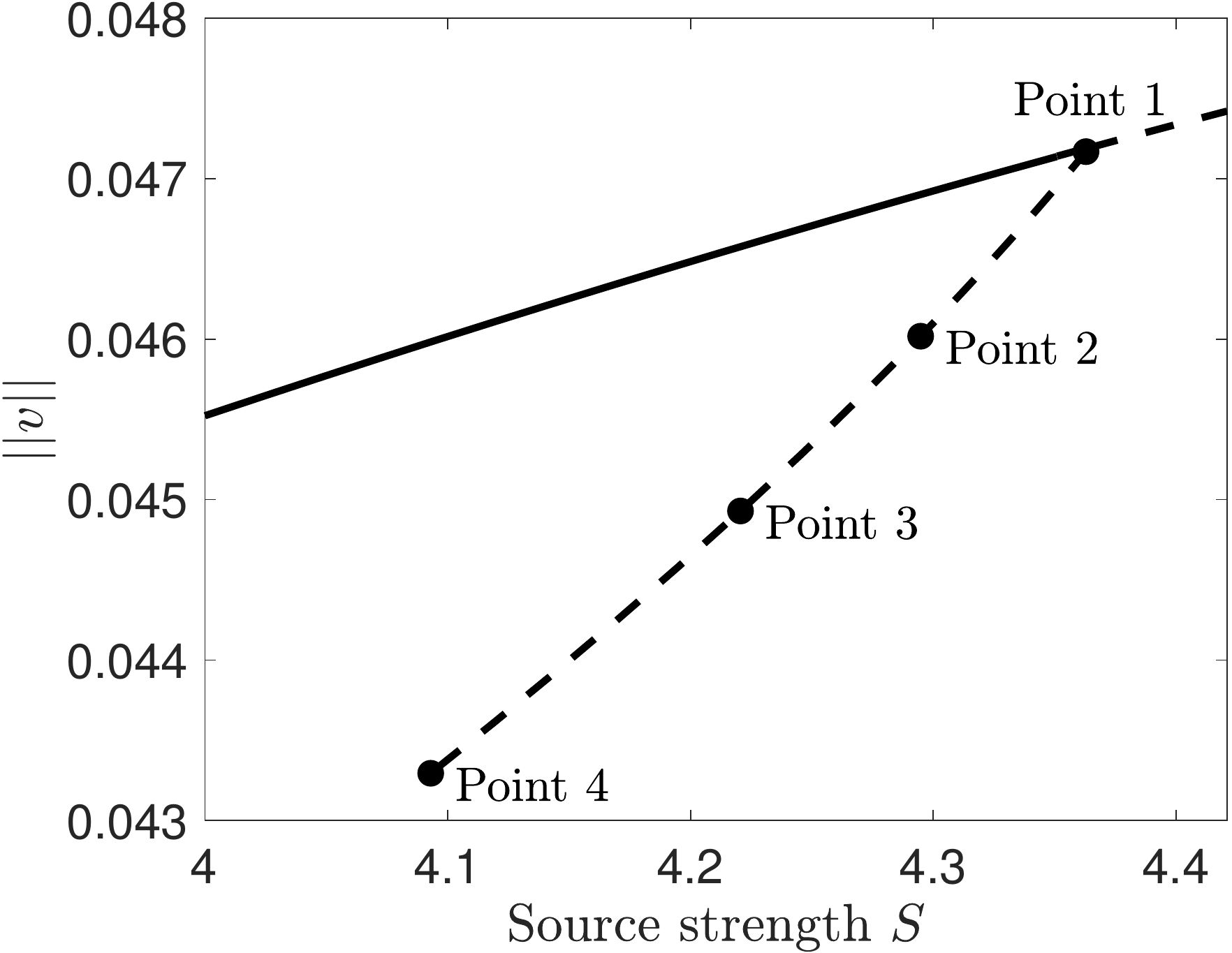} \hfill
\includegraphics[height=4.3cm,width=0.45\textwidth]{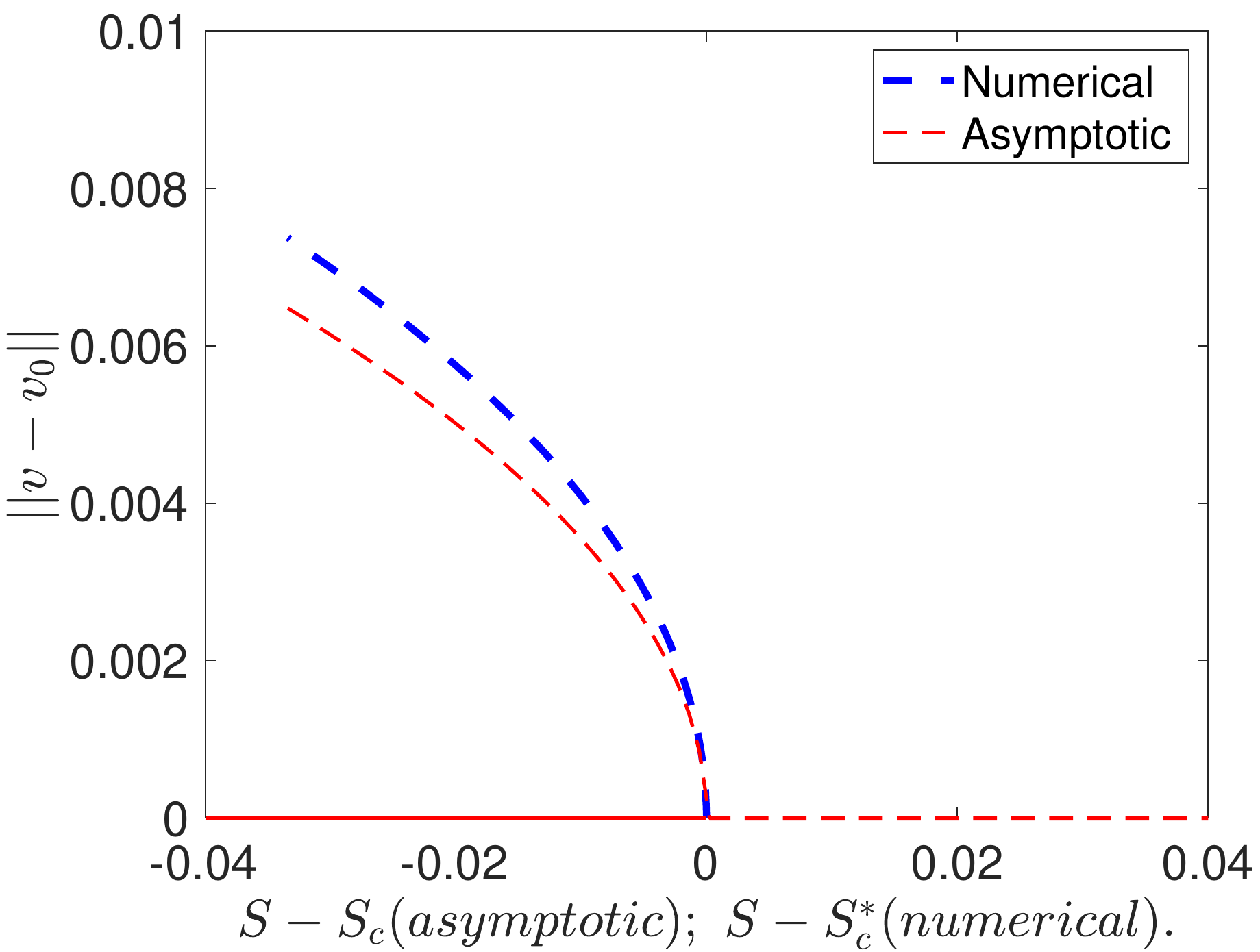}
\caption{Left panel: The $L_2$-norm of steady-state solution to
  \eqref{intro:pde} with $\eps=0.03$, as computed by the bifurcation
  software {\em pde2path} \cite{pde2path}. Numerically, the
  bifurcation occurs at $S_c^* \approx 4.3629$. The heavy solid curve
  is the radially symmetric spot solution branch. Right panel: Plot of
  $||v-v_0||$ from the numerically computed branches in the left
  panel versus $S-S_c^*$, where $S_c^* \approx 4.3629$ is the
  numerically computed bifurcation value. We compare it with the
  asymptotic result $\frac{\eps}{2}\sqrt{\frac{\pi c_1 (S_c-S)}{c_3}}$ in
  \eqref{amp:vnormdiff}, where $S_c \approx 4.3022$ is the asymptotic
  result computed from the eigenvalue problem \eqref{linstab:eig} for
  the mode $m=2$ peanut-shaped instability. The bifurcation is subcritical.}
\label{fig:bif_diag}
\end{figure}

\begin{figure}[htbp]
\centering
(a) \includegraphics[width=0.45\textwidth]{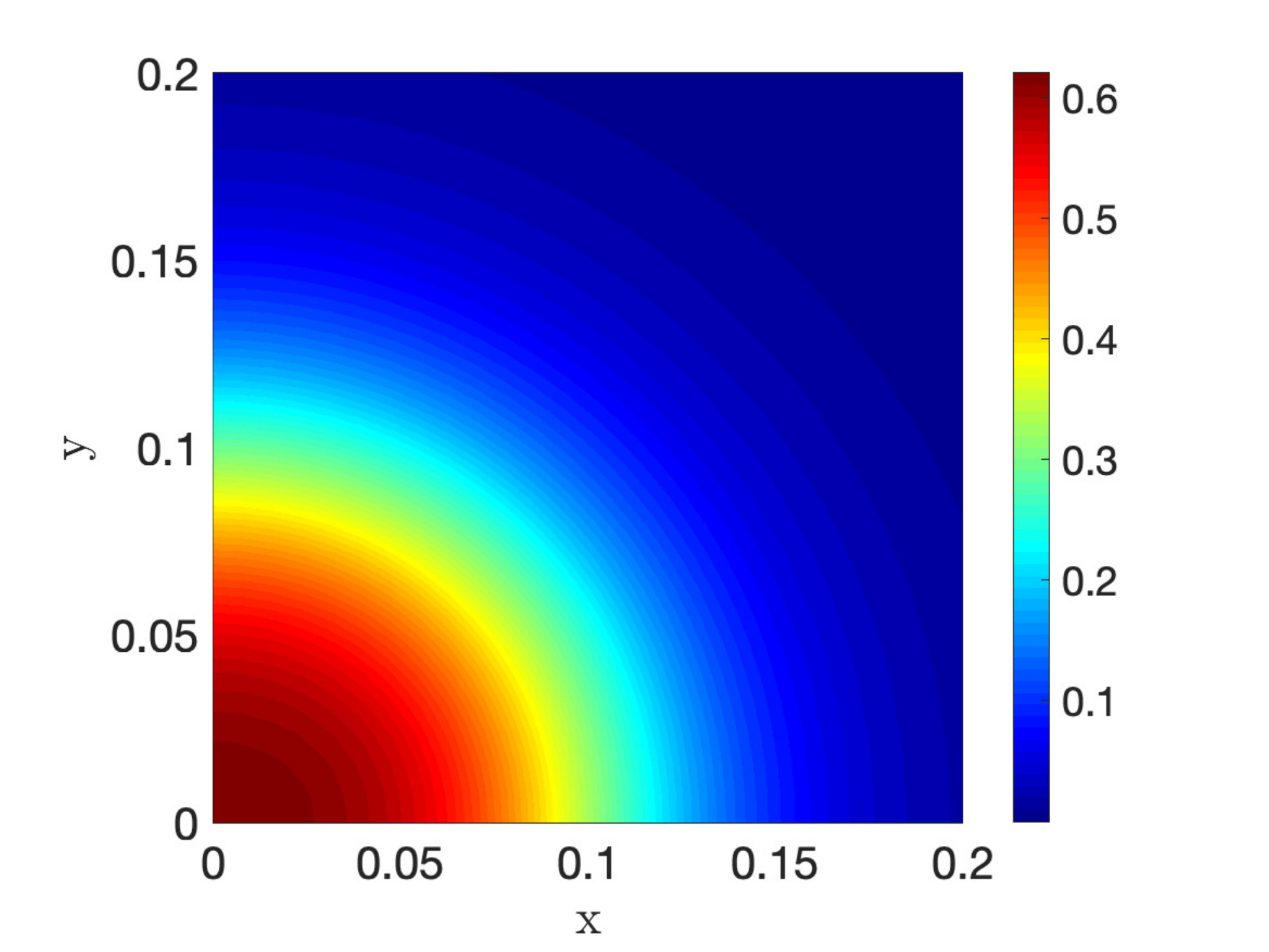} \hfill
(b) \includegraphics[width=0.45\textwidth]{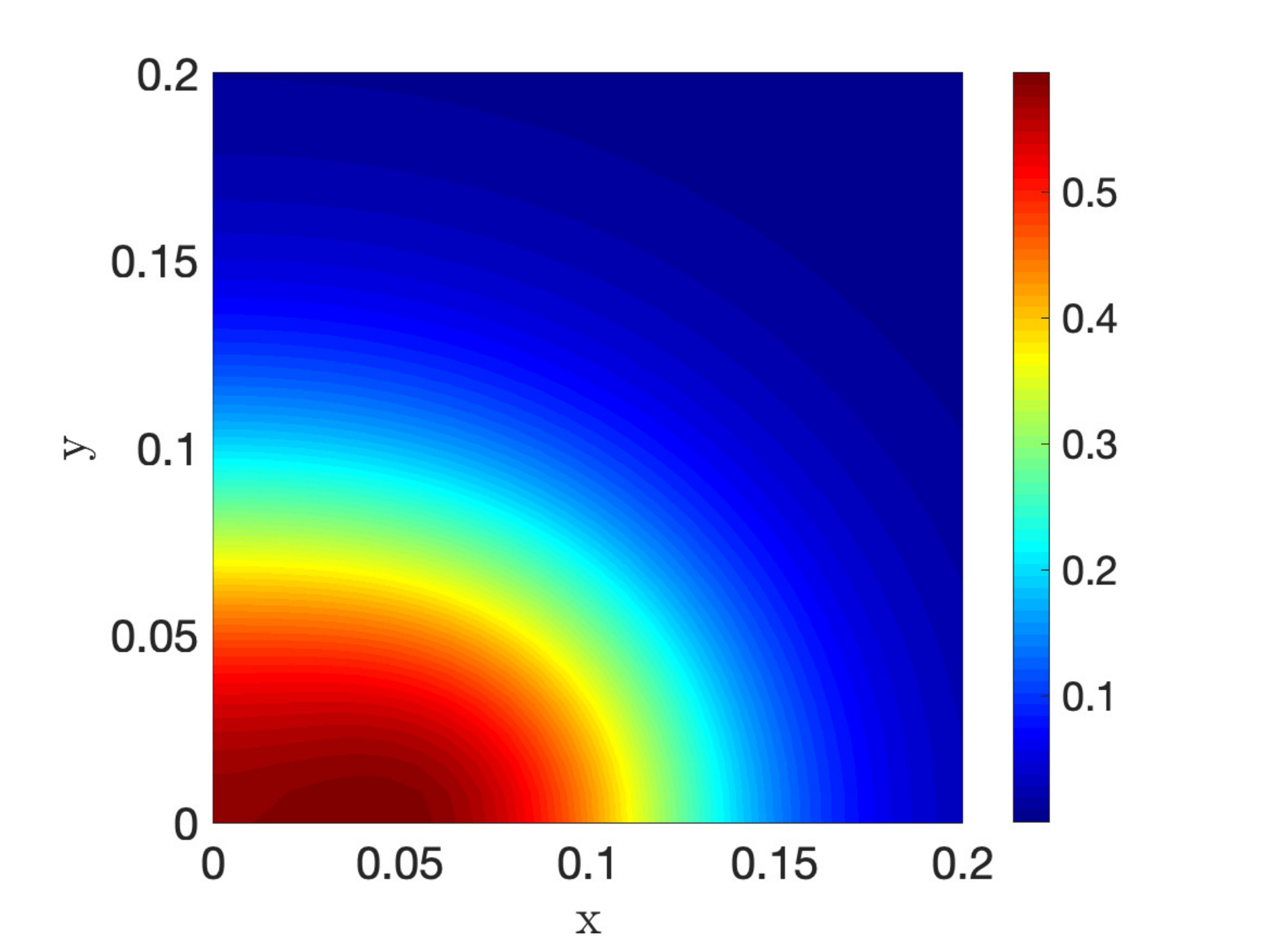}
\\
(c) \includegraphics[width=0.45\textwidth]{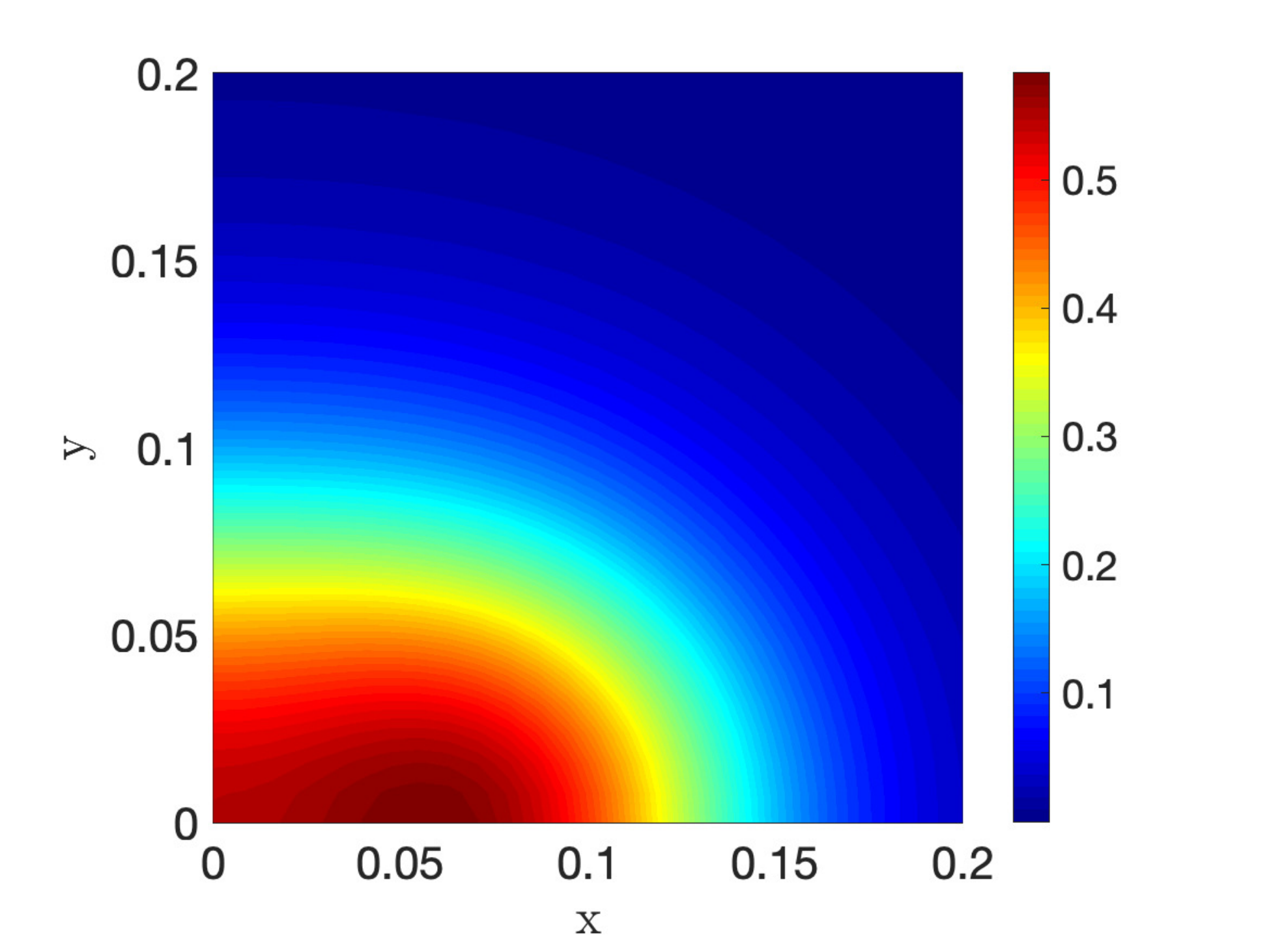} \hfill
(d) \includegraphics[width=0.45\textwidth]{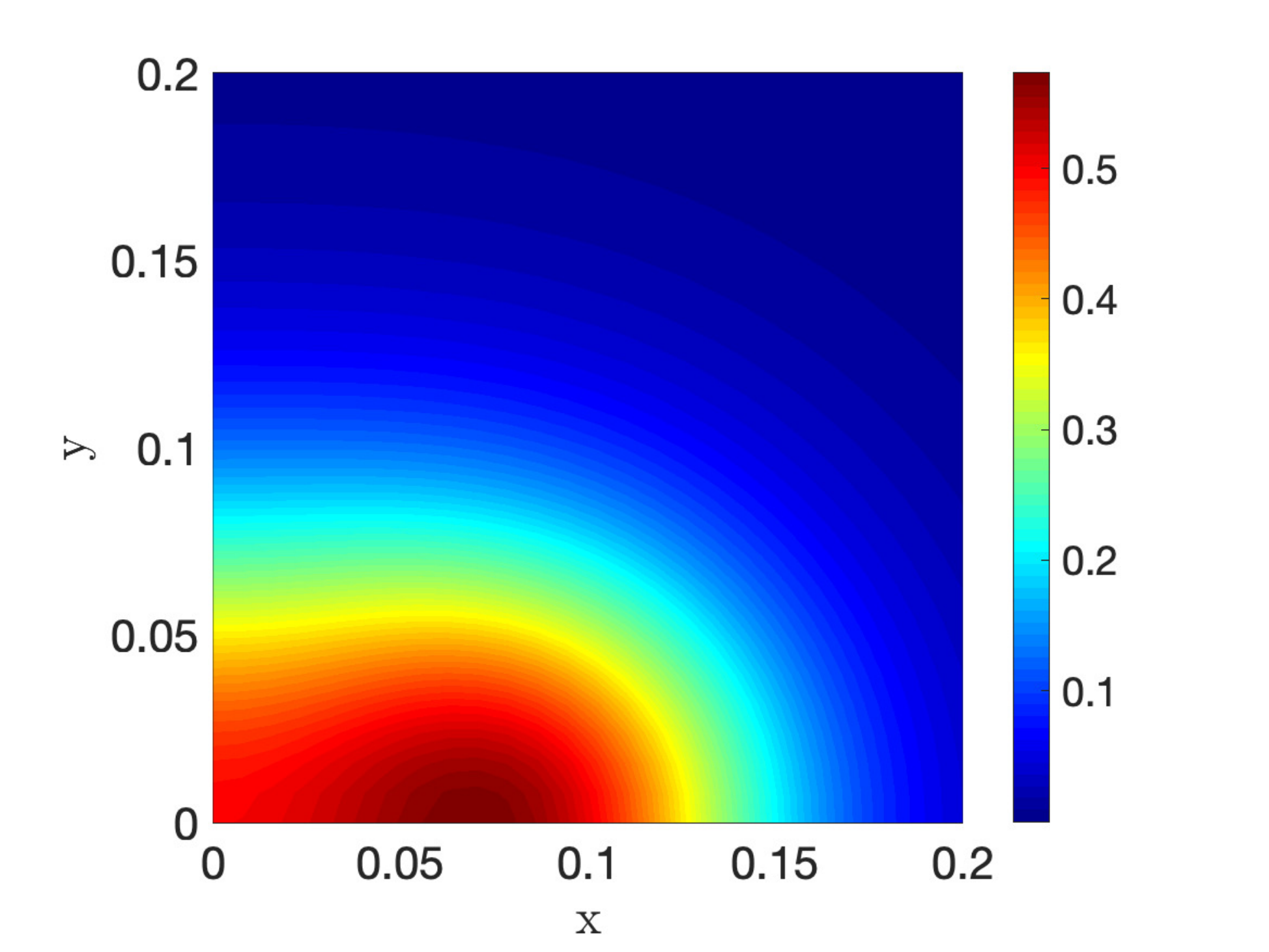}
\caption{Contour plot of the non-radially symmetric localized solution
  near the origin (zoomed) at the Points $1, 2, 3$ and $4$ as indicated
  in the bifurcation diagram in the left panel of
  Fig.\ref{fig:bif_diag}. }
\label{fig:bif_sol}
\end{figure}

In Fig.~\ref{fig:bif_diag} we show a favorable comparison of our
weakly nonlinear analysis result \eqref{amp:vnormdiff} with
corresponding full numerical results computed from the steady-state of
the Schnakenberg PDE system \eqref{intro:pde} with $\eps=0.03$ using
the bifurcation software {\em pde2path} \cite{pde2path}. The
computation is done in the quarter-disk geometry shown in the left panel of
Fig.~\ref{fig:eig_sc}. In Fig.~\ref{fig:bif_sol} we show contour
plots, zoomed near the origin, of the non-radially symmetric localized
steady-state at four points on the bifurcation diagram in
Fig.~\ref{fig:bif_diag}.

\section{Brusselator}\label{sec:bruss}

We now perform a similar weakly nonlinear analysis for the
Brusselator RD model. For this model, it is known that a localized
spot undergoes a peanut-shape deformation instability when the source
strength exceeds a threshold, with numerical evidence suggesting that
this linear instability is the trigger of a nonlinear spot-splitting
event (cf.~\cite{RRW}, \cite{TW}, \cite{TW2016}). Our weakly nonlinear
analysis will confirm that this peanut-shape symmetry-breaking bifurcation
is always subcritical.

The dimensionless Brusselator model in the two-dimensional unit disk $\Omega$
is formulated as (cf.~\cite{RRW})
\begin{equation}\label{bruss:pde}
  v_t = \eps^2 \Delta v + \eps^2 E - v + fuv^2\,, \quad
  \tau u_t = D \Delta u + \frac{1}{\eps^2} \left(v - uv^2\right)\,,
  \quad \v{x} \in \Omega\,,
\end{equation}
with no-flux boundary conditions $\partial_n u = \partial_n v = 0$ on
$\partial\Omega$. In \eqref{bruss:pde} the diffusivity $D$ and the feed-rate
$E$ are positive parameters, while the constant parameter $f$
satisfies $0<f<1$. Appendix A of \cite{RRW} provides the derivation of
\eqref{bruss:pde} starting from the form of the Brusselator model
introduced originally in \cite{PL}.

We first use the method of matched asymptotic expansions to construct
a one-spot steady-state solution centered at the origin of the unit
disk. In the inner region near $\v{x}=0$ we introduce $V$, $U$ and
$\v{y}$ by
\begin{equation}\label{b:inn_scale}
  v = \sqrt{D}\, V(\v{y})\,, \quad 
  u = {U(\v{y})/\sqrt{D}} \,, \quad \mbox{where} \quad
  \v{y} = \eps^{-1} \v{x} \,.
\end{equation}
In the inner region, for $\v{y}\in\R^2$, the steady-state problem obtained
from \eqref{bruss:pde} is
\begin{equation}
\Delta_{\v{y}} V - V + f U V^2 + \frac{\eps^2 E}{\sqrt{D}} = 0 \,, \qquad
\Delta_{\v{y}} U + V - U V^2 = 0 \,.
\end{equation}
Seeking a radially symmetric solution in the form
$V = V_0(\rho) + o(1)$ and $U = U_0(\rho)+o(1)$, with
$\rho = |\v{y}|$, we neglect the $\mc{O}(\eps^2)$ terms to obtain the
radially symmetric core problem
\begin{equation}\label{bruss:core_problem}
\begin{split}
  &\Delta_\rho V_0 - V_0 + f U_0 V_0^2 = 0\,, \quad
  \Delta_\rho U_0 = U_0 V_0^2 -V_0 \,, \quad \rho > 0\,, \\
  &V_0^{\prime}(0) = U_0^{\prime}(0) = 0\,; \quad V_0 \to 0\,, \quad
  U_0 \sim S \log \rho + \chi(S,f) + o(1)\,, \quad \mbox{as}
  \quad \rho \to \infty\,,
\end{split}
\end{equation}
where
$\Delta_{\rho}\equiv\partial_{\rho\rho}+\rho^{-1}\partial_{\rho}$.  We
observe that the $\mc{O}(1)$ term $\chi$, which must be
computed numerically, depends on the source strength $S$ and the
Brusselator parameter $f$, with $0<f<1$. By integrating the $U_0$
equation in \eqref{bruss:core_problem} we obtain the identity
\begin{equation}\label{bruss:S}
S = \int_0^\infty (U_0 V_0^2 - V_0) \rho \, d\rho\,.
\end{equation}

In the outer region, defined away from an $\mc{O}(\eps)$ region near
the origin, we obtain $v \sim \eps^2 E + \mc{O}(\eps^4)$ and that $u$
satisfies
\begin{equation}
D \Delta u + E + \frac{1}{\eps^2}( v - u v^2) = 0\,.
\end{equation}
Writing $v\sim\eps^2 E + \sqrt{D}V_0(\eps^{-1}|\v{x}|)$ and
$u\sim {U_0(\eps^{-1}|\v{x}|)/\sqrt{D}}$, we calculate in
the sense of distributions that, for $\eps\to 0$,
\begin{equation}
  \eps^{-2}\left( v - u v^2\right) \rightarrow E +
   2 \pi \sqrt{D} \int_0^\infty (V_0 - U_0 V_0^2) \rho \, d\rho
  = E -2 \pi \sqrt{D} S \delta(\v{x})\,,
\end{equation}
where we used \eqref{bruss:S} to obtain the last equality. Hence, upon
matching the outer to the inner solution for $u$, we obtain the
following outer problem:
\begin{equation}\label{bruss:outer}
\begin{split}
  &\Delta u = -\frac{E}{D} + \frac{2\pi S}{\sqrt{D}} \delta(\v{x})\,,
  \quad \v{x} \in \Omega\,, \quad \partial_n u = 0\,, \quad \v{x} \in
  \partial\Omega\,, \\
  &
  u \sim \frac{1}{\sqrt{D}} \left( S \log|\v{x}| +
    \frac{S}{\nu} + \chi \right) \quad \mbox{as} \quad \v{x} \to \v{0}\,,
  \quad \mbox{where} \quad \nu \equiv {-1/\log\eps}\,.
\end{split}
\end{equation}
By integrating \eqref{bruss:outer} over $\Omega$ and using the
Divergence theorem together with $|\Omega|=\pi$ we calculate $S$ as
\begin{equation}
S = \frac{E|\Omega|}{2 \pi \sqrt{D}} = \frac{E}{2\sqrt{D}} \,.
\end{equation}
The solution to \eqref{bruss:outer} is given by
\begin{equation}\label{bruss:outersol}
  u = \frac{1}{\sqrt{D}} \left(S \log |\v{x}| - \frac{E r^2}{4\sqrt{D}} +
    \frac{S}{\nu} + \chi\right)\,,
\end{equation}
where $r=|\v{x}|$.  Setting $|\v{x}| = \eps |\v{y}|$, and using $E=2S\sqrt{D}$,
we obtain that
\begin{equation}\label{bruss:inn_out}
u \sim \frac{1}{\sqrt{D}} \left(S \log|\v{y}| +\chi - \frac{S\eps^2 |\v{y}|^2}
    {2} \right)\,.
\end{equation}
This expression is identical to that derived in
\eqref{steady:u_sol_exp} for the Schnakenberg model, and shows that
there is an unmatched $\mc{O}(\eps^2 |\v{y}|^2)$ term feeding back
from the outer to the inner region (see Remark \ref{sch:equil_eps2}).

Next, we perform a linear stability analysis. Let $v_e, \, u_e$ denote
the steady-state spot solution centered at the origin. We introduce
the perturbation
\begin{equation}
v = v_e + e^{\lambda t}\phi\,,\quad u = u_e + e^{\lambda t} \eta \,,
\end{equation}
into \eqref{bruss:pde} and linearize. In this way, we obtain the eigenvalue
problem
\begin{equation}\label{bruss:full_eig}
  \eps^2 \Delta \phi - \phi + 2 f u_e v_e \phi + f v_e^2 \eta =
  \lambda \phi \,, \qquad
  D \Delta \eta + \frac{1}{\eps^2}(\phi - 2 u_e v_e \phi - v_e^2 \eta)
  = \tau \lambda \eta \,,
\end{equation}
with $\partial_n \phi = \partial_n \eta = 0$ on $\partial\Omega$. In the
inner region near $\v{x}=0$ we introduce
\begin{equation}
\begin{pmatrix} \phi \\ \eta\end{pmatrix} = \mrm{Re}(e^{im\theta}) \begin{pmatrix}
  \Phi(\rho) \\ N(\rho)/D \end{pmatrix}\,, \quad \mbox{where} \quad
\rho = |\v{y}| = \eps|\v{x}|\,, \quad \theta=\mbox{arg}(\v{y})\,,
\end{equation}
and $m=2,3,\ldots$.  With $v_e \sim \sqrt{D}V_0$ and
$u_e \sim {U_0/\sqrt{D}}$, we neglect the $\mc{O}(\eps^2)$ terms to
obtain the following spectral problem governing non-radially symmetric
instabilities of the steady-state spot solution:
\begin{subequations} \label{bruss:loc_eig}
\begin{equation}\label{bruss:eig}
  \mc{L}_m \begin{pmatrix} \Phi\\ N \end{pmatrix} + M \begin{pmatrix}
    \Phi \\ N \end{pmatrix} =
  \lambda \begin{pmatrix}  1 & 0 \\ 0 & 0 \end{pmatrix}
  \begin{pmatrix} \Phi \\ N \end{pmatrix} \,.
\end{equation}
Here we have defined
\begin{equation}\label{bruss:eig_mat}
  \mc{L}_m \v{\Phi} \equiv \partial_{\rho\rho}\v{\Phi} +
  \frac{1}{\rho} \partial_\rho \v{\Phi} - \frac{m^2}{\rho^2} \v{\Phi}\,,
  \qquad M \equiv \begin{pmatrix}
    2 f U_0 V_0 - 1 & f V_0^2 \\ 1 - 2 U_0 V_0 & -V_0^2 \end{pmatrix}\,.
\end{equation}
\end{subequations}
We seek eigenfunctions of \eqref{bruss:loc_eig} with $\Phi\to 0$ and
$N\to 0$ as $\rho\to\infty$. 

\begin{figure}[htbp]
	\centering
	\includegraphics[height=4.2cm, width=0.50\textwidth]{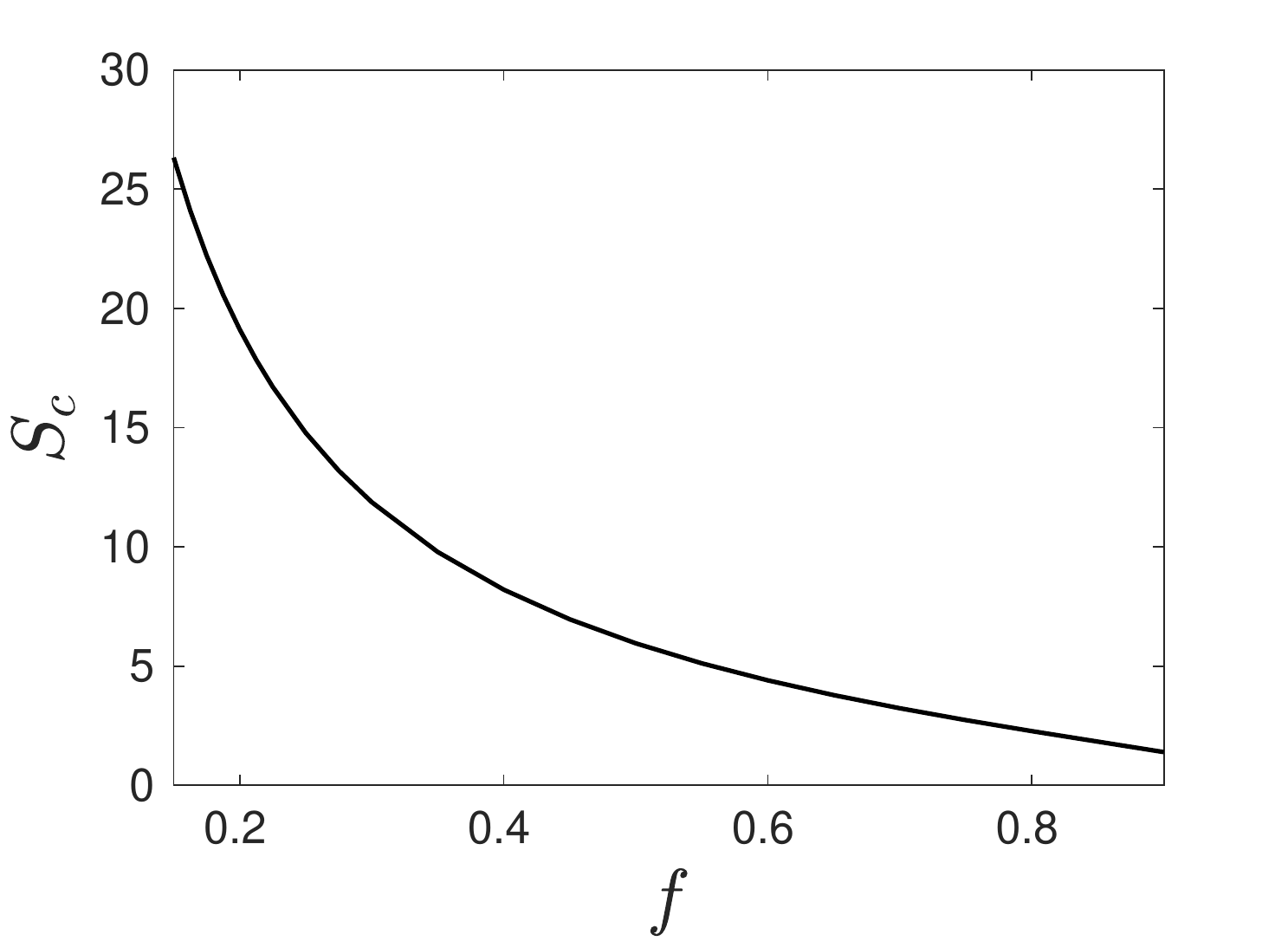} 
	\caption{Numerical results, computed from
          \eqref{bruss:loc_eig} with $m=2$, for the critical value
          $S_c$ of the source strength versus the Brusselator
          parameter $f$ on $0.15<f<0.9$ at which a one-spot solution
          first undergoes a peanut-shaped linear instability. The spot
          is unstable when $S>S_c$.}\label{fig:bruss_Sc}
\end{figure}

Next, we determine the stability threshold for a peanut-shape
deformation instability with angular mode $m=2$. For $m=2$, the
appropriate far-field condition is that $\Phi\to 0$ exponentially and
$\partial_{\rho} N \sim -{2N/\rho}$ for $\rho\to\infty$. As such, we
impose $N^{\prime} \sim -{2N/\rho}$ for $\rho\gg 1$. We denote
$\lambda_0$ as the eigenvalue of \eqref{bruss:loc_eig} with the
largest real part. Our numerical computations show that for fixed $f$
on $0<f<1$ we have $\mrm{Re}(\lambda_0)=0$ at some $S=S_c(f)$, and
that $\mrm{Re}(\lambda_0)>0$ for $S>S_c(f)$. In
Fig.~\ref{fig:bruss_Sc} we plot our results for $S_c(f)$ on
$0.15<f<0.9$. These results are consistent with the corresponding
thresholds first computed in \S 3 of \cite{RRW} at some specific
values of $f$. Moreover, as shown in Figure 4 of \cite{RRW},
  the peanut-splitting mode $m=2$ is the first mode to lose stability
  as $S$, or equivalently $E$, is increased.  Higher modes lose
  stability at larger value of $S$. Since in our weakly nonlinear
  analysis we will only consider the neighbourhood of the instability
  threshold for the peanut-splitting mode, the higher modes of spot-shape
  deformation are all linearly stable in this neighborhood.

\begin{figure}[htbp]
 \centering
 \includegraphics[width=0.48\textwidth]{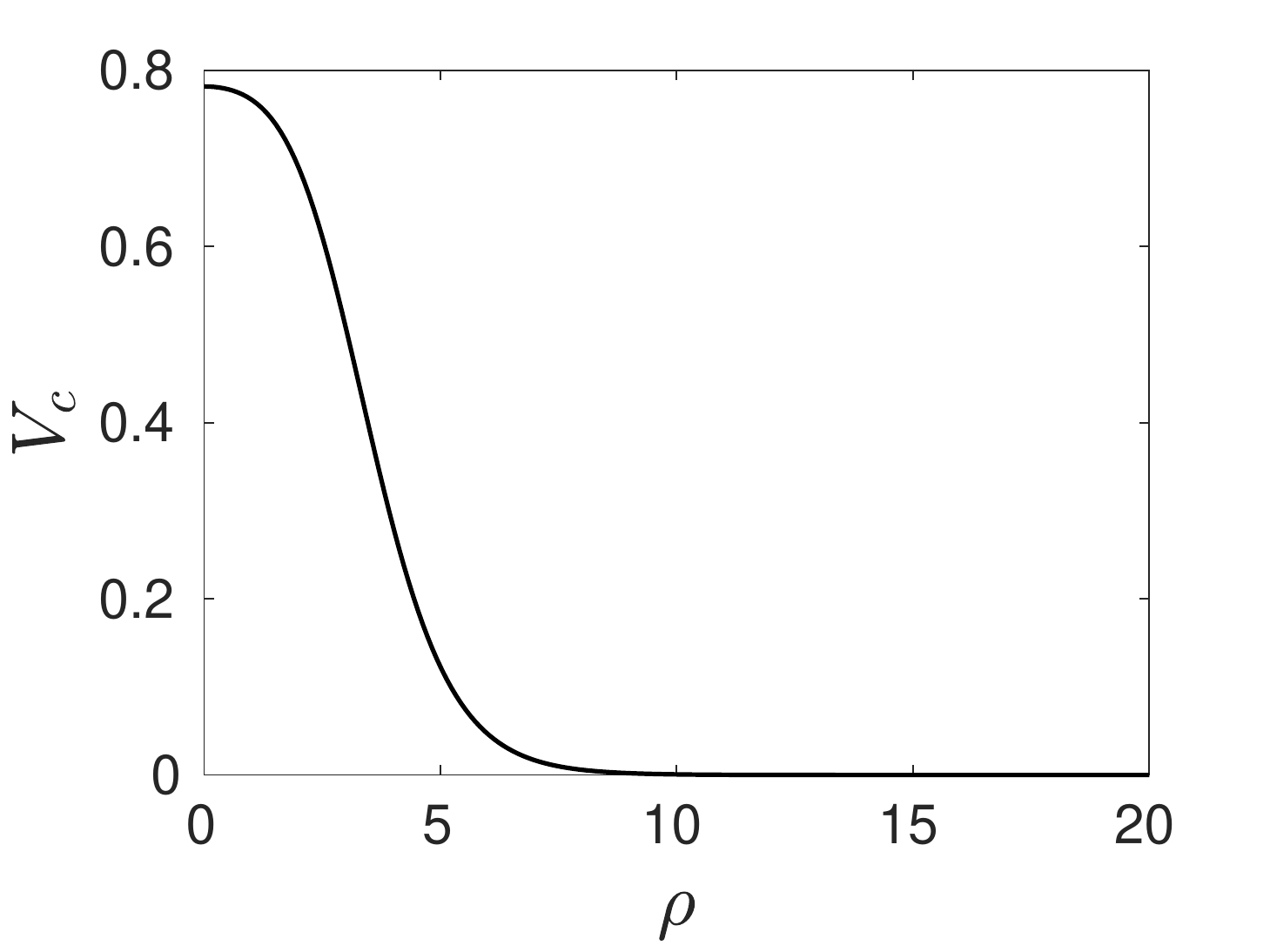} \hfill
 \includegraphics[width=0.48\textwidth]{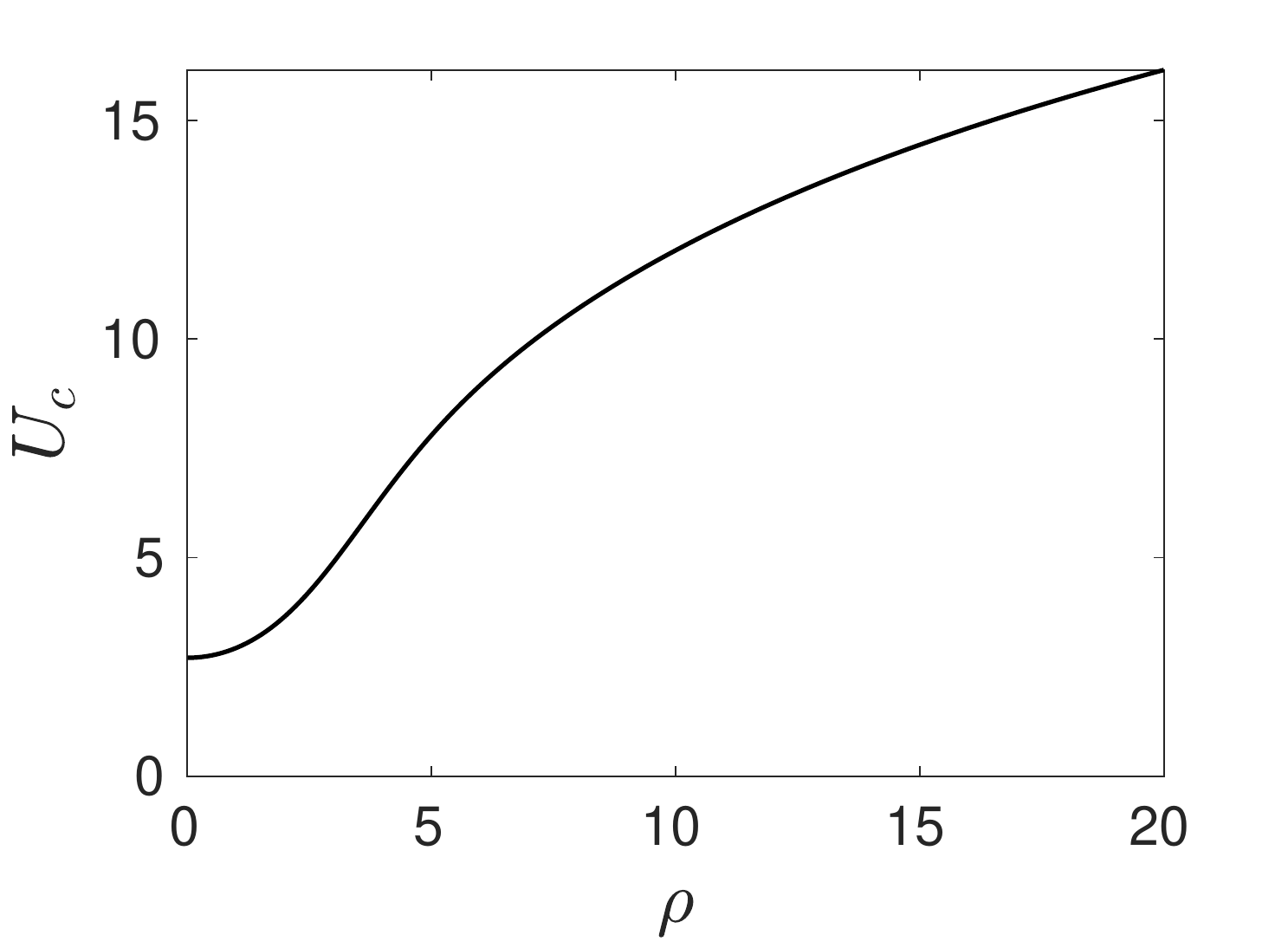}
 \caption{Plot of the core solution, computed numerically from
   \eqref{bruss:core_problem}, at $S=S_c(f)$ where the peanut-shape
   instability originates when $f=0.5$. Left panel: $V_c(\rho)$.
   Right panel: $U_c(\rho)$.}\label{fig:bruss_core}
\end{figure}

\begin{figure}[htbp]
	\centering
	\includegraphics[width=0.48\textwidth]{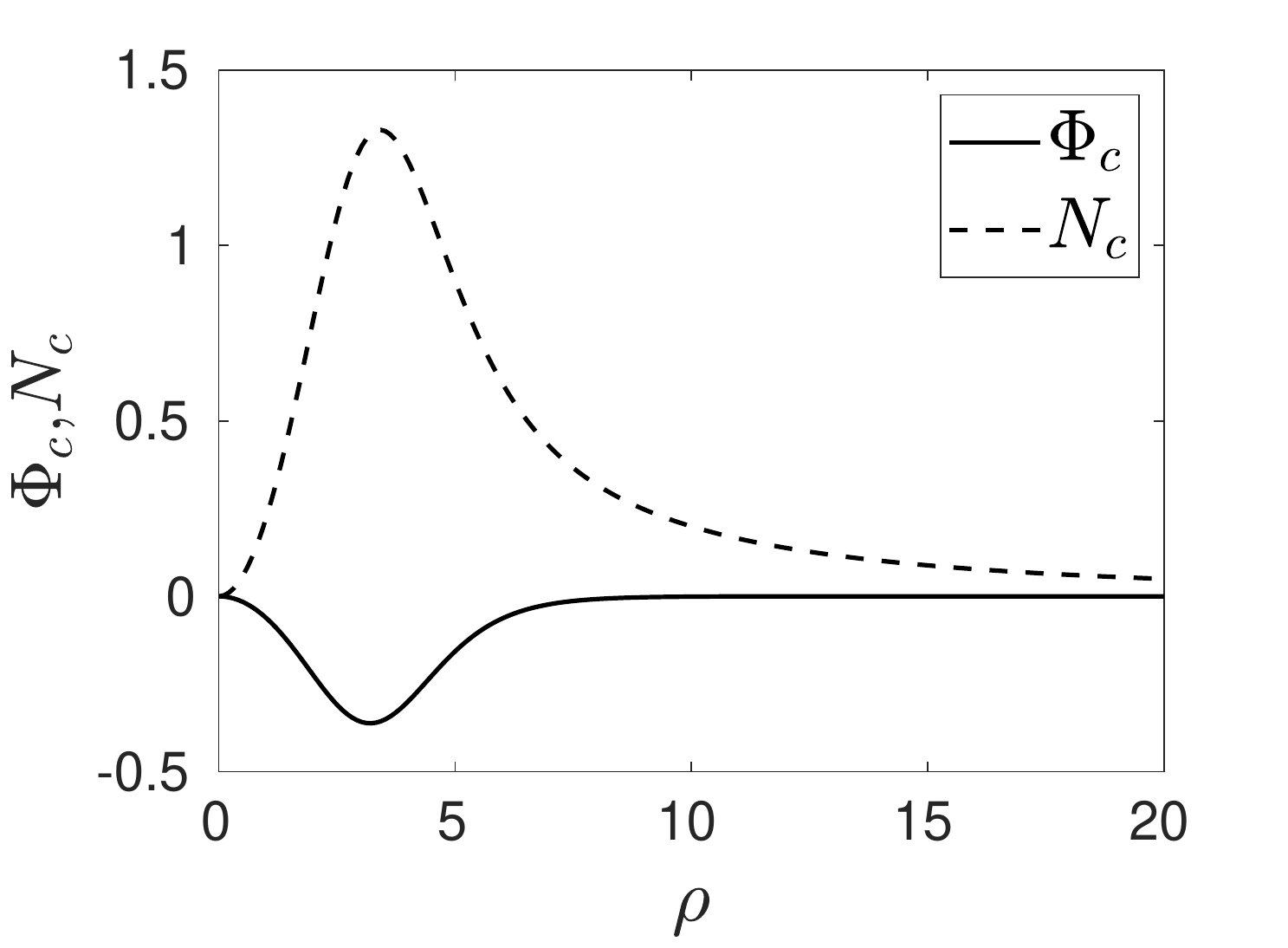} \hfill
	\includegraphics[width=0.48\textwidth]{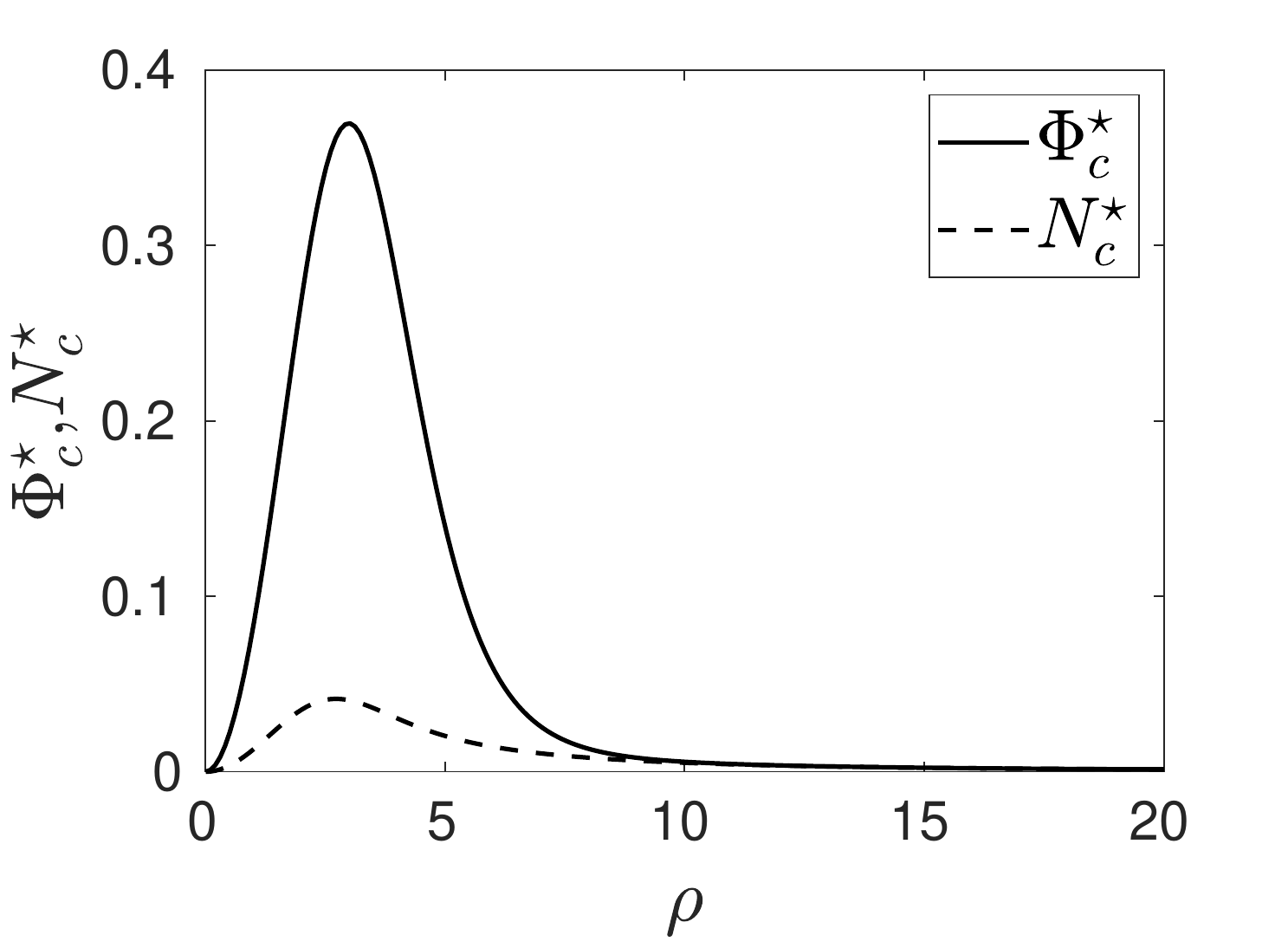}
	\caption{Left panel: Plot of $\Phi_c$ (solid curve) and $N_c$
          (dashed curve) for $f=0.5$, computed numerically from
          \eqref{bruss:null}. Right panel: Plot of $\Phi_c^{*}$ (solid
          curve) and $N_c^{*}$ (dashed curve) for $f=0.5$, computed
          numerically from
          \eqref{bruss:adj}.}\label{fig:bruss_nulladj}
\end{figure}

We denote $V_c(\rho)$ and $U_c(\rho)$ by $V_c \equiv V_0(\rho \,; \, S_c)$
and $U_c \equiv U_0(\rho \,; \, S_c)$, and we label $\v{\Phi}_c \equiv
(\Phi_c, N_c)$ as the normalized critical eigenfunction at $S=S_c$, which
satisfies
\begin{equation}\label{bruss:null}
  \mc{L}_2 \v{\Phi}_c  + M_c \v{\Phi_c}  = \v{0}\,, \quad
  M_c \equiv \begin{pmatrix} 2 f U_c V_c - 1 & f V_c^2 \\ 1 - 2 U_c V_c & -V_c^2
  \end{pmatrix}\,, \quad \mbox{with} \quad
  \int_0^\infty \Phi_c^2 \, \rho \, d\rho = 1\,.
\end{equation}
Likewise, at $S=S_c$, there exists a non-trivial normalized solution
$\v{\Phi}_c^* = (\Phi_c^*, N_c^*)$ to the homogeneous adjoint problem
\begin{equation}\label{bruss:adj}
  \mc{L}_2 \v{\Phi}_c^* + M_c^T \v{\Phi}_c^* = \v{0}\,, \quad \mbox{with}
  \quad \int_0^\infty (\Phi_c^*)^2 \, \rho \, d\rho = 1\,,
\end{equation}
where $\Phi_c^* \to 0$ and
$\partial_\rho N_c^{*\prime} \sim -{2N_c^{*}/\rho}$ as
$\rho \to \infty$. In Fig.~\ref{fig:bruss_core} we plot the core
solution $V_c$ and $U_c$ for $f=0.5$. In Fig.~\ref{fig:bruss_nulladj}
we plot the numerically computed eigenfunction $\Phi_c,\, N_c$ (left
panel) and adjoint eigenfunction $\Phi_c^{*},\, N_c^{*}$ (right panel)
when $f=0.5$.

\subsection{Amplitude equation for the Brusselator model}
\label{sec:bruss_amp_der}

We now derive the amplitude equation associated with the
peanut-splitting linear stability threshold for the Brusselator. Since
this analysis is very similar to that for the Schnakenberg model in
\S \ref{sec:amp} we only briefly outline the analysis.

We begin by introducing a neighborhood of $S_c$ and a slow time $T$
defined by
\begin{equation}\label{bruss:newvar}
  S = S_c + \kappa \sigma^2\,, \quad \kappa = \pm 1\,; \qquad T \equiv
  \sigma^2 t\,.
\end{equation}
In terms of the inner variables \eqref{b:inn_scale} and
\eqref{bruss:newvar}, we have
\begin{equation}\label{bruss:pde2}
\begin{split}
\sigma^2 V_T &= \Delta_\v{y} V - V + f U V^2 + \frac{\eps^2 E}{\sqrt{D}}\,, \\
\frac{\tau}{D} \eps^2 \sigma^2 U_T &= \Delta_\v{y} U + V - U V^2\,,
\end{split}
\end{equation}
with $V\to 0$ exponentially as $\rho\to\infty$ and
\begin{equation}
  \quad U \sim (S_c + \kappa \sigma^2) \log\rho + \chi(S_c) +
  \sigma^2\left[ \kappa \chi^{\prime}(S_c) + \mc{O}(1)\right]\,, \quad
  \mbox{as} \quad \rho=|\v{y}|\to \infty\,.
\end{equation}

We now use an approach similar to that in \S \ref{sec:amp} to derive
the amplitude equation for the Brusselator model. We substitute the
expansion \eqref{amp:expansion} into \eqref{bruss:pde2} and collect
powers of $\sigma$, and we assume that $\sigma^3\gg \mc{O}(\eps^2)$
as in \eqref{amp:sig_eps}. To leading order in $\sigma$, we obtain
that $V_0 = V_c$ and $U_0 = U_c$. The solution $(V_1,U_1)$ of the
$\mc{O}(\sigma)$ problem is 
\begin{equation}\label{bruss:v1}
  \begin{pmatrix} V_1 \\ U_1 \end{pmatrix} =
  A(T) \cos(2\phi) \begin{pmatrix} \Phi_c(\rho) \\ N_c(\rho) \end{pmatrix}\,,
\end{equation}
where $A(T)$ is the unknown amplitude and $\Phi_c, \, N_c$ is the
eigenfunction of \eqref{bruss:null}.

From our assumption that $\sigma^3 \gg \mc{O}(\eps^2)$, we can neglect the
$\mc{O}(\eps^2)$ terms in \eqref{bruss:pde2} as well as the
$\mc{O}(\eps^2)$ feedback term in \eqref{bruss:inn_out} arising from
the outer solution. In this way, the $\mc{O}(\sigma^2)$ problem for
$\v{V}_2 = (V_2, U_2)$ is given on $\v{y}\in \R^2$  by
\begin{equation}\label{bruss:amp_sigma2}
\begin{split}
  & \Delta_\v{y} \v{V}_2 + M_c \v{V}_2 = F_2 \, \v{q}\,, \quad \mbox{where}
  \quad F_2 \equiv U_c V_1^2 + 2 V_c V_1 U_1\,, \quad \v{q} \equiv
  \begin{pmatrix} -f \\ 1 \end{pmatrix}, \\
    & V_2 \to 0, \quad U_2 \sim \kappa \left[ \log\rho +
      \left.\frac{\partial\chi(S\,;f)}{\partial S}\right\vert_{S=S_c} +
      \mc{O}(1) \right]\,, \quad \mbox{as} \quad
  \rho \to \infty\,.
\end{split}
\end{equation}
Here $M_c$ is given in \eqref{bruss:null}. As we have shown in \S
\ref{sec:amp}, the solution to \eqref{bruss:amp_sigma2} can be
conveniently decomposed as
\begin{equation}\label{bruss:decv2}
  \v{V}_2 = \kappa \wtwoh + A^2 \hat{\v{V}}_{20}(\rho) +
  A^2 \v{V}_{24}(\rho)\cos(4\phi) \,,
\end{equation}
where $\wtwoh = (\partial_S V_c, \, \partial_S U_c)$. Here
$\hat{\v{V}}_{20} = (\hat{V}_{20}, \hat{U}_{20})$ and $\v{V}_{24}
= (V_{24}, U_{24})$ satisfy
\begin{subequations}
\begin{align}
  \Delta_\rho \hat{\v{V}}_{20} + M_c \hat{\v{V}}_{20} = F_{20} \, \v{q}\,;
  &\qquad \hat{V}_{20} \to 0\,, \quad \hat{U}_{20}^{\prime} \to 0\,,\quad
    \mbox{as} \quad \rho \to \infty\,, \label{bruss:vhat20}\\
  \mc{L}_4 \v{V}_{24} + M_c \v{V}_{24} = F_{20} \, \v{q}\,;
  &\qquad V_{24} \to 0\,, \quad U_{24}^{\prime} \sim -
    \frac{4U_{24}}{\rho}\,, \quad \mbox{as} \quad \rho \to \infty\,.
    \label{bruss:vhat24}
\end{align}
\end{subequations}
Here $F_{20}=F_{20}(\rho)$ is defined by
\begin{equation}\label{bruss:f20}
 F_{20} = \frac{1}{2}\left(U_c \Phi_c^2 + 2 V_c \Phi_c N_c\right) \,.
\end{equation}

\begin{figure}[htbp]
	\centering
	\includegraphics[width=0.48\textwidth]{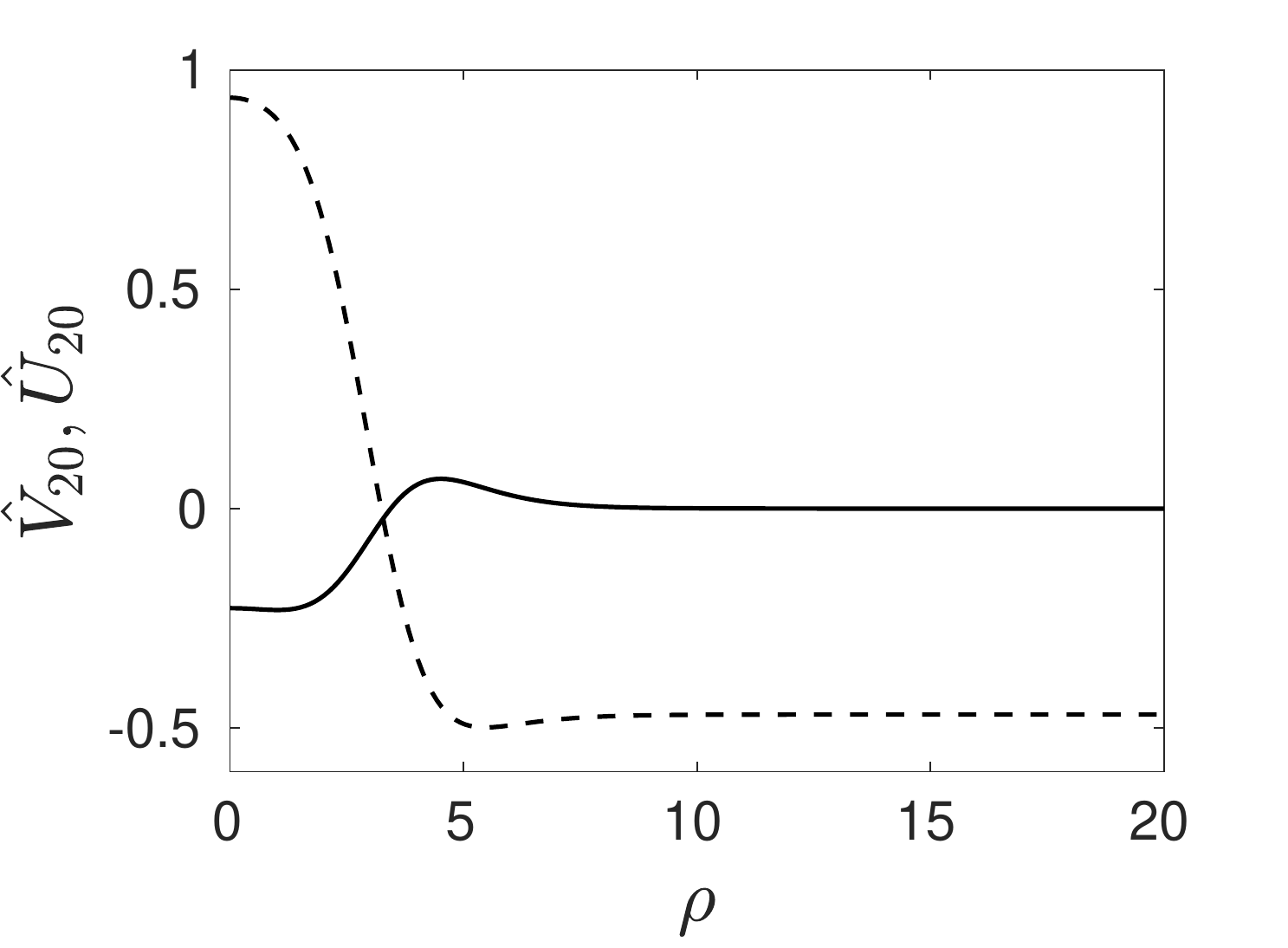} \hfill
	\includegraphics[width=0.48\textwidth]{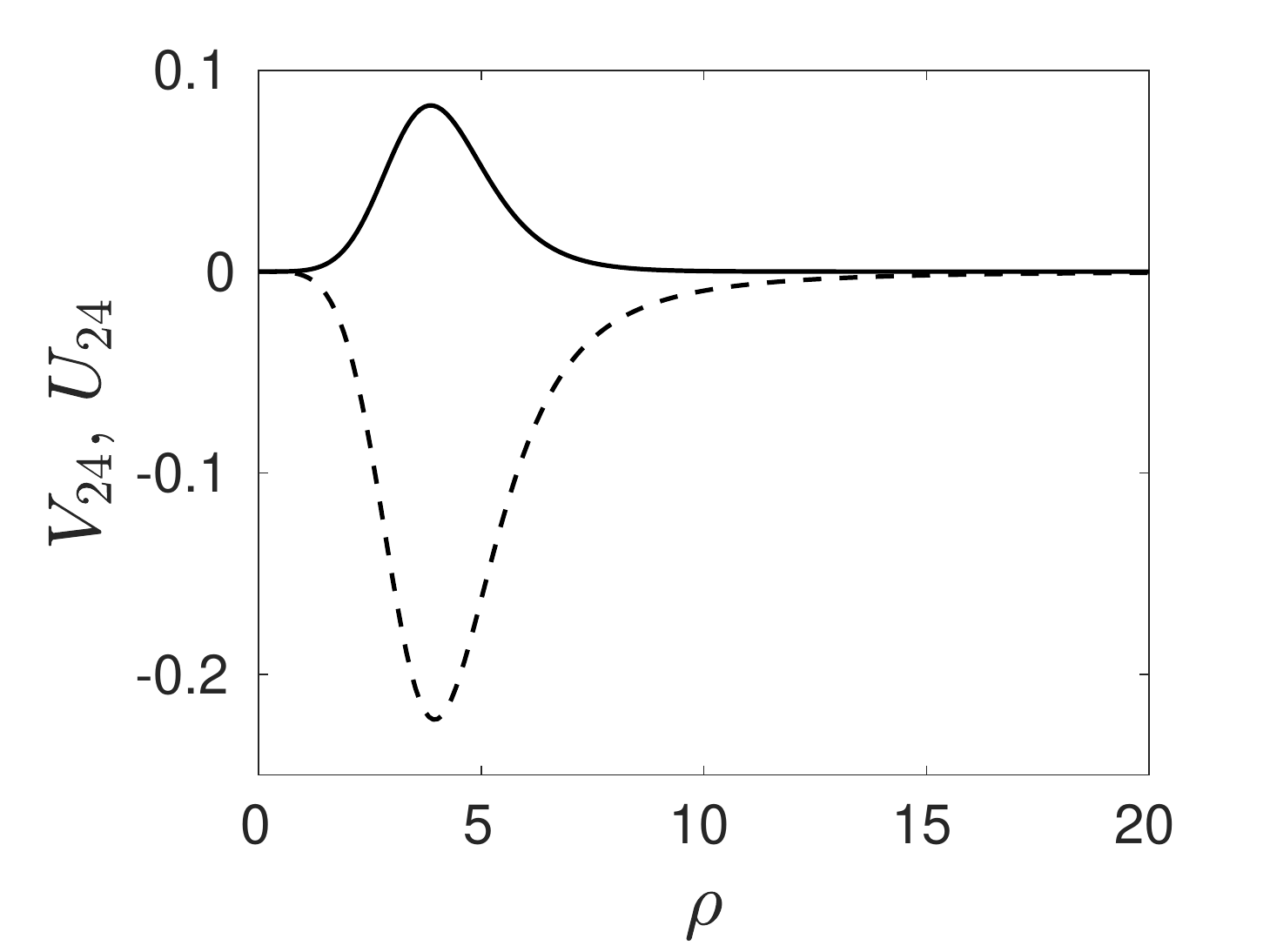}
	\caption{Left panel: $\hat{V}_{20}$ (solid curve) and
          $\hat{U}_{20}$ (dashed curve) for $f=0.5$ as computed numerically
          from \eqref{bruss:vhat20}. Right panel: $\hat{V}_{24}$ (solid curve)
          and $\hat{U}_{24}$ (dashed curve) for $f=0.5$ as computed numerically
          from \eqref{bruss:vhat24}.}\label{fig:bruss:2024}
\end{figure}

As in \S \ref{sec:amp}, we must numerically compute the solutions to
\eqref{bruss:vhat20} and \eqref{bruss:vhat24}. In Fig.~\ref{fig:bruss:2024}
we plot these solutions for $f=0.5$. We observe from the left panel of
Fig.~\ref{fig:bruss:2024} that $\hat{U}_{20}$ tends to a nonzero constant
for $\rho \gg 1$.

Next, by collecting the $\mc{O}(\sigma^3)$ terms in the weakly nonlinear
expansion, we find that $\v{V}_3 = (V_3, U_3)$ satisfies
\begin{subequations} \label{bruss:v3_all}
\begin{equation}\label{bruss:v3}
  \Delta_\v{y} \v{V}_3 + M_c \v{V}_3 = F_3 \, \v{q} + \partial_T V_1 \,
  \v{e}_1\,, \quad \v{V}_3 \to \v{0}\,, \quad \mbox{as} \quad \rho \to \infty
  \,.
\end{equation}
Here $\v{q}$ is defined in \eqref{bruss:amp_sigma2}, while $F_3$ and
$\v{e}_1$ are defined by
\begin{equation}\label{bruss:v3_def}
  F_3 \equiv 2 V_c V_1 U_2 + U_1 V_1^2 + 2 V_c U_1 V_2 + 2 U_c V_1 V_2\,,
  \qquad \v{e}_1 \equiv (1,0)\,.
\end{equation}
\end{subequations}
By using the expressions for $V_1,\, U_1$ and $V_2,\, U_2$ from
\eqref{bruss:v1} and \eqref{bruss:decv2}, respectively, we can obtain
a modal expansion of $F_3$ exactly as in \eqref{amp:F3_all} for the
Schnakenberg model. In this way, we obtain \eqref{amp:V3_nodal} in which
we replace $\v{q}$ by $\v{q} = (-f,1)$.

The remainder of the analysis involving the imposition of the
solvability condition to derive the amplitude equation exactly
parallels that done in \S \ref{sec:amp}. We conclude that the
amplitude equation associated with peanut-shape deformations of a a
spot is
\begin{subequations}\label{bruss:normal_all}
\begin{equation}\label{bruss:normalform0}
 \frac{dA}{dT} = \kappa c_1 A + c_3 A^3\,, \quad T = \sigma^2 t\,,
\end{equation}
where $c_1$ and $c_3$, which depend on the Brusselator parameter $f$,
are given by
\begin{equation}\label{bruss:coeff}
          c_1 = \frac{\int_0^\infty g_1 (f \Phi_c^* - N_c^*) \, \rho \,
    \mrm{d}\rho}{\int_0^\infty \Phi_c \Phi_c^* \, \rho \, \mrm{d}\rho}\,, \quad
  c_3 = \frac{\int_0^\infty g_2 (f \Phi_c^* - N_c^*) \, \rho \,\mrm{d} \rho}
  {\int_0^\infty \Phi_c \Phi_c^* \, \rho \, \mrm{d}\rho}\,.
	\end{equation}
\end{subequations}
Here $g_1$ and $g_2$ are defined in \eqref{amp:g1} and \eqref{amp:g2},
respectively, in terms of the Brusselator core solution $V_c,\,U_c$,
its eigenfunction $\Phi_c,\,N_c$ satisfying \eqref{bruss:null}, and
the solutions to \eqref{bruss:vhat20} and \eqref{bruss:vhat24}.

\begin{figure}[htbp]
	\centering
	\includegraphics[width=0.48\textwidth]{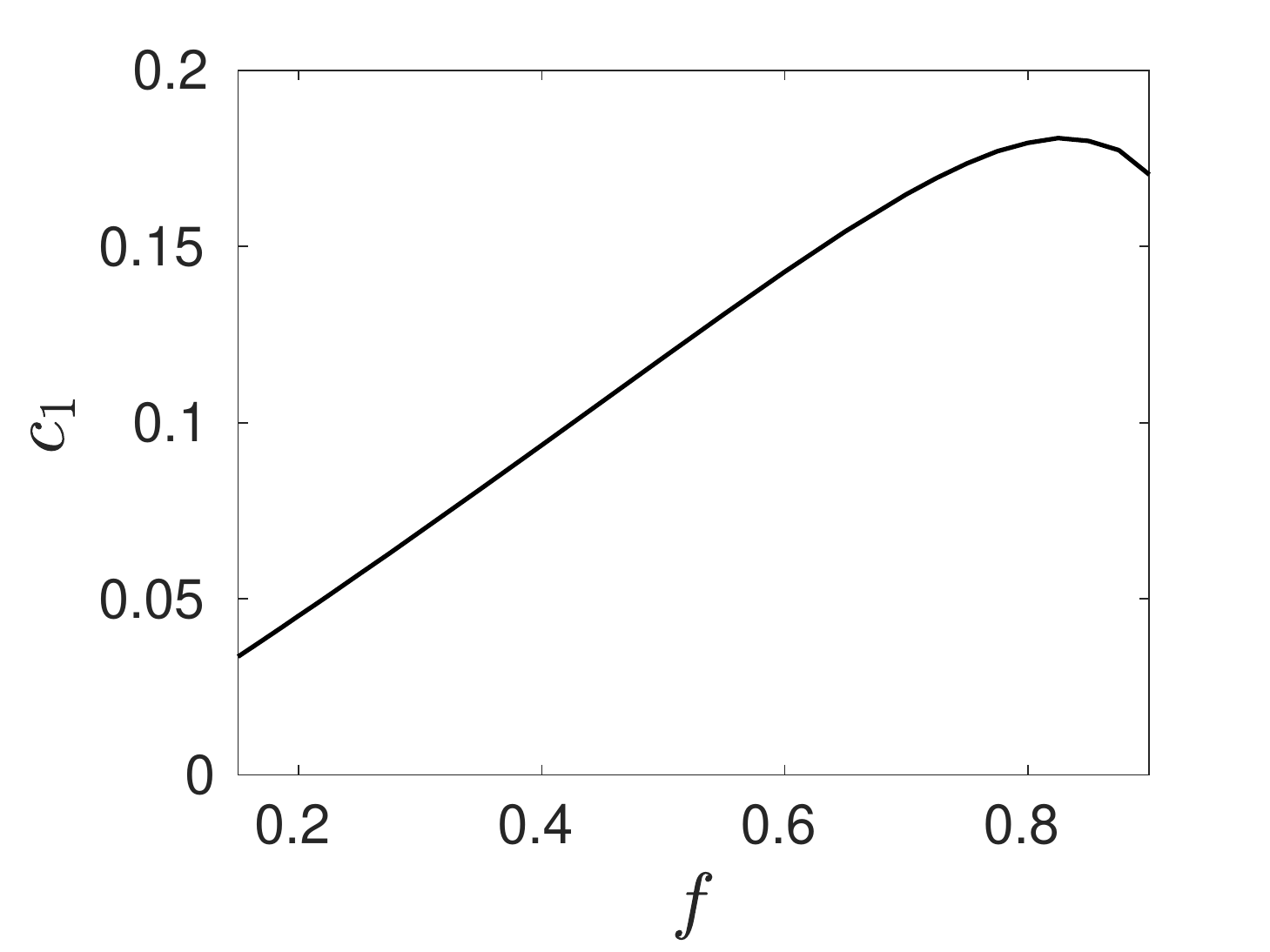} \hfill
	\includegraphics[width=0.48\textwidth]{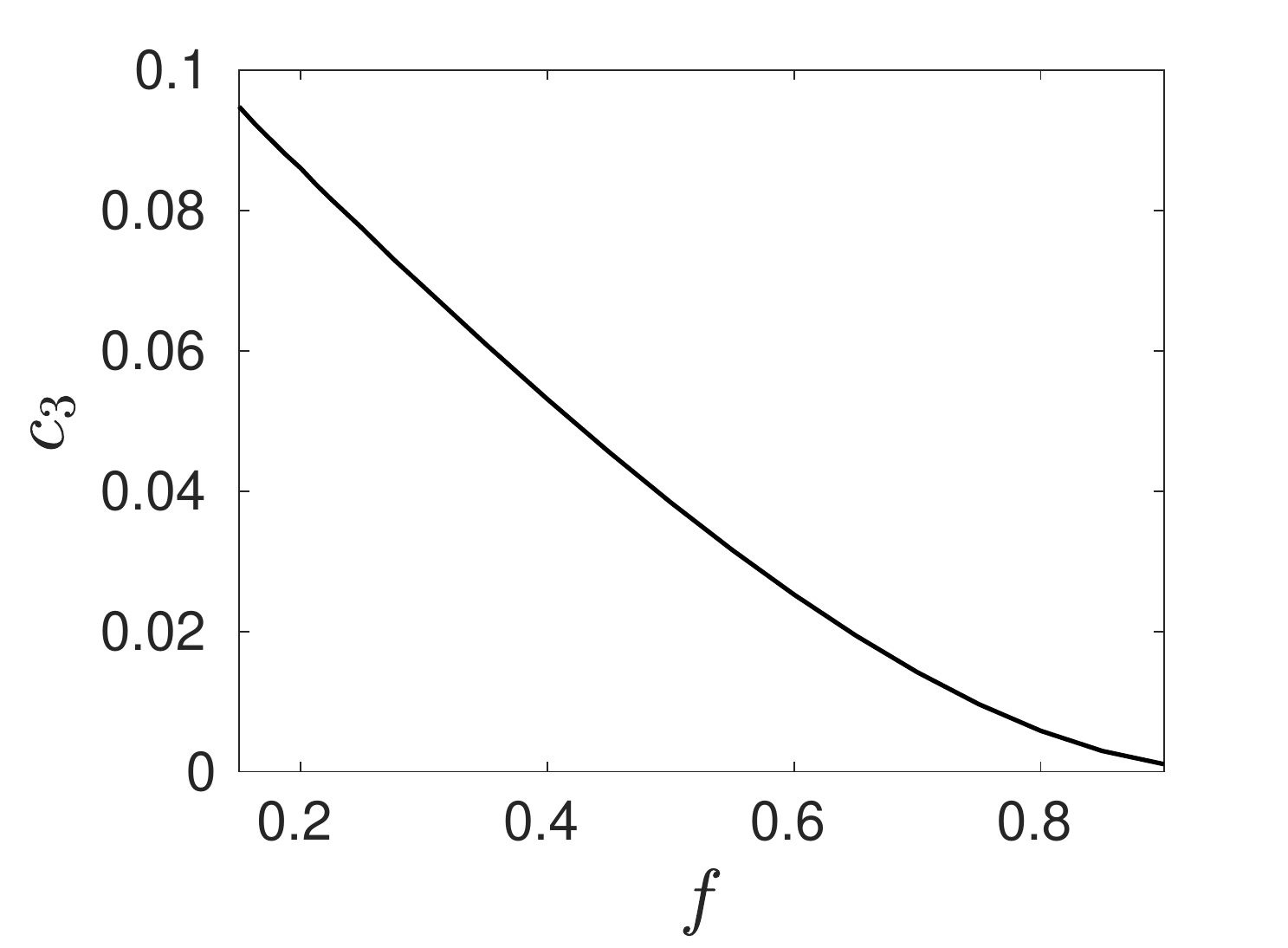} \hfill
	\caption{Numerical results for coefficients in the amplitude
          equation \eqref{bruss:coeff}. Left panel: $c_1$ versus
          $f$. Right panel: $c_3$ versus $f$. For
          $0.15 \leq f \leq 0.9$, we conclude that $c_1$
          and $c_3$ are positive. This shows that the peanut-shape
          deformation linear instability is subcritical on this
          range.}\label{fig:bruss_c1c3}
\end{figure}

In Fig.~\ref{fig:bruss_c1c3} we plot the numerically computed
coefficients $c_1$ and $c_3$ in the amplitude equation
\eqref{bruss:normalform0} versus the Brusselator parameter $f$ on
$0.15<f<0.9$. We observe that both $c_1>0$ and $c_3>0$ on this range. This
establishes that the peanut-shaped deformation of a steady-state spot is
always subcritical, and that the emerging solution branch of non-radially
symmetric spot equilibria, which exists only if $\kappa=-1$, is linearly
unstable. The steady-state amplitude of this bifurcating non-radially symmetric
solution branch is
\begin{equation}\label{bruss:small_amplitude}
  \tilde{A}_0 = \sqrt{\frac{c_1(S_c - S)}{c_3}} \,, \quad
  \mbox{valid for} \quad S_c - S = \sigma^2 \gg \mc{O}(\eps^{4/3}) \,.
\end{equation}

\begin{figure}[htbp]
	\centering
	\includegraphics[width=0.32\textwidth]{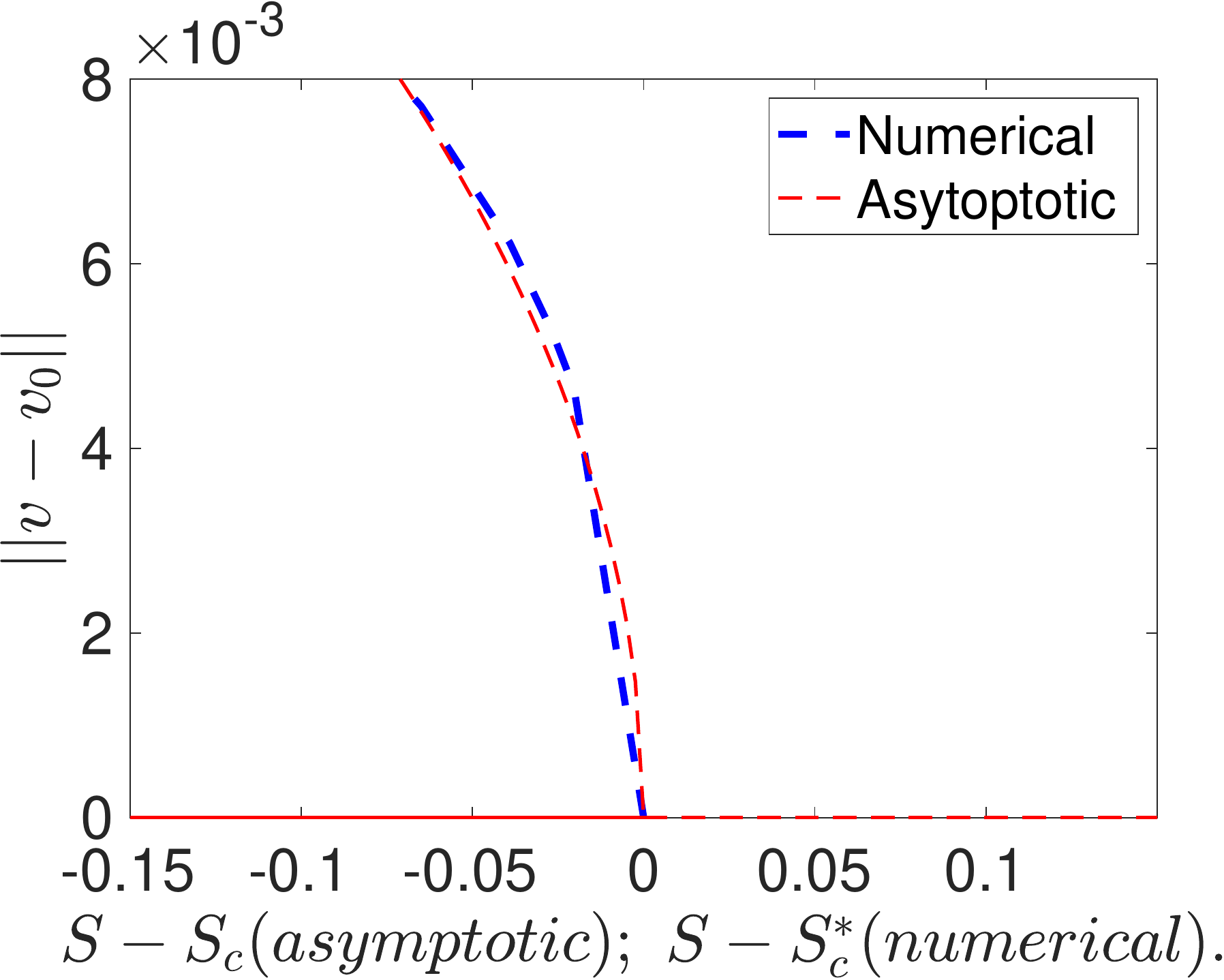} \hfill
	\includegraphics[width=0.32\textwidth]{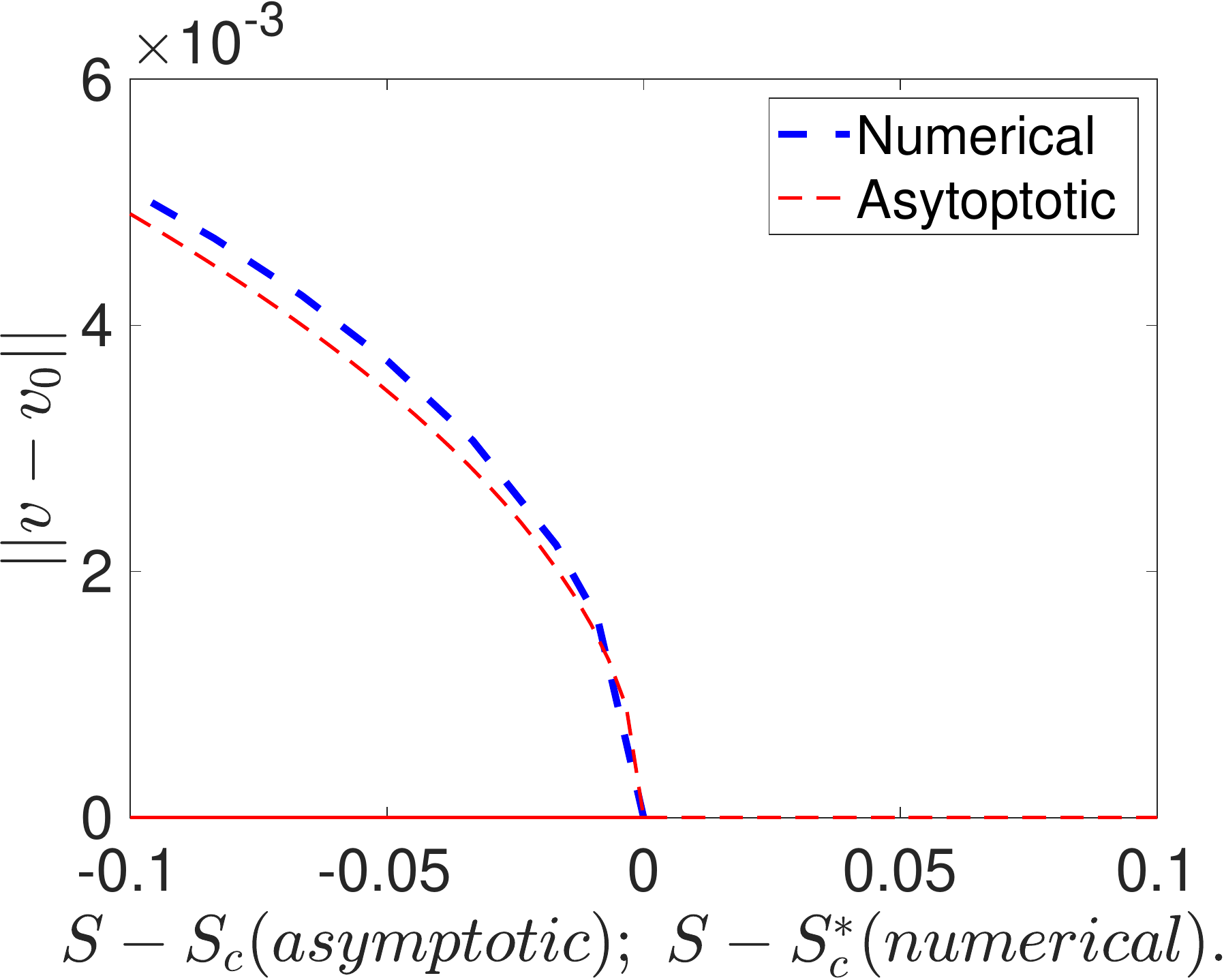} \hfill
	\includegraphics[width=0.32\textwidth]{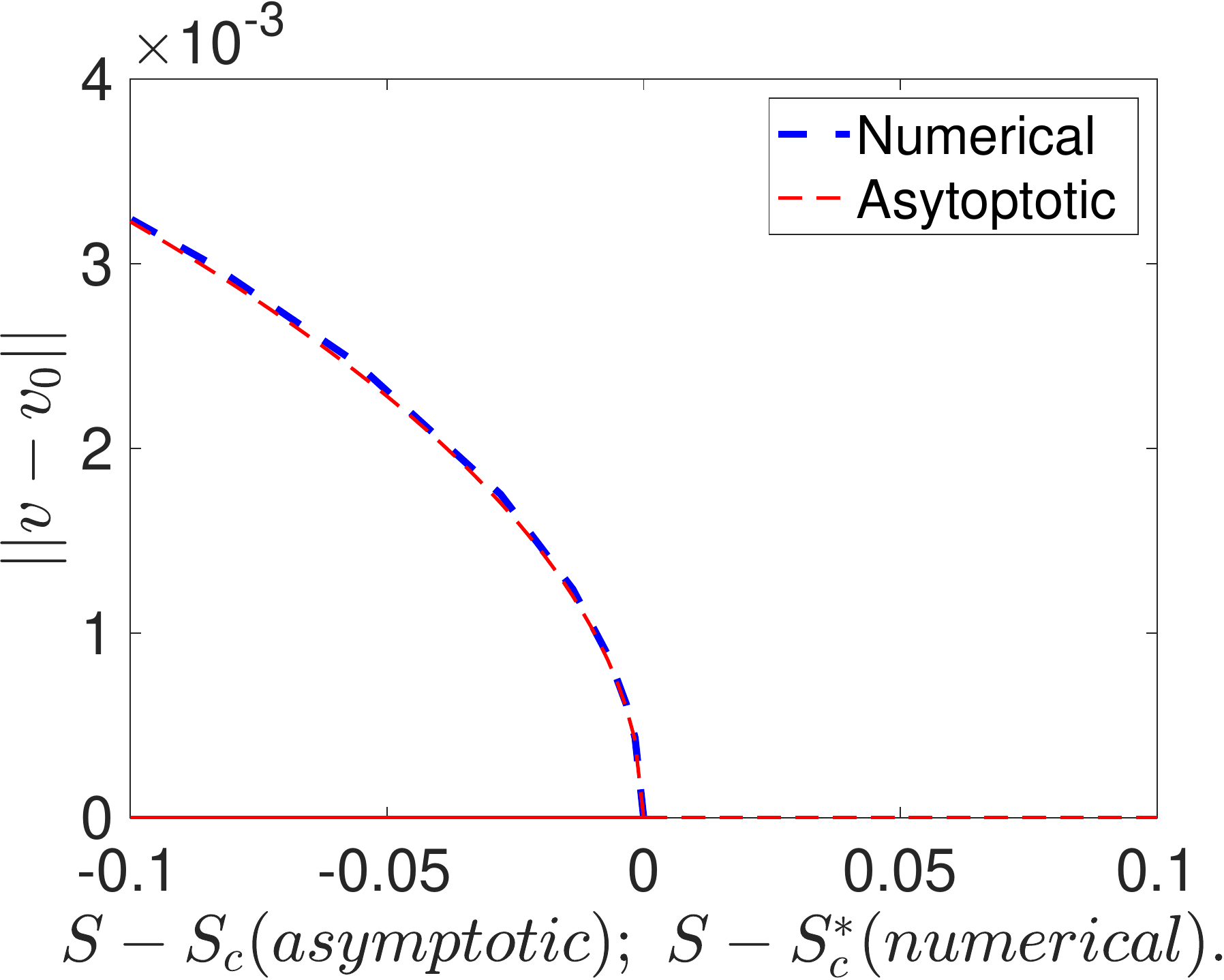} \hfill
	\caption{Plot of $||v-v_0||$ versus $S-S_c^*$ computed
          numerically from the full PDE \eqref{bruss:pde} with
          $\eps=0.01$ using {\em pde2path} \cite{pde2path}. Here
          $S_c^*$ is the numerically computed bifurcation
          value. Numerical results are compared with the asymptotic
          result $\frac{\eps\sqrt{\pi}}{2}\tilde{A}_0 =
          \frac{\eps}{2}\sqrt{\frac{\pi c_1 (S_c-S)}{c_3}}$ (see
          \ref{amp:vnormdiff}) for the steady-state amplitude, as
        given in \eqref{bruss:small_amplitude}, where $S_c$ is the
        asymptotic result computed from the eigenvalue problem
        \eqref{bruss:loc_eig} for the onset of the mode $m=2$
        peanut-shaped instability. Left panel: $f = 0.7$. Middle
        panel: $f=0.5$. Right panel: $f=0.35$.}\label{fig:bruss_valid}
\end{figure}

For three values of $f$, in Fig.~\ref{fig:bruss_valid} we favorably
compare our weakly nonlinear analysis result
\eqref{bruss:small_amplitude} with corresponding full numerical
results computed from the steady-state of the Brusselator
\eqref{bruss:pde} with $\eps=0.01$ in a quarter-disk geometry (see
Fig.~\ref{fig:eig_sc}). The full numerical results are obtained using
the continuation software {\em pde2path} \cite{pde2path}, and in
Fig.~\ref{fig:bruss_valid} we plot the norm of the deviation from the
radiallly symmetric steady state (see \eqref{sc:dnorm_all}).

\section{Discussion}\label{sec:disc}

We have developed and implemented a weakly nonlinear theory to derive
a normal form amplitude equation characterizing the branching behavior
associated with peanut-shaped non-radially symmetric linear
instabilities of a steady-state spot solution for both the
Schnakenberg and Brusselator RD systems. From a numerical computation
of the coefficients in the amplitude equation we have shown that such
peanut-shaped linear instabilities for these specific RD systems are
always subcritical. A numerical bifurcation study using pde2path
\cite{pde2path} of a localized steady-state spot was used to validate
the weakly nonlinear theory, and has revealed the existence of a
branch of unstable non-radially symmetric steady-state localized spot
solutions. Our weakly nonlinear theory provides a theoretical basis
for the observations in \cite{KWW}, \cite{RRW} and \cite{TW2016} (see
also \cite{TW}) obtained through full PDE simulations that a linear
peanut-shaped instability of a localized spot is the mechanism
triggering a fully nonlinear spot self-replication event.

We remark that instabilities resulting from non-radially symmetric
shape deformations of a steady-state localized spot solution are
localized instabilities, since the associated eigenfunction for shape
instabilities decays rapidly away from the center of a spot. As a
result, our weakly nonlinear analysis predicting a subcritical
peanut-shape instability also applies to steady-state spot patterns of
the 2-D Gray-Scott model analyzed in \cite{CW2011}, which has the same
nonlinear kinetics near a spot as does the Schnakenberg RD system.

However, an important technical limitation of our analysis is that our
weakly nonlinear theory is restricted to the consideration of
steady-state spot patterns, and does not apply to quasi-equilibrium
spot patterns where the centers of the spots evolve dynamically on
asymptotically long $\mc{O}(\eps^{-2})$ time intervals towards a
steady-state spatial configuration of spots. For such
quasi-equilibrium spot patterns there is a non-vanishing
$\mc{O}(\eps)$ feedback from the outer solution that results from the
interaction of a spot with the domain boundary or with the other spots
in the pattern. This $\mc{O}(\eps)$ feedback term then violates the
asymptotic ordering of the correction terms in our weakly nonlinear
perturbation expansion. For steady-state spot patterns there is an
asymptotically smaller $\mc{O}(\eps^2)$ feedback from the outer
solution, and so our weakly nonlinear analysis is valid for
$|S-S_c|={\mathcal O}(\sigma^2)$, under the assumption that
$\sigma^3\gg \mc{O}(\eps^2)$ (see Remark \ref{sch:equil_eps2}). Here
$S_c$ is the spot source strength at which a zero-eigenvalue crossing
occurs for a small peanut-shaped deformation of a localized spot. In
contrast, for a quasi-equilibrium spot pattern, it was shown for the
Schnakenburg model in \S 2.4 of \cite{KWW} that, when
$S-S_c=\mc{O}(\eps)$, the direction of the bulge of a peanut-shaped
linear instability is perpendicular to the instantaneous direction of
motion of a spot. This result was based on a simultaneous linear
analysis of mode $m=1$ (translation) and mode $m=2$ (peanut-shape)
localized instabilities near a spot. The full PDE simulations in
\cite{KWW} indicate that this linear instability triggers a fully
nonlinear spot-splitting event where the spot undergoes a splitting
process in a direction perpendicular to its motion. To provide a
theoretical understanding of this phenomena it would be worthwhile to
extend this previous linear theory of \cite{KWW} for quasi-equilibrium
spot patterns to the weakly nonlinear regime.

Although our weakly nonlinear theory of spot-shape deformation
instabilities has only been implemented for the Schnakenberg and
Brusselator RD systems, the hybrid analytical-numerical theoretical
framework presented herein applies more generally to other reaction
kinetics where a localized steady-state spot solution can be
constructed. It would be interesting to determine whether one can
identify other RD systems where the branching is supercritical,
thereby allowing for the existence of linearly stable non-radially
symmetric localized spot steady-states.

In another direction, for the Schnakenberg model in a 3-D spatial
domain, it was shown recently in \cite{TXKW} through PDE simulations
that a peanut-shaped linear instability is also the trigger for a
nonlinear spot self-replication event. It would be worthwhile to
extend our 2-D weakly nonlinear theory to this more intricate 3-D
setting.

\section{Acknowledgements}\label{sec:ak}
Tony Wong was supported by a UBC Four-Year Graduate
Fellowship. Michael Ward gratefully acknowledges the financial support
from the NSERC Discovery Grant program. The authors would like to thank Fr\'{e}d\'{e}ric Paquin-
	Lefebvre for some insightful discussions.

\begin{appendix}
\renewcommand{\theequation}{\Alph{section}.\arabic{equation}}
\setcounter{equation}{0}

\section{Far-field condition for $\hat{U}_{20}$ for
 the Schnakenberg model}\label{sec:far_field}

We derive the far-field condition for $\hat{U}_{20}$ used in
\eqref{amp:Vhat20} in the derivation of the amplitude equation for
peanut-splitting instabilities for the Schnakenberg model. We first
observe that the second component $\hat{U}_{20}$ of \eqref{amp:Vhat20}
satisfies
\begin{equation}\label{app:u20}
  \hat{U}_{20}^{\prime\prime} + \frac{1}{\rho} \hat{U}_{20}^{\prime} -
  V_c^2 \hat{U}_{20}
  = F_{20} + 2 U_c V_c \hat{V}_{20}\,, \quad \mbox{for} \quad \rho\ge 0\,,
\end{equation}
where $F_{20}$ is defined in \eqref{amp:F20} and where primes indicate
derivatives in $\rho$. For $\rho\to \infty$, we have from the first
equation in \eqref{amp:Vhat20} that $\Delta_\rho V_c - V_c \sim 0$
with $V_c\to 0$ as $\rho\to \infty$. This yields the asymptotic decay
behavior
\begin{equation}\label{far_field:Vc}
  V_c \sim \alpha \rho^{-1/2} e^{-\rho} \,, \quad \mbox{so that} \quad
  V_c^{\prime} \sim - \left(1 + \frac{1}{2\rho}\right) V_c\,, \quad
  \mbox{as} \quad \rho \to \infty\,,
\end{equation}
for some $\alpha>0$. As such, we impose
$V_{c}^{\prime}=-\left[1+{1/(2\rho)}\right] V_c$ at
$\rho=\rho_{m}\approx 20$ in solving \eqref{amp:Vhat20}
numerically. The constant $\alpha$ in \eqref{far_field:Vc} can be
calculated from the limit
$\alpha=\lim_{\rho\to\infty} \sqrt{\rho} \, e^{\rho}V_{c}(\rho)$. Our
numerical solution of the BVP problem \eqref{amp:Vhat20} with
$\rho_{m}=20$ yields $\alpha \approx 32.5$.

To find the asymptotic behavior for $\hat{U}_{20}$ in \eqref{app:u20}
we decompose it into homogeneous and inhomogeneous parts as
\begin{subequations}
\begin{equation}
	\hat{U}_{20} = \hat{U}_h + \hat{U}_p\,,
\end{equation}
where $\hat{U}_h$ and $\hat{U}_p$ satisfies
\begin{equation}\label{app:uh_decomp}
  \hat{U}_h^{\prime\prime} + \frac{1}{\rho} \hat{U}_h^{\prime} -
  V_c^2 \hat{U}_h = 0\,, \qquad
  \hat{U}_p^{\prime\prime} + \frac{1}{\rho} \hat{U}_p^{\prime} - V_c^2 \hat{U}_p
                  = F_{20} +  2 U_c V_c \hat{V}_{20} \,.
\end{equation}
\end{subequations}

We first estimate $\hat{U}_h$ for $\rho \to \infty$. By using
\eqref{far_field:Vc} for $V_c$, and using the dominant balance ansatz
$\hat{U}_{h} = e^R$, we obtain that \eqref{app:uh_decomp} transforms
exactly to
\begin{equation}\label{far_field:R}
  \frac{1}{\rho} \left(\rho R^{\prime}\right)^{\prime}
  + \frac{1}{\rho} R^{\prime} + (R^{\prime})^2 \sim
  \frac{\alpha^2 e^{-2\rho}}{\rho}\,,  \quad \mbox{as} \quad \rho \to \infty\,.
\end{equation} 
To estimate the asymptotic behavior of $R^{\prime}$ we apply the method
of dominant balance. The appropriate balance for $\rho\gg 1$ is found to be
$\left(\rho\,R^{\prime}\right)^{\prime} \sim \alpha^2 e^{-2\rho}$, which
yields
\begin{equation}\label{far_field:balance3}
  R^{\prime} \sim -\frac{\alpha^2 e^{-2\rho}}{2\rho} \,, \quad \mbox{for} \quad
  \rho \gg 1\,.
\end{equation}
Our leading-order balance is self-consistent since we have
$\left(R^{\prime}\right)^2 \ll \rho^{-1} \alpha^2 e^{-2\rho}$ for $\rho\gg 1$.
By integrating $R^{\prime}$ in \eqref{far_field:balance3}, we get
\begin{equation}
  R \sim \frac{\alpha^2 e^{-2\rho}}{4\rho}
  \left[1 + \mc{O}\left(\frac{1}{\rho}\right)\right] + \mbox{constant}\,, \quad
  \mbox{as} \quad \rho \to \infty\,.
\end{equation}
Therefore, we have
\begin{equation}\label{far_field:Uh}
  \hat{U}_h \sim K \left(1 + \frac{\alpha^2 e^{-2\rho}}{4\rho} \right)\,,
  \quad \mbox{as} \quad \rho \to \infty \,,
\end{equation}
for some constant $K>0$.
By differentiating the ansatz $\hat{U}_h = e^R$, followed by using the
estimates \eqref{far_field:balance3} and \eqref{far_field:Uh}, we obtain
\begin{equation}
  \hat{U}_h^{\prime} = R^{\prime}
  \, \hat{U}_h \sim -K \left(\frac{\alpha^2 e^{-2\rho}}{2\rho}
  \right) \left(1 + \frac{\alpha^2 e^{-2\rho}}{4\rho} \right)\,, \quad
  \mbox{as} \quad \rho \to  \infty\,.
\end{equation}
As a result, we conclude for the homogeneous solution $\hat{U}_h$ that
\begin{equation}\label{far_field:Uh_far_field}
  \hat{U}_h^{\prime} \to 0 \quad\mbox{exponentially as}\quad \rho \to
  \infty\,.
\end{equation}

Next, we consider the particular solution $\hat{U}_p$ satisfying
\eqref{app:uh_decomp}. We use the far field behavior
$\hat{V}_{20} = \mc{O}\left(\rho^{-1/2}e^{-\rho}\right)$,
$V_c = \mc{O}\left(\rho^{-1/2}e^{-\rho}\right)$, $U_{c}=\mc{O}(\log \rho)$,
$\Phi_c = \mc{O}\left(\rho^{-1/2}e^{-\rho}\right)$ and
$N_c =\mc{O}\left(\rho^{-2}\right)$ for $\rho\gg 1$, to deduce from
\eqref{bruss:f20} that
\begin{equation}
  F_{20} = \mc{O}\left(\rho^{-1}e^{-2\rho}\log\rho\right)\,, \quad\mbox{and}
  \quad U_c V_c \hat{V}_{20} = \mc{O}\left(\rho^{-1}e^{-2\rho}\log\rho\right)\,,
  \quad \mbox{as} \quad \rho \to \infty\,.
\end{equation}
Therefore, from \eqref{app:uh_decomp}, for $\rho\gg 1$ the particular
solution $\hat{U}_p$ satisfies
\begin{equation}
  \frac{(\rho\,\hat{U}_p^{\prime})^{\prime}}{\rho} - \mc{O}(\rho^{-1} e^{-2\rho})
  \hat{U}_p = \mc{O}\left(\rho^{-1} e^{-2\rho}\log\rho\right)\,.
\end{equation}
By balancing the first and third terms in this expression we get
\begin{equation}
  (\rho \, \hat{U}_p^{\prime})^{\prime}
  = \mc{O}(e^{-2\rho} \log \rho)\,, \quad \mbox{as}\quad \rho \to \infty\,.
\end{equation}
From this expression, we readily derive that
\begin{equation}
  \hat{U}_p^{\prime} = \mc{O}\left(\rho^{-1}e^{-2\rho}\log\rho\right)\,,
  \quad \mbox{as} \quad \rho \to \infty\,.
\end{equation}
This shows that $\hat{U}_p^{\prime} \to 0$ exponentially as
$\rho \to \infty$.  Upon combining this result with
\eqref{far_field:Uh_far_field} we conclude that
\begin{equation}
  \hat{U}_{20}^{\prime} = \hat{U}_h^{\prime} + \hat{U}_p^{\prime} \to 0\,, \quad
  \mbox{as} \quad   \rho \to \infty\,.
\end{equation}
This dominant balance analysis justifies our imposition of the
homogeneous Neumann far-field condition for $\hat{U}_{20}$ in
\eqref{amp:Vhat20} for the Schnakenberg model. An identical argument
can be performed to justify the far-field condition in
\eqref{bruss:vhat20} for the Brusselator model.

From our numerical computation of $\hat{U}_{20}$ from
\eqref{amp:Vhat20}, shown in Fig.~\ref{fig:order_two}, we observe that
$\hat{U}_{20}\to U_{20\infty}\neq 0$ as $\rho\to \infty$. We now show
how this non-vanishing limit can be accounted for in a modified outer
solution.  From \eqref{amp:expansion} we have for
$S=S_c + \kappa \sigma^2$ that
\begin{equation}
U = U_c + \sigma U_1 + \sigma^2 U_2 + \sigma^3 U_3 + \ldots\,,
\end{equation}
where $U_1=A\cos(2\phi)N_c$ from \eqref{amp:matching1}, while
$U_{2}=\kappa\, \partial_{S} U_c + A^2 \hat{U}_{20} +
A^2U_{24}\cos(4\phi)$ from \eqref{amp:V2_decomposition} and
\eqref{amp:V20_decomposition}. Since
$U_c\sim S_c\log\rho + \chi(S_c) + o(1)$ as $\rho\to \infty$, while
$N_c\to 0$ and $U_{24}\to 0$ as $\rho\to\infty$, we obtain that the
far-field behavior of $U$ is
\begin{equation}\label{app:pde_far_field}
  U \sim S_c \log \rho + \chi(S_c) + \sigma^2 \left[\kappa \log\rho +
    \kappa \chi^{\prime}(S_c) + A^2 \hat{U}_{20\infty}\right] +
  \ldots \,, \quad \mbox{as} \quad \rho = |\v{y}| \to \infty\,,
\end{equation}
which specifies the $\mc{O}(1)$ term in \eqref{amp:pde_far_field}. Since
$u={U/\sqrt{D}}$ and $S = {a/(2\sqrt{D})}$ from \eqref{steady:S_2}, the
modified outer solution has the form
\begin{equation}\label{app:uall}
  u = \frac{1}{\sqrt{D}} \left(S_c \log|\v{x}| - \frac{S_c|\v{x}|^2}{2} +
    \chi(S_c) + \frac{S_c}{\nu}\right) + \sigma^2 u_1 + o(\sigma^2) \,,
\end{equation}
where, in the unit disk $\Omega$, $u_1$ satisfies
\begin{subequations}\label{app:u1}
\begin{align}
  \Delta u_1 &= - \frac{2\kappa}{\sqrt{D}} \,, \quad \mbox{in} \quad \v{x}\in
               \Omega\backslash \lbrace{\v{0}\rbrace} \,; \qquad
               \partial_n u_1 = 0 \,, \quad \v{x}\in
  \partial\Omega \,,\\
  u_1 &\sim \frac{1}{\sqrt{D}} \left( \kappa \log|\v{x}| + \frac{\kappa}{\nu}
  + \kappa \chi^{\prime}(S_c) + A^2 \hat{U}_{20\infty} \right) + o(1) \,, \quad
  \mbox{as} \quad \v{x}\to \v{0} \,,
\end{align}
\end{subequations}
where $\nu={-1/\log\eps}$. To complete the expansion in \eqref{app:uall}
we solve \eqref{app:u1} to get
\begin{equation}\label{app:u1sol}
  u_1 =\frac{1}{\sqrt{D}} \left( \kappa \log|\v{x}| + \frac{\kappa}{\nu}
    -\frac{\kappa |\v{x}|^2}{2} + \kappa \chi^{\prime}(S_c) +
    A^2 \hat{U}_{20\infty}  \right) \,.
\end{equation}
In this way, the non-vanishing limiting behavior of $\hat{U}_{20}$ as
$\rho\to \infty$ leads to only a simple modification of the outer
solution as given in \eqref{steady:u_sol}.

Finally, we remark that an identical modification of the outer
expansion for the Brusselator model can be done when deriving the
amplitude equation for peanut-shaped instability of a localized spot.

\end{appendix}


\begin{thebibliography}{99}

\bibitem{as1} Y.~A.~Astrov, H.~G.~Purwins, {\em Spontaneous division
    of dissipative solitons in a planar gas-discharge system with high
    ohmic electrode}, Phys. Lett. A, {\bf 358}(5-6), (2006),
  pp.~404--408.

\bibitem{brena} D.~Avitabile, V.~Brena-Medina, M.~J.~Ward, {\em
Spot dynamics in a plant hair initiation model}, SIAM J. Appl. Math.,
{\bf 78}(1), (2018), pp.~291--319.

\bibitem{callahan} T.~K.~Callahan, {\em Turing patterns
  with O(3) symmetry}, Physica D, {\bf 188}(1), (2004), pp.~65--91.

\bibitem{CW2011} W.~Chen, M.~J.~Ward, {\em The stability and dynamics of 
localized spot patterns in the two-dimensional Gray-Scott model}, 
SIAM J. Appl. Dyn. Sys., {\bf 10}(2), (2011), pp.~582--666.

\bibitem{ch} M.~Cross, P.~Hohenburg, {\em Pattern formation outside of
equilibrium}, Rev. Mod. Physics, {\bf 65}, (1993), pp.~851-1112.

\bibitem{blobs} P.~W.~Davis, P.~Blanchedeau, E.~Dullos and P.~De Kepper
(1998), {\em Dividing blobs, chemical flowers, and patterned
islands in a reaction-diffusion system}, J.~Phys.~Chem. A, {\bf
102}(43), pp.~8236--8244.

\bibitem{dgk1} A.~Doelman, R.~A.~Gardner, T.~J.~Kaper, {\em Stability 
analysis of singular patterns in the 1D Gray-Scott model: A matched
asymptotics  approach}, Physica D, {\bf 122}(1-4), (1998), pp.~1-36.

\bibitem{enu} S.~Ei, Y.~Nishiura, K.~Ueda, {\em $2^{n}$ splitting or
edge splitting?: A manner of splitting in dissipative systems},
Japan. J. Indus. Appl. Math., {\bf 18}, (2001), pp.~181-205.

\bibitem{gomez} D.~Gomez, L.~Mei, J.~Wei, {\em Stable and unstable periodic
    spiky solutions for the Gray-Scott system and the Schnakenberg system},
  to appear, J. Dyn. Diff. Eqns. (2020).

\bibitem{K} E.~Knobloch, {\em Spatial localization in dissipative systems},
Annu. Rev. Cond. Mat. Phys., {\bf 6}, (2015), pp.~325--359.

\bibitem{kww_gs} T.~Kolokolnikov, M.~Ward, J.~Wei, {\em The stability
    of spike equilibria in the one-dimensional Gray-Scott model: the
    pulse-splitting regime}, Physica D, {\bf 202}(3-4), (2005),
  pp.~258--293.

\bibitem{kww_gm} T.~Kolokolnikov, M.~J.~Ward, J.~Wei, {\em 
  Pulse-splitting for some reaction-diffusion systems in one-space dimension}, 
  Studies in Appl. Math., {\bf 114}(2), (2005), pp.~115--165.

\bibitem{KWW} T.~Kolokolnikov, M.~J.~Ward, J.~Wei, {\em Spot
self-replication and dynamics for the Schnakenberg model in a
two-dimensional domain}, J. Nonlinear Sci., {\bf 19}(1), (2009),
pp.~1--56.

\bibitem{LMPS} K.~J.~Lee, W.~D.~McCormick, J.~E.~Pearson, H.~L.~Swinney,
{\em Experimental observation of self-replicating spots in a 
reaction-diffusion system}, Nature, {\bf 369}, (1994), pp.~215-218.

\bibitem{ls} K.~J.~Lee, H.~Swinney, {\em Lamellar structures and 
self-replicating spots in a reaction-diffusion system}, Phys. Rev. E, 
{\bf 51}(3), (1995), pp.~1899--1915.

\bibitem{comp} F.~Paquin-Lefebvre, T.~Kolokolnikov, M.~J.~Ward, {\em
    Competition instabilities of pulse patterns for the 1-d
    Gierer-Meinhardt model are subcritical}, to be submitted, SIAM J.
   Appl. Math., (2020).

\bibitem{M1} P.~C.~Matthews, {\em Transcritical bifurcation with
$O(3)$ symmetry}, Nonlinearity, {\bf 16}(4), (2003), pp.~1449--1471.

\bibitem{M2} P.~C.~Matthews, {\em Pattern formation on a sphere},
  Phys. Rev. E., {\bf 67}(3), (2003), pp.~036206.

\bibitem{MO1} C.~Muratov, V.~V.~Osipov, {\em Static spike autosolitons in
the Gray-Scott model}, J. Phys. A: Math Gen. {\bf 33}, (2000),
pp.~8893--8916.

\bibitem{MO2} C.~Muratov, V.~V.~Osipov, {\em Spike autosolitons and pattern
formation scenarios in the two-dimensional Gray-Scott model}, Eur.~Phys.~J.~B.
{\bf 22}, (2001), pp.~213--221.

\bibitem{nishiura} Y.~Nishiura, {\em Far-from equilibrium dynamics,
  translations of mathematical monographs}, Vol.~{\bf 209}, (2002),
  AMS Publications, Providence, Rhode Island.

\bibitem{nu1} Y.~Nishiura, D.~Ueyama, {\em A skeleton structure of 
self-replicating dynamics}, Physica D, {\bf 130}(1-2), (1999), pp.~73-104.

\bibitem{NTU}, Y.~Nishiura, T.~Teramoto, K.~I.~Ueda, {\em Scattering
    of traveling spots in dissipative systems}, Chaos {\bf 15}(4),
  047509 (2005).
  
\bibitem{bulk_memb} F.~Paquin-Lefebrve, W.~Nagata, M.~J.~Ward, {\em
Pattern formation and oscillatory dynamics in a 2-d coupled bulk-surface
reaction-diffusion system}, SIAM J. Appl. Dyn. Sys., {\bf 18}(3), (2019),
pp.~1334-1390.

\bibitem{pearson} J.~E.~Pearson, {\em Complex patterns in a simple system},
Science, {\bf 216}, (1993), pp.~189--192.

\bibitem{PL} I.~Prigogine, R.~Lefever, {\em Symmetry breaking
  instabilities in dissipative systems. {II}}, J. Chem. Physics, {\bf
  48}, (1968), pp.~1695.

\bibitem{rpd} W.~N.~Reynolds, S.~Ponce-Dawson, J.~E.~Pearson, {\em
    Self-replicating spots in reaction-diffusion systems},
  Phys. Rev. E, {\bf 56}(1), (1997), pp.~185-198.

\bibitem{RRW} I.~Rozada, S.~Ruuth, M.~J.~Ward, {\em The stability of
 localized spot patterns for the Brusselator on the sphere}, SIAM J. Appl. 
Dyn. Sys., {\bf 13}(1), (2014), pp.~564--627.

\bibitem{TSN} T.~Teramoto, K.~Suzuki, Y.~Nishiura, {\em Rotational
    motion of traveling spots in dissipative systems}, Phys. Rev. E.
  {\bf 80}(4):046208, (2009).
  
\bibitem{TW} P.~Trinh, M.~J.~Ward, {\em The dynamics of localized
spot patterns for reaction-diffusion systems on the sphere}, 
Nonlinearity, {\bf 29}(3), (2016), pp.~766--806.

\bibitem{turing} A.~Turing, {\em The chemical basis of morphogenesis},
 Phil. Trans. Roy. Soc. B, {\bf 327}, (1952), pp.~37--72.

\bibitem{TW2016} J.~Tzou, M.~J.~Ward, {\em Effect of open systems on the
existence, stability, and dynamics of spot patterns in the 2D Brusselator
model}, Physica D, {\bf 373}, (2018),  pp.~13--37.

\bibitem{TXKW} J.~Tzou, S.~Xie, T.~Kolokolnikov, M. J. Ward, {\em The
stability and slow dynamics of localized spot patterns for the 3D
Schnakenberg reaction-diffusion model}, SIAM J. Appl. Dyn. Sys.,
{\bf 16}(1), (2017), pp.~294--336.

\bibitem{pde2path} H.~Uecker, D.~Wetzel, J.~D.~Rademacher, {\em
    Pde2path-A Matlab package for continuation and bifurcation in 2D elliptic
    systems}, Numerical Mathematics: Theory, Methods and Applications,
    {\bf 7}(1), (2014), pp.~58--106.

\bibitem{u} D.~Ueyama, {\em Dynamics of self-replicating patterns in the
one-dimensional Gray-Scott model}, Hokkaido Math J., {\bf 28}(1),
(1999), pp.~175-210.

\bibitem{vanag} V.~K.~Vanag, I.~R.~Epstein, {\em Localized patterns in
reaction-diffusion systems}, Chaos {\bf 17}(3), 037110, (2007).

\bibitem{veer} F.~Veerman, {\em Breathing pulses in singularly perturbed
reaction-diffusion systems}, Nonlinearity, {\bf 28}, (2015), pp.~2211-2246.

\bibitem{w} D.~Walgraef, {\em Spatio-temporal pattern formation, with
    examples from physics, chemistry, and materials science}, {\em in book
  series, Partially ordered systems},  Springer, New York, (1997), 306 p.

\bibitem{ward} M.~J.~Ward, {\em Spots, traps, and patches:
asymptotic analysis of localized solutions to some linear and nonlinear
diffusive processes}, Nonlinearity, {\bf 31}(8), (2018), R189 (53 pages).

\bibitem{wgs3} J.~Wei, M.~Winter, {\em Existence and stability of
  multiple spot solutions for the Gray-Scott model in $\mathbb{R}^2$},
  Physica D, {\bf 176}(3-4), (2003), pp.~147-180.

\bibitem{ww_schnak} J.~Wei, M.~Winter, {\em Stationary multiple
  spots for reaction-diffusion systems}, J. Math. Biol., {\bf 57}(1),
  (2008), pp.~53--89.

\bibitem{wei-book} J.~Wei, M.~Winter, {\em Mathematical aspects of pattern
formation in biological systems}, Applied Mathematical Science Series,
Vol. 189, Springer, (2014).

\bibitem{wh} R.~Wittenberg, P.~Holmes, {\em The limited effectiveness of
normal forms: a critical review and extension of local bifurcation studies
of the Brusselator PDE}, Physica D, {\bf 100}(1-2), (1997), pp.~1--40.

\end{thebibliography}
\end{document}